\documentclass[12pt,dvipsnames]{article}
\usepackage[top=80pt,bottom=85pt,left=85pt,right=85pt]{geometry}

\pdfoutput=1
\usepackage[utf8]{inputenc}
\usepackage[T1]{fontenc} 
\usepackage[english]{babel} 
\usepackage{braket}
\usepackage[dvipsnames]{xcolor} 
\usepackage{physics}
\usepackage{tabularx} 
\usepackage{ragged2e}
\usepackage{booktabs} 
\usepackage{rotating}
\usepackage{makecell,booktabs}
\usepackage{multirow} 
\usepackage{diagbox}
\usepackage{comment}
\usepackage{mathtools}
\usepackage{amsmath}
\usepackage{cite}
\usepackage{pifont}
\usepackage{bbding}
\usepackage{amssymb}
\usepackage[all]{xy}

\usepackage{tabularray}

\usepackage{tikz}

\usetikzlibrary{math} 
\usetikzlibrary{3d} 
\usepackage{tikz-3dplot}
\usepackage{tikz-cd} 
\usetikzlibrary{decorations.pathreplacing}
\usepackage{microtype}
\usepackage{amsthm}
\usepackage{mathrsfs} 
\usepackage[strict]{changepage}
\newcommand\scalemath[2]{\scalebox{#1}{\mbox{\ensuremath{\displaystyle #2}}}}

\tikzcdset{
  cells={font=\everymath\expandafter{\the\everymath\displaystyle}},
}

\newcommand{\xdownarrow}[1]{%
  {\left\downarrow\vbox to #1{}\right.\kern-\nulldelimiterspace}
}

\usepackage{amssymb} 
\usepackage{subcaption} 
\DeclareCaptionFormat{custom}

\usepackage[pdftex,colorlinks=true]{hyperref} 
\hypersetup{urlcolor=MidnightBlue, citecolor=burntorange, linkcolor=darkmagenta}

\usepackage{float} 
\usepackage{todonotes}

\usepackage[capitalise]{cleveref}

\newcommand*\circled[1]{\tikz[baseline=(char.base)]{
            \node[shape=circle,draw,inner sep=2pt] (char) {#1};}}

\newcolumntype{E}{>{\hfil$}p{1.65cm}<{$\hfil}}

\newcolumntype{L}{>{\hfil$}p{16cm}<{$\hfil}}

\newcolumntype{D}{>{\hfil$}p{7.4cm}<{$\hfil}}
\newcolumntype{C}{>{\hfil$}p{3cm}<{$\hfil}}
\newcolumntype{P}{>{\hfil$}p{7.7cm}<{$\hfil}}
\newcolumntype{F}{>{\hfil$}p{5.7cm}<{$\hfil}}
\newcolumntype{S}{>{\hfil$}p{1.8cm}<{$\hfil}}
\newcolumntype{R}{>{\hfil$}p{5.2cm}<{$\hfil}}
\newcolumntype{U}{>{\hfil$}p{4.2cm}<{$\hfil}}
\newcolumntype{Q}{>{\hfil$}p{6.4cm}<{$\hfil}}
\newcolumntype{T}{>{\hfil$}p{1.9cm}<{$\hfil}}
\newcolumntype{V}{>{\hfil$}p{5.8cm}<{$\hfil}}
\newcolumntype{H}{>{\hfil$}p{1.8cm}<{$\hfil}}
\newcolumntype{A}{>{\hfil$}p{6cm}<{$\hfil}}
\newcolumntype{B}{>{\hfil$}p{2cm}<{$\hfil}}

\numberwithin{equation}{section}

\definecolor{byzantine}{rgb}{0.74, 0.2, 0.64}
\definecolor{burntorange}{rgb}{0.8, 0.39, 0.0}
\definecolor{cambridgeblue}{rgb}{0.64, 0.76, 0.68}
\definecolor{caribbeangreen}{rgb}{0.0, 0.8, 0.6}
\definecolor{celadon}{rgb}{0.67, 0.88, 0.69}
\definecolor{champagne}{rgb}{0.97, 0.91, 0.81}
\definecolor{cream}{rgb}{1.0, 0.99, 0.82}
\definecolor{cyan(process)}{rgb}{0.0, 0.72, 0.92}
\definecolor{brilliantlavender}{rgb}{0.96, 0.73, 1.0}
\definecolor{candypink}{rgb}{0.89, 0.44, 0.48}
\definecolor{darkmagenta}{rgb}{0.55, 0.0, 0.55}

\begin{document}

\begin{titlepage}

\phantom{wowiezowie}

\vspace{-1cm}

\begin{center}

{\Huge {\bf 5d Trinions and Tetraons}}

\vspace{1cm}

{\Large  Mario De Marco,$^{\sharp}$ Michele Del Zotto,$^{\dagger\ddagger*}$}\\ 

\medskip

{\Large  Michele Graffeo,$^{\star}$ and Andrea Sangiovanni $^{\dagger\ddagger*}$}\\

\vspace{1cm}

{\it
{\small

{\footnotesize$^\dagger$ Mathematics Institute, Uppsala University, \\ Box 480, SE-75106 Uppsala, Sweden}\\
\vspace{.25cm}
{\footnotesize$^\ddagger$ Department of Physics and Astronomy, Uppsala University,\\ Box 516, SE-75120 Uppsala, Sweden}\\
\vspace{.25cm}
{\footnotesize$^{*}$ Center for Geometry and Physics, Uppsala University,\\ Box 516, SE-75120 Uppsala, Sweden}\\
\vspace{.25cm}
{\footnotesize $^{\star}$ Dipartimento di Matematica, Politecnico di Milano \\
Piazza Leonardo da Vinci 32, 20133 Milano, Italy} \\
\vspace{.25cm}
{\footnotesize $\sharp$ Physique Th\'eorique et Math\'ematique and International Solvay Institutes\\
Universit\'e Libre de Bruxelles, C.P. 231, 1050 Brussels, Belgium}
\vspace{.25cm}
}}

\vskip .5cm
{\footnotesize \tt mario.de.marco@ulb.be \hspace{1cm} michele.delzotto@math.uu.se } \\
{\footnotesize \tt    michele.graffeo@polimi.it \hspace{1cm} andrea.sangiovanni@math.uu.se}

\vskip 1cm
     	{\bf Abstract }
\vskip .1in

\end{center}

\noindent 
Recently, an atomic classification scheme of 5d SCFTs has been proposed, relying on the identification of indecomposable building blocks that can be fused together to produce large classes of 5d SCFTs. These novel SCFTs are known as bifundamental 5d conformal matter theories, and their fusion produces \textit{linear} generalized quivers. We generalize such picture employing M-theory geometric engineering to construct trinions and tetraon 5d SCFTs with flavor symmetry of type D. They correspond to \textit{non-linear} generalized quivers displaying novel patterns of instantonic symmetry enhancement, and their fusion produces singular geometries that often are non-toric non-complete intersections. Finally, within our setup, we rule out trinions and tetraons of type E.

\eject

\end{titlepage}

{
  \hypersetup{linkcolor=black}
  \tableofcontents
}

\section{Introduction}
\label{sec:intro}
The discovery in the mid-nineties of non-trivial interacting superconformal fixed points (SCFTs) in spacetime dimension greater than four \cite{Witten:1995ex,Strominger:1995ac,Witten:1995em,Ganor:1996mu,Seiberg:1996qx,Seiberg1996,Morrison_1997,Douglas:1996xp} (and up to six \cite{Nahm:1977tg}) has initiated an intense effort devoted to understanding their properties and features. Yet remaining elusive and mysterious, these models have many consequences and implications --- e.g. for lower dimensional field theories obtained via dimensional reduction --- which motivates work towards their classification. For theories with 8 supercharges, an effective approach is to identify a set of irreducible SCFTs that act as elementary building blocks out of which all other SCFTs in the same dimension can be obtained. These endeavours, that can be subsumed under the name of \textit{atomic classifications} \cite{Heckman:2015bfa},
 have met with varying degrees of success depending on the spacetime dimension. This paper is in the context of the atomic classification of 5d SCFT proposed in \cite{DeMarco:2023irn}\footnote{An all-embracing classification of 5d SCFTs has been a highly coveted prize since the seminal papers \cite{Ganor:1996mu,Seiberg:1996qx,Seiberg1996,Ganor:1996pc,Douglas:1996xp,Aharony:1997ju,Aharony:1997bh,DeWolfe:1999hj,Morrison_1997,Intriligator_1997,leung1997branes}. A wide set of technologies has been developed to tackle this classification problem, including bottom-up constructions of consistent shrinkable local CY3 geometries \cite{Jefferson:2017ahm,
Jefferson:2018irk}, the circle reduction of 6d $\mathcal{N}=(1,0)$ SCFTs with an F-theory origin \cite{DelZotto:2017pti,Bhardwaj:2018yhy,Bhardwaj:2018vuu,Bhardwaj:2019xeg,Bhardwaj:2020gyu,Apruzzi:2019opn,
Apruzzi_2020}, 5-brane webs and GTPs \cite{Benini:2009gi,Bergman:2013aca,Zafrir:2014ywa,Hayashi:2015zka,Hayashi:2015fsa,Bergman:2015dpa,Hayashi:2018lyv,Hayashi:2018bkd,Hayashi:2019yxj,Hayashi_2020,Bergman:2020myx,Alexeev:2024bko,Franco:2023flw,Arias-Tamargo:2024fjt,CarrenoBolla:2024fxy,Kim:2025qaf}, as well as M-theory geometric engineering on canonical Calabi-Yau threefolds (CY3) \cite{Xie:2017pfl,
Closset:2020scj,
Closset:2020afy,
Closset:2021lwy,
Collinucci:2021ofd,
Collinucci:2021wty,
DeMarco:2021try,
Tian:2021cif,
Closset:2022vjj,
Collinucci:2022rii,DeMarco:2022dgh,
Closset:2023pmc,
Mu:2023uws,
Bourget:2023wlb,Najjar:2024vmm}.}. The main objective of this work is to give a construction and study some basic properties of specific families of 5d SCFTs, called \textit{5d  trinions} and \textit{5d tetraons}, exploiting their geometric realization in M-theory. Our readers might wonder what is special about these 5d SCFTs and why it is interesting to know about them. To address these questions requires a little detour in the general structure of atomic classification, to which we now turn.
 
 \medskip
 
 As the name suggests, atomic classifications rely on the identification of SCFTs that behave as elementary irreducible building blocks, that (following \cite{Bourget:2026ono}) we distinguish in \textit{atoms} and \textit{hybrids}. Our working definition of  superconformal atoms and hybrids is as follows. Fix an unordered collection of $n \geq 1$ simple Lie algebras $(\mathfrak{g}_1,...,\mathfrak{g}_n)$. 
 \begin{itemize}
 \item We say that an SCFT with flavor symmetry $\prod_{i=1}^n \mathfrak{g}_i$ has \textit{valency} $n$ and we call $(\mathfrak{g}_1,...,\mathfrak{g}_n)$ its \textit{type};\footnote{\, For the scope of this work, these are 5d SCFTs arising at the collision of a collection of $n$ non-compact singular lines, each supporting a singularity of type $\mathfrak{g}_i$. When $\mathfrak{g}_i$ is not a simply-laced Lie algebra a discrete flux for $C_3$ needs to be specified as well -- see \cite{Tachikawa:2015wka,Cvetic:2024mtt} for details. In this work we focus on the simply laced case, the non-simply laced cases are the subject of a follow-up work. In the algebras defining the type of a valency we omit possible abelian factors. The rationale for this definition is that we will be primarily interested in flavor algebras that can be constructed colliding singularities from a CY3 point of view, i.e.\ non-abelian algebras.}
 \item Fix $1\leq \ell \leq n$. An \textit{irreducible SCFT of type} $(\mathfrak{g}_1,...,\mathfrak{g}_\ell)$ is a SCFT with valency $n$ of type $(\mathfrak{g}_1,...,\mathfrak{g}_n)$ that it does not exhibit any gauge group (in a generalized gauge theory phase in 5d or along a 6d tensor branch) with a Lie algebra among $\mathfrak{g}_1,...,\mathfrak{g}_\ell$. 
 \item We call a non-irreducible SCFT with valency $n$ of type $(\mathfrak{g}_1,...,\mathfrak{g}_n)$  
 a \textit{molecule}. 
 \item We call \textit{superconformal hybrid of type} $(\mathfrak{g}_1,...,\mathfrak{g}_\ell)$ an irreducible SCFT that has a Higgs branch RG flow to a molecule. 
\item We call \textit{superconformal atom of type} $(\mathfrak{g}_1,...,\mathfrak{g}_\ell)$ an irreducible SCFT that has \textit{no} Higgs branch RG flow to a molecule.
\end{itemize}
In other words, an atom or a hybrid of type $(\mathfrak{g}_1,...,\mathfrak{g}_\ell)$ cannot be obtained by a diagonal gauging --- or, more generally, by a diagonal fusion \cite{Heckman:2018pqx,DelZotto:2018tcj} --- of other SCFTs that involve a group with Lie algebra among $\mathfrak{g}_1,...,\mathfrak{g}_\ell$. Superconformal molecules can be described in terms of suitable generalized quivers that have edges corresponding to atoms and hybrids, and nodes corresponding to diagonal gaugings/fusions. Once a superconformal atom or hybrid has been identified, one has to study its \textit{chemistry}, \textit{i.e.} its fusion properties with other atoms, hybrids or molecules. 

\medskip

The 6d SCFTs have an atomic classification where molecules are described by generalized linear quivers \cite{Heckman:2015bfa} (see also \cite{Bhardwaj:2015oru,Bhardwaj:2019hhd}) with 6d conformal matter building blocks \cite{DelZotto:2014hpa,Mekareeya:2017sqh}. The 4d SCFTs of class ${\mathcal S}$ also have an atomic classification where atoms of valency three (known as trinions or fixtures) are key --- the corresponding generalized quivers can have intriguing non-trivial topologies leading to $\mathcal N=2$ dualities \cite{Gaiotto:2009we,Gaiotto:2009hg,Chacaltana:2012ch,Chacaltana:2011ze,Chacaltana:2012zy,Chacaltana:2013oka,Chacaltana:2015bna,Chacaltana:2017boe,Chacaltana:2018vhp}. In an atomic classification it is important to establish whether superconformal molecules described by generalized quivers with more interesting topologies are allowed. The latter are often realized using SCFTs that have valency $\geq 3$ as well as an interesting chemistry. For this reason establishing the existence of SCFTs with higher valency is among the key questions for atomic classifications. These SCFTs are called \textit{trinions} (valency 3) and \textit{tetraons} (valency 4). For 5d SCFTs we have examples in type $A$ where trinions arise \cite{Benini:2009gi} leading to generalized quivers with trivalent vertices \cite{Hayashi:2018iqb,Eckhard:2020jyr} but there are no known trinions of types $D$ and $E$, and it is therefore interesting to try to construct them and, when successful, study the properties of the resulting theories. This is the main aim of this paper. The interested readers can find below a concise summary of our main results.

\subsection{Summary of results}





The main tool we use to attack the question of the existence of 5d trinions and tetraons is the M-theory geometric engineering of 5d SCFTs \cite{Seiberg:1996bd,Morrison:1996xf,Douglas:1996xp}. In that context, often a model with flavor symmetry of type $\mathfrak g$ corresponds to a 3-fold canonical singularity with a non-compact curve supporting an ADE singularity of type $\mathfrak g$. 5d bifundamental conformal matter theories of type $(\mathfrak g,\mathfrak g)$ indeed arise at the collision of two such loci \cite{DeMarco:2023irn}. Searching for a geometric realization of 5d trinions and 5d tetraons it is therefore natural to look at geometric singularities at finite distance in CY moduli space corresponding to the collision of three, four or more singularities of $A$, $D$ and $E$ types. When one such singularity is at finite distance in moduli space \cite{Gukov_2000}, it corresponds to a geometric realization of a 5d trinion SCFT (collision of 3 curves of singularities) or of a 5d tetraon SCFT (collision of 4 curves of singularities). We find the following infinite families of 5d trinions and tetraons that are not of type $A$:
\begin{itemize}
\item 5d trinion SCFTs of types $(D_{2j + 1},D_{k},D_{k})$, $(A_{2j + 1},D_{k},D_{k})$ and $(E_7,D_{k},D_{k})$, and 
\item 5d tetraon SCFTs of type $(D_{2j},D_{k},D_{k},D_{k})$,
\end{itemize}
where $k \geq 4$ and $j\geq 2$ are positive integers. We also find an isolated sporadic 5d trinion of type $(D_4,D_4,D_4)$ which is a canonical singularity, and two geometries that would correspond to sporadic 5d trinions of types $(E_6,E_6,E_6)$ and of type $(D_5,D_5,D_5)$, but are not canonical singularities. Within the reach of our techniques, trinions or tetraons of type $E$ do not appear (nor do SCFTs of high valency and exceptional flavor symmetry types).\footnote{\ This result can be explained (at a naive heuristic level) with the conjecture that 5d SCFTs have a 6d $\mathcal{N}=(1,0)$ origin. Indeed, from the 6d atomic classification, the largest exceptional factor of the symmetry group is $E_8 \times E_8$ \cite{Heckman:2015bfa}. By exploiting a circle reduction to 5d, the resulting theory will have a global symmetry which is at most a subgroup of $E_8 \times E_8$ corresponding to a maximal subalgebra (obtained by breaking the global symmetry via a flavor Wilson line along the KK circle). This argument rules out the possibility of having a 5d flavor symmetry consisting of three exceptional groups generated in a standard way. We thank Kantaro Ohmori for this remark.}

\medskip

\begin{figure}
    \centering
    \scalebox{0.8}{
    \begin{tikzpicture}
        \draw[thick] (0,0) circle (0.65);
        \node at (0,0) {\small$\mathfrak{g}$};
        \draw[thick,double] (0.7,0)--(1.3,0);
        \draw[thick] (2,0) circle (0.65);
        \node at (2,0) {\small$\mathfrak{g}$};
        \draw[thick,double] (-0.7,0)--(-1.3,0);
        \draw[thick,double] (2.7,0)--(3.3,0);
        \node at (4,0) {$\ldots$};
        \draw[thick,double] (4.7,0)--(5.3,0);
        \draw[thick] (6,0) circle (0.65);
        \draw[thick,double] (6.7,0)--(7.4,0);
        \draw[thick] (7.5,-0.5)--(8.5,-0.5)--(8.5,0.5)--(7.5,0.5)--cycle ;
        \node at (6,0) {\small${\mathfrak g}$};
        \node at (8,0) {\small${\mathfrak g}$};
        \draw[thick] (-2,0) circle (0.65);
        \draw[thick] (0,2) circle (0.65);
        \node at (-2,0) {\small $\emptyset$};
        \node at (0,2) {\small $\emptyset$};
        \node at (0.5,1) {\small$\mathcal{T}_{\mathfrak{g}}$};
        \node at (-1,-0.5) {\small$\mathcal{T}_{\mathfrak{g}}$};
         \node at (1,-0.5) {\small$X_{\mathfrak{g}}^{(1)}$};
        \node at (3,-0.5) {\small$X_{\mathfrak{g}}^{(1)}$};
        \node at (5,-0.5) {\small$X_{\mathfrak{g}}^{(1)}$};
        \node at (7,-0.5) {\small$X_{\mathfrak{g}}^{(1)}$};
        \draw[thick,double] (0,0.7)--(0,1.3);
        \end{tikzpicture}}
    \caption{Gauge theory phase for a trinion (resp. tetraon) theory with flavor symmetry $(\mathfrak{g},D_k^{\oplus (\nu + 1)})$, with $\nu = 1$ (resp. $\nu =2$). Here $D_k$ corresponds to the Dynkin diagram realized by the gauge and the empty nodes. Notice that $\nu$ corresponds to the rank of $\mathfrak g_{extra}$ in Table \ref{tab:lacompletezzaperdio}.}
    \label{fig:mannaggia}
    \end{figure}
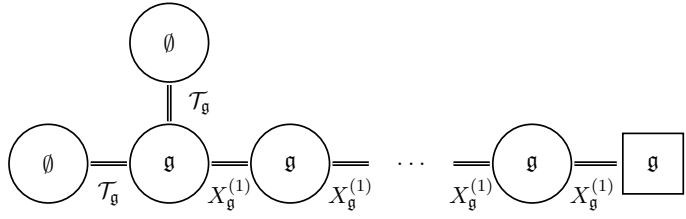

    \begin{table}
\begin{center}
\begin{tabular}{c||c|c|c|c|c|c|c}
   $\mathfrak{g}$ & $A_{2j+1}$ & $A_{2j}$ & $D_{2j+2}$ & $D_{2j+3}$ & $E_6$ & $E_7$ & $E_8$\\\hline
     $\mathfrak{g}_{extra}$ & $\mathfrak{su}(2)$ & - & $\mathfrak{u}(1) \times \mathfrak{u}(1)$ & $\mathfrak{u}(1)$ &  -  & $\mathfrak{u}(1)$ &  - \\
\end{tabular}
\caption{Extra flavor symmetry for the bifundamental conformal matter theories $X_{\mathfrak{g}}^{(1)}$ \cite{DeMarco:2023irn}.}\label{tab:lacompletezzaperdio}
\end{center}
\end{table}

The next question is whether these SCFTs are irreducible. By studying suitable partial resolutions of the corresponding CY3 singularities we find that both the novel trinion and tetraon SCFTs of type $(\mathfrak g, D_k^{\oplus (1+\nu)})$ have gauge theory phases with generalized quivers of Dynkin type $D_k$ as in Figure \ref{fig:mannaggia}. Edges labeled $X^{(1)}_\mathfrak{g}$ correspond to 5d $(\mathfrak g,\mathfrak g)$ atoms of \cite{DeMarco:2023irn}. The edges labeled $\mathcal T_\mathfrak{g}$ correspond to 5d SCFTs with unit valency and type $\mathfrak g$ -- possibly a 5d version of a 4d $\mathcal{N}=2$ $D_p(G)$ theory \cite{Cecotti:2012jx,Cecotti:2013lda,giaco1} of the kind that appeared in \cite{DelZotto:2015rca} --- we present more details about these models in Section \ref{sec:gluingtrinionstetraonsdnode}. Since these gauge theory phases have indeed gauge groups of the same type of the 5d trinion and tetraon SCFTs, the latter are not irreducible.

\medskip

It is remarkable that these gauge theory phases allow to confirm the M-theoretic expectation on the UV flavor symmetry with a slight generalization of the UV symmetry enhancement argument of \cite{Tachikawa,Yonekura}. Indeed, since this quiver is $D_k$-shaped and balanced, we expect a flavor enhancement of the topological symmetries to a non-abelian $D_k$ global symmetry group. However, the edges of this quiver consist of 5d conformal matter theories, which therefore can also contribute to the flavor enhancement. In particular, the bifundamental conformal matter theories $X_{\mathfrak g}^{(1)}$ have flavor symmetry $(\mathfrak g, \mathfrak g, \mathfrak g_{extra})$ where $\mathfrak{g}_{extra}$ is listed in Table \ref{tab:lacompletezzaperdio}. We find that these generalized edges contribute an extra factor $D_k^{\oplus \nu}$ to the non-abelian enhancement, where $\nu = \text{rank } \mathfrak g_{extra}$, resulting in a global symmetry $$(\mathfrak g, \underbrace{D_k \oplus \phantom{\big|}\dots \phantom{\big|} \oplus D_k }_{\nu+1 \text{ times}}\,)\,\,\,\,,$$ matching precisely the one expected for trinions and tetraons.

\medskip

Based on this result one is led to claim that there are further 5d SCFT molecules with the same quiver as in Figure \ref{fig:mannaggia} and with gauge nodes of the missing types, namely $\mathfrak g = A_{2j},E_6$ and $E_8$. In all these cases $\nu = 0$, and therefore one would expect no extra flavor enhancement, ending up with infinite families of SCFTs molecules of types $(A_{2j}, D_k)$, $(E_6, D_k)$ and $(E_8,D_k)$. We confirm this prediction by building the corresponding canonical singularities.

\medskip



\medskip

Finally, one has to study the chemistry of these new 5d SCFTs. As we will show in Section \ref{sec:gluingrules}, the new trinion and tetraon theories introduced in this work are subject to a set of \textit{fusion rules}, that describe the possible (generalized) diagonal gaugings that the 5d trinion and tetraon can undergo. The latter turn out to be rather limited, the interested readers can find our characterization in Section \ref{sec:gluingrules}.

\medskip

Despite the above-mentioned restrained behaviour, some wiggle room is left to perform a limited set of fusions, that do not produce new independent trinions and tetraons. As a straightforward mathematical by-product of our constructions we find SCFTs geometrically engineered using singular CY3 phases which are \textit{non-toric non-complete intersections} correspond to trinion, tetraon and pentaon 5d conformal matter of type $(D_k^{\oplus 2},A_n)$, $(D_k^{\oplus 2},A_n,A_m)$, $(D_k^{\oplus 2},A_n,A_m,A_k)$. The explicit appearance of this class of singularities, and the identification of some features of the corresponding SCFT, is along the lines of recent results (obtained in a different context involving orbifold singularities) \cite{Collinucci:2025rrh,Dramburg:2025tlb}. We summarize our findings schematically in Table \ref{table:oscar}.

 \renewcommand{\arraystretch}{1.1}
\begin{table}
\centering
$
\begin{array}{|C||l|l|l|l|l|}
\hline
\text{Valency} & \boldsymbol{A} & \boldsymbol{D}  & \boldsymbol{E_6} & \boldsymbol{E_7} & \boldsymbol{E_8} \\
 \hline
 \hline
 2 & A^{\oplus 2}  & \begin{array}{l}D_k\oplus A \\ D_k^{\oplus 2}\\\textcolor{red}{D_k^{\oplus 2}}\end{array} & \begin{array}{l}E_6^{\oplus 2}\\ \textcolor{red}{E_6\oplus D_k}\end{array}& E_7^{\oplus 2} & \begin{array}{l}E_8^{\oplus 2}\\ \textcolor{red}{E_8\oplus D_k}\end{array}\\
  \hline
 3 & A^{\oplus 3} & \begin{array}{l}D_k \oplus A^{\oplus 2} \\ D_k^{\oplus 2}\oplus A\\\textcolor{red}{D_k^{\oplus 2}\oplus A} \\ \textcolor{red}{D_{2j+1}\oplus D_k^{\oplus 2}}\end{array}& E_6^{\oplus 2} \oplus A & \begin{array}{l} E_7^{\oplus 2}\oplus A\\ \textcolor{red}{E_7\oplus D_k^{\oplus 2}} \end{array}& E_8^{\oplus 2}\oplus A \\
  \hline
 4 &A^{\oplus 4}& \begin{array}{l}D_k\oplus A^{\oplus 3}\\ D_k^{\oplus 2} \oplus A^{\oplus 2}\\\textcolor{red}{D_k^{\oplus 2} \oplus A^{\oplus 2}} \\ \textcolor{red}{D_{2j}\oplus D_k^{\oplus 3}}\end{array} & E_6^{\oplus 2} \oplus A^{\oplus 2}& E_7^{\oplus 2 \oplus A^{\oplus 2}}& E_8^{\oplus 2} \oplus A^{\oplus 2}\\
  \hline
5 &A^{\oplus 5} & \begin{array}{l}D_k\oplus A^{\oplus 4}\\ D_k^{\oplus 2}\oplus A^{\oplus 3}\\ \textcolor{red}{D_k^{\oplus 2}\oplus A^{\oplus 3}}\\\end{array} &E_6^{\oplus 2} \oplus A^{\oplus 3} &E_7^{\oplus 2} \oplus A^{\oplus 3} &E_7^{\oplus 2} \oplus A^{\oplus 3}  \\
  \hline
  6 &A^{\oplus 6} & \begin{array}{l}D_k\oplus A^{\oplus 5}\\D_k^{\oplus 2}\oplus A^{\oplus 4}\end{array} &E_6^{\oplus 2} \oplus A^{\oplus 4}  &? &?  \\
  \hline
   7 &A^{\oplus 7} & \begin{array}{l}D_k\oplus A^{\oplus 6}\end{array} &?&? &?  \\
  \hline
 \vdots &\vdots & \vdots &? &? &?  \\
  \hline
 n &A^{\oplus n} & D_k \oplus A^{\oplus n-1} &? &? &?  \\
  \hline
    \end{array}
    $
    \caption{Coarse summary of known infinite families of 5d conformal matter theories with valency $n$. The entries in \textcolor{red}{red} are SCFTs introduced in this work. In this Table we omit ranks for $A$ type flavor symmetries, moreover each entry corresponds to one or more families with the given valency. Some entries are repeated, as some cases were previously constructed in \cite{DeMarco:2023irn} and in this work we add further independent instances.}
    \label{table:oscar}
\end{table}

\medskip

Therefore, the trinion/tetraon theories of type $D$ that we discover in this work behave more like exotic theories, that enrich the landscape of 5d SCFTs without dramatically increasing its size:  the chemistry of these theories is quite restricted, as they can be gauged with bifundamental conformal matter theories, but not with other trinions/tetraons of the same type. In summary, while the 6d atomic classification leads to linear quivers, and the 4d atomic classification leads to trivalent quivers of general types, the 5d atomic classification seems to showcase qualitatively more interesting features, with many generalized linear quivers with gauge nodes of A, D or E types, many generalized trivalent quivers of A-type, few generalized trivalent quivers of D-type, and no generalized trivalent quivers of E-type.

\subsection{Organization of the work}

The work is structured as follows: in Section \ref{sec:1classification}, after a brief review of $n$-valent 5d conformal matter theories in both non-toric and toric setups, we introduce novel 5d trinion and tetraon SCFTs, which constitute the foundation of this work. In Section \ref{sec:CBdata} we collect a few basic physical features of the newly-constructed trinions and tetraons, such as their Coulomb Branch dimension. In Section \ref{sec:gluingrules} we show that trinions and tetraons of type $D$ are CM molecules, and we lay down their fusion rules. The outcome is a restricted set of 5d generalized quivers admitting a 5d SCFT phase, which we classify in section \ref{sec: blowdowns}. As a byproduct, this also shows a geometric generalization of the criterion of \cite{Yonekura} for the instantonic enhancement of the flavor symmetry, to  intrinsically non-Lagrangian SCFTs. We concisely exhibit a sporadic trinion case of type $D_4$ in Section \ref{sec:sporadic}.  We present a discussion of our results in Section \ref{sec:conclusions}. We quickly review why a specific class of geometries and generalized quivers \textit{cannot} engineer trinions and tetraons, and we collect the singular geometries corresponding to the generalized quiver phases in Appendix \ref{app:sing geometries}. Finally, in Appendix \ref{app:vertical}, we present an explicit resolution of the $\mathcal T_{\mathfrak g}$ geometries introduced in this paper, and we compute the rank of their flavor group.

\section{\texorpdfstring{$n$-valent 5d conformal matter}{n-valent 5d conformal matter}}
\label{sec:1classification}
In this Section we will list the possible $n$-valent theories that can be obtained from M-theory geometric engineering on hypersurface singularities.  We organize the theories according to the increasing number of $D$-factors, as in general the number of (simple) flavor factors is not monotonically decreasing with respect to the 5d  RG flow. Indeed, if a UV theory has a number $n_{UV}$ of $A$ flavor factors, a second IR theory might have $n_{IR} > n_{UV}$ flavor factors. Instead, the number of $D$-type (and $E$-type) factors can only \textit{decrease} along  RG flows. The reason is that, according to the versal deformation/ resolution theory of Du Val singularities \cite{Katz:1992aa,KatzVafa}, a line of $A_n$ singularities can be split into multiple lines $A_{n_1}\oplus\ldots\oplus A_{n_k}$ through a complex deformation/blow-up\footnote{From here on, we employ the standard physics notation "$\oplus$" denoting the presence of more than one simple factor in the flavor symmetry also in a geometric sense. In the latter language, $\mathfrak{g}\oplus\mathfrak{g}'$ means that the CY3 at hand displays two singular lines of type $\mathfrak{g}$ and $\mathfrak{g}'$, respectively.}:  
\begin{equation}
    A_n \to  A_{n_1} \oplus ... \oplus A_{n_k},
\end{equation}
such that $n_{1}+ ... +n_k \leq n$. Instead, an analogous splitting 
\begin{equation}
\label{eq:splitting}
    D_{n} \to D_{n_1} \oplus ... \oplus D_{n_k},
\end{equation} 
with $k>1$ is \textit{not} allowed by the versal deformation/resolution theory of $D_n$.\footnote{Notice that the $E_6,E_7,E_8$ cases behave exactly as the $D_n$. Furthermore, the obstruction highlighted in \eqref{eq:splitting} is a purely 5d phenomenon. In fact, in going from 6d to 5d, a splitting of a $D_n$ factor of the 6d flavor group can be achieved as the class of the elliptic curve of the CY3 permits to break $D_n$ to any of its regular subalgebras.} The argument clearly translates to the context of canonical threefolds, which will be of interest in this work: indeed, the non-compact singular lines of canonical CY3 are (apart from dissident points \cite{reid1980canonical}) always a direct product between $\mathbb{C}$ and a Du Val singularity. Therefore, the number of $D$ flavor factors in a 5d SCFT engineered by a canonical CY3 is a piece of data that is robust along RG flow (as it can only decrease), furnishing the underpinning for the structure of this work.\\ 

\indent We start by reviewing $n$-valent CM of type $A$ in \ref{sec:5dtoric}, and $n$-valent 5d SCFTs with \textit{one} $D$-factor in the flavor group in \cref{sec:trinions and tetraons with one D factor}. We proceed with the construction of 5d CM with at most \textit{two} $D,E$ factors in \ref{sec:5dbifundamental}. This sets the stage for the introduction of the novel trinion and tetraon 5d CM theories of type $D$ in \ref{sec:5dtriniontetraon}. 

\subsection{5d conformal matter with no \texorpdfstring{$D,E$}{D,E} factors}
\label{sec:5dtoric}
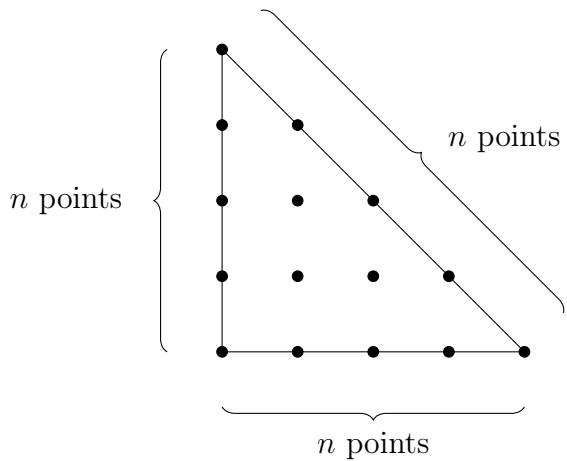
\begin{figure}[H]
\centering
    \scalebox{1.}{
    \begin{tikzpicture}
        \draw (0,4)--(1,3)--(2,2)--(3,1)--(4,0)--(0,0)--(0,4);
        %
        \filldraw[black] (0,0) circle (2pt);
        \filldraw[black] (0,1) circle (2pt);
        \filldraw[black] (0,2) circle (2pt);
        \filldraw[black] (0,3) circle (2pt);
        \filldraw[black] (0,4) circle (2pt);
        \filldraw (1,0) circle (2pt);
        \filldraw[black] (1,1) circle (2pt);
        \filldraw[black] (1,2) circle (2pt);
        \filldraw (1,3) circle (2pt);
        \filldraw (2,0) circle (2pt);
        \filldraw[black] (2,1) circle (2pt);
        \filldraw (2,2) circle (2pt);
        \filldraw (3,0) circle (2pt);
        \filldraw (3,1) circle (2pt);
        \filldraw[black] (4,0) circle (2pt);
        \draw [decorate,decoration={brace,amplitude=5pt,mirror,raise=4ex}]
  (0,0) -- (4,0) node[midway,yshift=-3em]{$n$ points};
   \draw [decorate,decoration={brace,amplitude=5pt,mirror,raise=4ex}]
  (0,4) -- (0,0) node[midway,xshift=-5em]{$n$ points};
   \draw [decorate,decoration={brace,amplitude=5pt,mirror,raise=4ex}]
  (4,0) -- (0,4) node[midway,xshift=4.2em,yshift=2em]{$n$ points};
    \end{tikzpicture}}
    \caption{Toric diagram for the $T_n$ theories.}
    \label{fig:torictrinion}
    \end{figure}

In our gentle progression towards $n$-valent 5d CM with larger and larger $n$, the toric cases play a well-established role. The elementary building blocks for toric conformal matter are given by the theories engineered by the threefolds:
\begin{equation}\label{A building blocks}
    \mathbb{C}^3/(\mathbb{Z}_p\times\mathbb{Z}_q),
\end{equation}
with $p$ and $q$ positive integers. An interesting subset of these is given by the $T_n$ theories \cite{Benini:2009gi}. These are engineered by M-theory on the non-compact CY3 encoded by the toric diagrams in Figure \ref{fig:torictrinion}. Algebraically, they are given by:
\begin{equation}
    T_n \text{ theory:} \quad xyz = w^{n}.
\end{equation}

The corresponding 5d SCFTs have tri-valent $A_{n-1}^{\oplus 3}$ UV flavor symmetry (with possible non-abelian enhancement) and non-trivial gauge content. This is a clear consequence of the pattern of non-compact singular lines in the CY3.\\

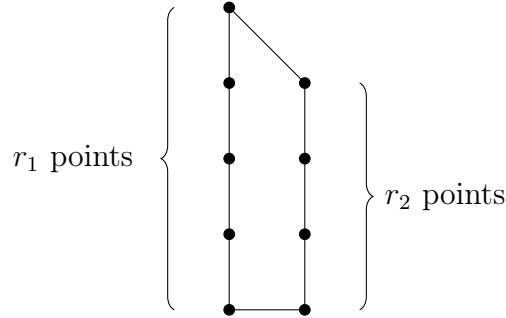
\begin{figure}[H]
\centering
    \scalebox{1.}{
    \begin{tikzpicture}
       \draw (8,4)--(8,0)--(9,0)--(9,3)--(8,4);
        \filldraw[black] (8,0) circle (2pt);
        \filldraw[black] (8,1) circle (2pt);
       \filldraw[black] (8,2) circle (2pt);
        \filldraw[black] (8,3) circle (2pt);
       \filldraw[black] (8,4) circle (2pt);
       \filldraw (9,0) circle (2pt);
       \filldraw[black] (9,1) circle (2pt);
        \filldraw[black] (9,2) circle (2pt);
       \filldraw (9,3) circle (2pt);
         \draw [decorate,decoration={brace,amplitude=5pt,mirror,raise=4ex}]
  (8,4) -- (8,0) node[midway,xshift=-5em]{$r_1$ points};
         \draw [decorate,decoration={brace,amplitude=5pt,mirror,raise=4ex}]
  (9,0) -- (9,3) node[midway,xshift=4.5em]{$r_2$ points};
    \end{tikzpicture}}
    \caption{Toric diagram for the $SPP$ theory.}
    \label{fig:SPP}
    \end{figure}

\indent An additional non-toric family (with one toric family member) of singularities that engineers 5d CM theories with $n$ flavor factors of type $A$ is given by a generalization of the construction of \cite{KatzVafa}. They are given by the collision of $n$ non-compact lines of type $A$:
\begin{equation}\label{genSPP}
    xy = (z+c_1t)^{r_1}\cdot\cdots\cdot(z+c_nt)^{r_n},
\end{equation}
where the case $n=2$ is toric and is the famous toric suspended pinch point singularity, depicted in Figure \ref{fig:SPP}. \\
\indent As we have briefly mentioned, it has been conjectured \cite{Xie_2017} that all 5d $\mathcal{N}=1$ SCFTs can be engineered via M-theory on canonical CY3\footnote{For the definition of canonical singularities, we refer to \cite{reidyoungpersonsing}.}. To show that a quasi-homogeneous hypersurface singularity is canonical, we can employ a criterion based on quasi-homogeneous weights due to \cite{reidyoungpersonsing}. It reads as follows: consider a complete intersection $n$-fold singularity defined by 
\begin{equation}\label{complete intersection}
    f_i(x_1,\ldots,x_k) = 0 \quad \subset \mathbb{C}^k, \quad i = 1,\ldots,k-n,
\end{equation}
and admitting a quasi-homogeneous action:
\begin{equation}
     f_i(\lambda^{x_1}x_1,\ldots,\lambda^{x_k}x_k) = \lambda^{f_i} f_i(x_1,\ldots,x_k) 
\end{equation}
Then the complete intersection \eqref{complete intersection} is \textit{canonical} if and only if:
\begin{equation}\label{canonicity}
    \sum_{j=1}^k \lambda_{x_j} > \sum_{i=1}^{k-n} \lambda_{f_i}
\end{equation}
We can check that the threefolds \eqref{genSPP} satisfy the bound \eqref{canonicity} for any choice of $r_j$.

The singularities \eqref{genSPP} lead to rank-zero 5d SCFTs: as they are of compound Du Val (cDV) form, it is well-known \cite{reid1983minimal} that they admit at most \textit{small resolutions}. This implies that the corresponding SCFTs have an empty Coulomb branch, and a non-empty Higgs branch: they are a collection of hypermultiplets charged under the suitable representation of $\oplus_{i=1,\ldots,n}A_{r_i-1}$.

\begin{figure}[H]
\centering
    \scalemath{0.7}{
    \begin{tikzpicture}
       \scalebox{0.6}{\draw (0,4)--(1,3)--(2,2)--(3,1)--(4,0)--(0,0)--(0,4);
        %
        \filldraw[black] (0,0) circle (2pt);
        \filldraw[black] (0,1) circle (2pt);
        \filldraw[black] (0,2) circle (2pt);
        \filldraw[black] (0,3) circle (2pt);
        \filldraw[black] (0,4) circle (2pt);
        \filldraw (1,0) circle (2pt);
        \filldraw[black] (1,1) circle (2pt);
        \filldraw[black] (1,2) circle (2pt);
        \filldraw (1,3) circle (2pt);
        \filldraw (2,0) circle (2pt);
        \filldraw[black] (2,1) circle (2pt);
        \filldraw (2,2) circle (2pt);
        \filldraw (3,0) circle (2pt);
        \filldraw (3,1) circle (2pt);
        \filldraw[black] (4,0) circle (2pt);}
        \scalebox{0.6}{\draw (6,0)--(6,4)--(2,4)--(6,0);
        \filldraw[black] (6,0) circle (2pt);
        \filldraw[black] (6,1) circle (2pt);
        \filldraw[black] (6,2) circle (2pt);
        \filldraw[black] (6,3) circle (2pt);
        \filldraw[black] (6,4) circle (2pt);
        \filldraw[black] (5,4) circle (2pt);
        \filldraw[black] (5,3) circle (2pt);
        \filldraw[black] (5,2) circle (2pt);
        \filldraw[black] (5,1) circle (2pt);
        \filldraw[black] (4,4) circle (2pt);
        \filldraw[black] (4,3) circle (2pt);
        \filldraw[black] (4,2) circle (2pt);
        \filldraw[black] (3,4) circle (2pt);
        \filldraw[black] (3,3) circle (2pt);
        \filldraw[black] (2,4) circle (2pt);}
        \draw [decorate,decoration={brace,amplitude=5pt,mirror,raise=4ex}]
  (0,0) -- (4,0) node[midway,yshift=-3em]{glue};
  \draw[thick,->] (4.5,1.2) -- (7.5,1.2);
    \scalebox{0.6}{  
    \draw (14,0)--(18,0)--(18,4)--(14,4)--(14,0);
        \filldraw[black] (14,1) circle (2pt);
        \filldraw[black] (14,2) circle (2pt);
        \filldraw[black] (14,3) circle (2pt);
        \filldraw[black] (14,4) circle (2pt);
        \filldraw[black] (14,0) circle (2pt);
        \filldraw[black] (15,1) circle (2pt);
        \filldraw[black] (15,2) circle (2pt);
        \filldraw[black] (15,3) circle (2pt);
        \filldraw[black] (15,4) circle (2pt);
        \filldraw[black] (15,0) circle (2pt);
        \filldraw[black] (16,1) circle (2pt);
        \filldraw[black] (16,2) circle (2pt);
        \filldraw[black] (16,3) circle (2pt);
        \filldraw[black] (16,4) circle (2pt);
        \filldraw[black] (16,0) circle (2pt);
        \filldraw[black] (17,1) circle (2pt);
        \filldraw[black] (17,2) circle (2pt);
        \filldraw[black] (17,3) circle (2pt);
        \filldraw[black] (17,4) circle (2pt);
        \filldraw[black] (17,0) circle (2pt);
        \filldraw[black] (18,1) circle (2pt);
        \filldraw[black] (18,2) circle (2pt);
        \filldraw[black] (18,3) circle (2pt);
        \filldraw[black] (18,4) circle (2pt);
        \filldraw[black] (18,0) circle (2pt);}
    \end{tikzpicture}}
    
    \vspace{-1cm}
    
    \begin{tikzpicture}
       \scalebox{0.6}{\draw (5,0)--(5,4)--(9,0)--(5,0);
        \filldraw[black] (5,0) circle (2pt);
        \filldraw[black] (5,1) circle (2pt);
        \filldraw[black] (5,2) circle (2pt);
        \filldraw[black] (5,3) circle (2pt);
        \filldraw[black] (5,4) circle (2pt);
        \filldraw[black] (6,0) circle (2pt);
        \filldraw[black] (6,1) circle (2pt);
        \filldraw[black] (6,2) circle (2pt);
        \filldraw[black] (6,3) circle (2pt);
        \filldraw[black] (7,0) circle (2pt);
        \filldraw[black] (7,1) circle (2pt);
        \filldraw[black] (7,2) circle (2pt);
        \filldraw[black] (8,0) circle (2pt);
        \filldraw[black] (8,1) circle (2pt);
        \filldraw[black] (9,0) circle (2pt);}
       \scalebox{0.6}{  \draw (4,0)--(4,4)--(0,0)--(4,0);
        \filldraw[black] (4,0) circle (2pt);
        \filldraw[black] (4,1) circle (2pt);
        \filldraw[black] (4,2) circle (2pt);
        \filldraw[black] (4,3) circle (2pt);
        \filldraw[black] (4,4) circle (2pt);
        \filldraw[black] (3,0) circle (2pt);
        \filldraw[black] (3,1) circle (2pt);
        \filldraw[black] (3,2) circle (2pt);
        \filldraw[black] (3,3) circle (2pt);
        \filldraw[black] (2,0) circle (2pt);
        \filldraw[black] (2,1) circle (2pt);
        \filldraw[black] (2,2) circle (2pt);
        \filldraw[black] (1,0) circle (2pt);
        \filldraw[black] (1,1) circle (2pt);
        \filldraw[black] (0,0) circle (2pt);}
        \scalebox{0.6}{\draw (4,-5)--(4,-1)--(0,-1)--(4,-5);
        \filldraw[black] (4,-5) circle (2pt);
        \filldraw[black] (4,-4) circle (2pt);
        \filldraw[black] (4,-3) circle (2pt);
        \filldraw[black] (4,-2) circle (2pt);
        \filldraw[black] (4,-1) circle (2pt);
        \filldraw[black] (3,-1) circle (2pt);
        \filldraw[black] (3,-2) circle (2pt);
        \filldraw[black] (3,-3) circle (2pt);
        \filldraw[black] (3,-4) circle (2pt);
        \filldraw[black] (2,-1) circle (2pt);
        \filldraw[black] (2,-2) circle (2pt);
        \filldraw[black] (2,-3) circle (2pt);
        \filldraw[black] (1,-1) circle (2pt);
        \filldraw[black] (1,-2) circle (2pt);
        \filldraw[black] (0,-1) circle (2pt);}
         \scalebox{0.6}{\draw (5,-5)--(5,-1)--(9,-1)--(5,-5);
        \filldraw[black] (5,-5) circle (2pt);
        \filldraw[black] (5,-4) circle (2pt);
        \filldraw[black] (5,-3) circle (2pt);
        \filldraw[black] (5,-2) circle (2pt);
        \filldraw[black] (5,-1) circle (2pt);
        \filldraw[black] (6,-1) circle (2pt);
        \filldraw[black] (6,-2) circle (2pt);
        \filldraw[black] (6,-3) circle (2pt);
        \filldraw[black] (6,-4) circle (2pt);
        \filldraw[black] (7,-1) circle (2pt);
        \filldraw[black] (7,-2) circle (2pt);
        \filldraw[black] (7,-3) circle (2pt);
        \filldraw[black] (8,-1) circle (2pt);
        \filldraw[black] (8,-2) circle (2pt);
        \filldraw[black] (9,-1) circle (2pt);}
         \scalebox{0.6}{\draw (10,0)--(10,4)--(6,4)--(10,0);
        \filldraw[black] (10,0) circle (2pt);
        \filldraw[black] (10,1) circle (2pt);
        \filldraw[black] (10,2) circle (2pt);
        \filldraw[black] (10,3) circle (2pt);
        \filldraw[black] (10,4) circle (2pt);
        \filldraw[black] (9,4) circle (2pt);
        \filldraw[black] (9,3) circle (2pt);
        \filldraw[black] (9,2) circle (2pt);
        \filldraw[black] (9,1) circle (2pt);
        \filldraw[black] (8,4) circle (2pt);
        \filldraw[black] (8,3) circle (2pt);
        \filldraw[black] (8,2) circle (2pt);
        \filldraw[black] (7,4) circle (2pt);
        \filldraw[black] (7,3) circle (2pt);
        \filldraw[black] (6,4) circle (2pt);}
   
        \scalebox{0.6}{  
    \draw (22,4)--(26,4)--(26,0)--(22,-4)--(18,0)--(22,4);
        \filldraw[black] (18,0) circle (2pt);
        \filldraw[black] (19,-1) circle (2pt);
        \filldraw[black] (19,0) circle (2pt);
        \filldraw[black] (19,1) circle (2pt);
        \filldraw[black] (20,-2) circle (2pt);
        \filldraw[black] (20,-1) circle (2pt);
        \filldraw[black] (20,0) circle (2pt);
        \filldraw[black] (20,1) circle (2pt);
        \filldraw[black] (20,2) circle (2pt);
        \filldraw[black] (21,-3) circle (2pt);
        \filldraw[black] (21,-2) circle (2pt);
        \filldraw[black] (21,-1) circle (2pt);
        \filldraw[black] (21,0) circle (2pt);
        \filldraw[black] (21,1) circle (2pt);
        \filldraw[black] (21,2) circle (2pt);
        \filldraw[black] (21,3) circle (2pt);
        \filldraw[black] (22,-4) circle (2pt);
        \filldraw[black] (22,-3) circle (2pt);
        \filldraw[black] (22,-2) circle (2pt);
        \filldraw[black] (22,-1) circle (2pt);
        \filldraw[black] (22,0) circle (2pt);
        \filldraw[black] (22,1) circle (2pt);
        \filldraw[black] (22,2) circle (2pt);
        \filldraw[black] (22,3) circle (2pt);
        \filldraw[black] (22,4) circle (2pt);
        \filldraw[black] (23,-3) circle (2pt);
        \filldraw[black] (23,-2) circle (2pt);
        \filldraw[black] (23,-1) circle (2pt);
        \filldraw[black] (23,0) circle (2pt);
        \filldraw[black] (23,1) circle (2pt);
        \filldraw[black] (23,2) circle (2pt);
        \filldraw[black] (23,3) circle (2pt);
        \filldraw[black] (23,4) circle (2pt);
        \filldraw[black] (24,-2) circle (2pt);
        \filldraw[black] (24,-1) circle (2pt);
        \filldraw[black] (24,0) circle (2pt);
        \filldraw[black] (24,1) circle (2pt);
        \filldraw[black] (24,2) circle (2pt);
        \filldraw[black] (24,3) circle (2pt);
        \filldraw[black] (24,4) circle (2pt);
        \filldraw[black] (25,-1) circle (2pt);
        \filldraw[black] (25,0) circle (2pt);
        \filldraw[black] (25,1) circle (2pt);
        \filldraw[black] (25,2) circle (2pt);
        \filldraw[black] (25,3) circle (2pt);
        \filldraw[black] (25,4) circle (2pt);
        \filldraw[black] (26,0) circle (2pt);
        \filldraw[black] (26,1) circle (2pt);
        \filldraw[black] (26,2) circle (2pt);
        \filldraw[black] (26,3) circle (2pt);
        \filldraw[black] (26,4) circle (2pt);
        }
         \draw [decorate,decoration={brace,amplitude=5pt,mirror,raise=4ex}]
  (0,-3) -- (6,-3) node[midway,yshift=-3em]{glue};
  \draw[thick,->] (7,0) -- (9.5,0);
    \end{tikzpicture}
    \caption{Gluing procedures to obtain 4-valent (top) and 5-valent (bottom) 5d toric CM.}
    \label{fig:toricgluings}
    \end{figure}
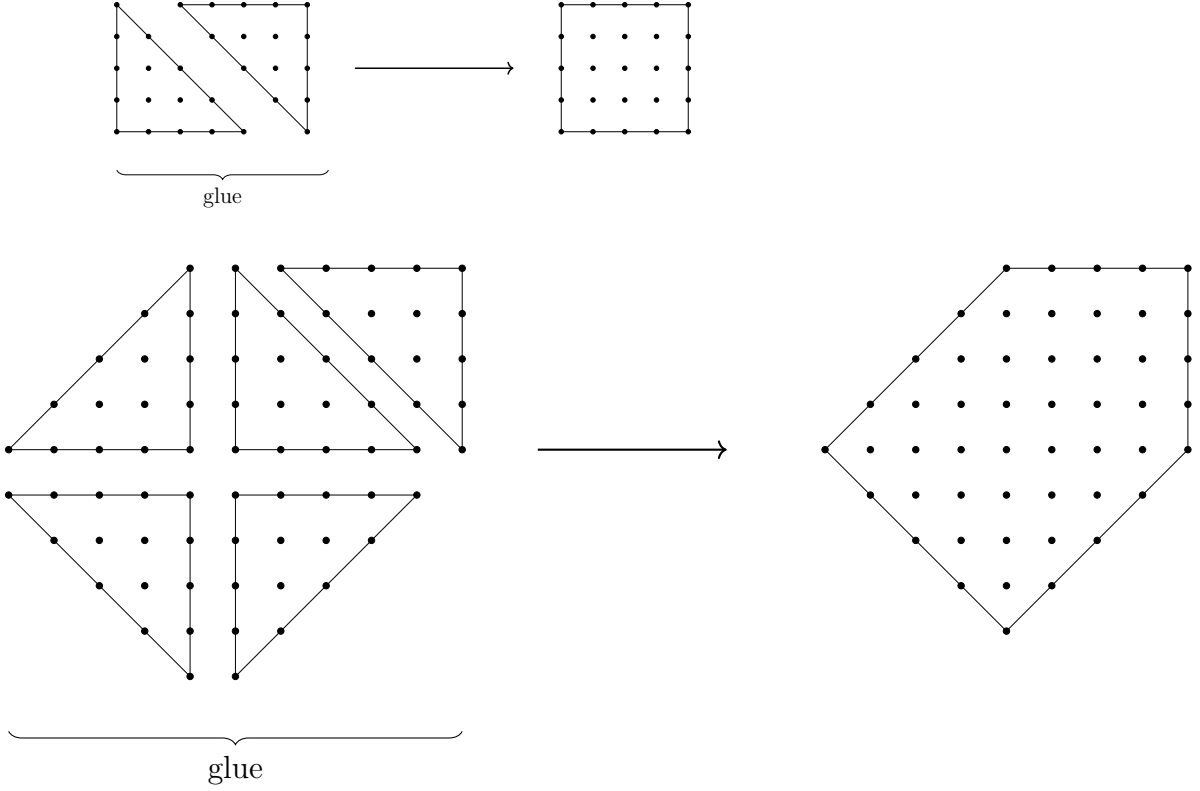

Nonetheless, it is straightforward to observe that non-trivial interacting $n$-valent 5d SCFTs with $\oplus_{i=1,\ldots,n}A_{r_i-1}$ flavor symmetry (and arbitrary positive $n$) can be constructed, once the elementary building blocks (atoms and hybrids) \eqref{A building blocks} are suitably glued along non-compact lines. This operation preserves the convexity of the toric diagram, and thus guarantees that the end result is a CY3 that can be consistently shrunk to a singular phase, corresponding to the $n$-valent 5d SCFT. E.g.\ 4-valent and 5-valent 5d SCFTs with flavor $A_3^{\oplus 4}$ and $A_3^{\oplus 5}$ can be produced via the gluings of $T_n$ theories in Figure \ref{fig:toricgluings}.
    In this section we have reviewed the atomic classification of toric 5d CM, whose irreducible SCFT blocks can be easily molded in such a way as to produce $n$-valent CM with arbitrary $n$. Namely, we have reproduced the first column of Table \ref{table:oscar}, that we report in Table \ref{table:n-fundamental toric CM} for convenience. Notice that 2-valent 5d CM of type $A$ can be both toric and non-toric, as we will review in Section \ref{sec:5dbifundamental}.
    In the next sections, we further complicate the picture by considering $n$-valent 5d SCFTs with $D$ flavor factors.
  \renewcommand{\arraystretch}{1.1}
\begin{table}[H]
\centering
$
\begin{array}{|C||l|}
\hline
\text{Valency} & \boldsymbol{A_n}  \\
 \hline
 \hline
 2 & A^{\oplus 2}  \\
  \hline
 3 & A^{\oplus 3} \\
  \hline
 4 &A^{\oplus 4}\\
  \hline
 \vdots & \vdots \\
  \hline
 n&A^{\oplus n}  \\
  \hline
    \end{array}
    $
    \caption{$n$-valent 5d conformal matter theories that can be engineered via toric methods, or via the techniques of \cite{KatzVafa}.}
    \label{table:n-fundamental toric CM}
\end{table}

\subsection{5d conformal matter with at most one \texorpdfstring{$D$}{D} flavor factor}
\label{sec:trinions and tetraons with one D factor}
The simplest and well-known technology to engineer 5d SCFTs with at least one flavor symmetry factor of type $D$ is offered by the construction of \cite{KatzVafa}, that involves a collision of two non-compact lines: one of type $D$, and one of type $A$. A straightforward generalization of this method allows the construction of $n$-valent conformal matter theories with flavor symmetry of the form:
\begin{equation}\label{DAA flavor}
    F_{UV} \supseteq D\oplus \underbrace{A \oplus A\oplus\ldots }_{n-1\text{ factors}}.
\end{equation}
The corresponding threefold is a hypersurface equation:
\begin{equation}\label{DAA threefold}
    x^2+zy^2+z^{k-1}(z+c_1t)^{r_1}\cdot \cdots \cdot (z+c_{n-1}t)^{r_{n-1}},
\end{equation}
where the $c_i$ are distinct coefficients\footnote{This does not imply any loss of generality: if two or more coefficients coincide, they lead to an enhancement of an $A$ factor, that can be subsumed in a redefinition of the $r_i$.}.
The threefold \eqref{DAA threefold} correctly displays one line of $D_k$ singularities, along with $n-1$ lines of $A_{r_i-1}$ singularities. It is easy to check that \eqref{DAA threefold} satisfies the canonicity condition \eqref{canonicity} for every positive integer choice of $k$, $n$ and $r_i$. As usual, the non-compact singular lines are responsible for the presence of the symmetry factor \eqref{DAA flavor} in the UV flavor symmetry.\\
\indent From the physics perspective, the singularities \eqref{DAA threefold} give rise to rank-zero 5d SCFTs, since they are cDV threefolds. Thus, their only content is a collection of hypermultiplets charged under the suitable representation of \eqref{DAA flavor}. Examples with non-trivial gauge content can be produced via orbifolds of $\mathbb{C}^3$, as in \cite{Tian:2021cif}.\\
\indent From the point of view of Table \ref{table:oscar}, in this Section we have reviewed the construction of the 5d CM theories highlighted in Table \ref{tab:DAA CM}.
 \renewcommand{\arraystretch}{1.1}
\begin{table}[H]
\centering
$
\begin{array}{|C||l|}
\hline
 \text{Valency} & \boldsymbol{D_n}  \\
 \hline
 \hline
 2 & D\oplus A  \\
  \hline
 3 &D\oplus A^{\oplus 2} \\
  \hline
 4 &D\oplus A^{\oplus 3}\\
  \hline
 \vdots & \vdots \\
  \hline
 n &D\oplus A^{\oplus n-1}  \\
  \hline
    \end{array}
    $
    \caption{$n$-valent 5d conformal matter with only one $D$ factor, constructed with the methods of \cite{KatzVafa}.}
    \label{tab:DAA CM}
\end{table}

\subsection{5d conformal matter with at most two \texorpdfstring{$D,E$}{D,E} factors}
\label{sec:5dbifundamental}
In the works \cite{DeMarco:2023irn,DeMarco:2025ugw}, a set of irreducible 5d CM SCFTs with at most two $D,E$ factors were constructed, employing M-theory geometric engineering on non-compact singular canonical Calabi-Yau threefolds (CY3). The sought-after 5d CM SCFT inhabits the five spacetime directions transverse to the CY3. As per the consolidated dictionary between CY3 geometry and 5d SCFTs, a flavor group of type $\mathfrak{g}\oplus\mathfrak{g}$ can be engineered via singular non-compact complex lines supporting a Du Val singularity of type $\mathfrak{g}$. The  form of Du Val singularities that we are going to adopt in this work is
\begin{equation}\label{ADE sing}
 \begin{cases}
P_{A_k}(x_1,x_2,x_3) = x_1^2+x_2^2-x_3^{k+1}, \\
P_{D_k}(x_1,x_2,x_3) = x_1^2 +x_3x_2^2+x_3^{k-1}, \\
P_{E_6}(x_1,x_2,x_3) =x_1^2+x_2^3+x_3^4, \\
P_{E_7}(x_1,x_2,x_3) = x_1^2+x_2^3+x_2x_3^3, \\
P_{E_8}(x_1,x_2,x_3) = x_1^2+x_2^3+x_3^5.
\end{cases}
\end{equation}
The reduction of the M-theory 3-form $C_3$ on the Poincaré duals of the resolved lines provides the degrees of freedom of the background flavor vector multiplets. In order to produce a non-trivial 5d bifundamental fixed point, \textit{two non-compact lines} must be intersected, hence giving rise to a flavor group of type $\mathfrak{g}\oplus \mathfrak{g}$.
Following this logic, 5d CM theories are constructed employing the following elementary building blocks:
\begin{equation}\label{elementary 3folds}
    X_{\mathfrak{g}}^{(i)}: \quad \begin{cases}
        P_{\mathfrak{g}}^{(i)}(x_1,x_2,x_3) = 0, \\
        uv = x_i.
    \end{cases}
\end{equation}
 It is easy to check that the threefolds \eqref{bifund 3folds} possess two non-compact singular lines supporting a singularity of Du Val type $\mathfrak{g}$, as desired. The lines are:
\begin{equation}
    x_1=x_2=x_3 = u = 0, \quad\quad x_1=x_2=x_3= v = 0.
\end{equation}
In general, additional singular lines can be present in \eqref{bifund 3folds}, further enhancing the UV flavor group.\\
In the wake of the atomic classification program, the elementary building blocks \eqref{elementary 3folds} can be glued together along one non-compact singular line of type $\mathfrak{g}$, giving rise to novel bifundamental 5d SCFTs. Thus, they rightfully earn the name of atom (or hybrid) SCFTs. Only 2-valent gluings preserve the Calabi-Yau condition and, in the SCFT limit, give rise to  the following singularities:
\begin{equation}\label{bifund 3folds}
    X_{\mathfrak{g}}^{({n_1},{n_2},{n_3})}: \quad \begin{cases}
        P_{\mathfrak{g}}(x_1,x_2,x_3) = 0,\\
        uv = w(x,y,z),
    \end{cases}
\end{equation}
where $w(x,y,z)$ is a generic polynomial in the given complex variables. As shown in \cite{Bourget:2026ono}, the Calabi-Yau threefolds in \eqref{bifund 3folds} give rise to 5d conformal matter atoms, hybrids and molecules, and are encoded by the coweights of the corresponding Lie algebra $\mathfrak{g}$. We refer to the aforementioned work for further details on their classification.

\begin{figure}[H]
    \centering
   \scalemath{0.65}{  \begin{tikzpicture}
        \draw[thick] (-2.5,-3)--(2.5,3);
        \draw[thick] (-2.5,3)--(2.5,-3);
        \draw[thick,dashed,<-] (0.5,0)--(1.5,0);
        \node at (3.7,0) {Enhanced singularity};
        \draw[fill=red] (0,0) circle (0.13);    
        \node[rotate=49] at (-1.5,-2.2) {$\boldsymbol{u=0}$};
        \node[rotate=-49] at (1.5,-2.2) {$\boldsymbol{v=0}$};
\draw[thick,->] (0.8,-4) -- (6,-10);
\draw[thick,->] (-0.8,-4) -- (-6,-7);
 \draw [decorate,decoration={brace,amplitude=5pt,mirror,raise=4ex}]
  (-4,3) -- (-4,-3) node[midway,xshift=-6em]{SCFT phase};
    \end{tikzpicture}}

    \vspace{-1.5cm}
\scalemath{0.5}{
  \begin{tikzpicture}
        \draw[thick] (0,0) circle (0.7);
        \node at (0,0) {\small$\mathfrak{su}$};
        \draw[thick] (0.8,0)--(1.4,0);
        \draw[thick] (2.2,0) circle (0.7);
        \node at (2.2,0) {\small$\mathfrak{su}$};
        \draw[thick] (-0.8,0)--(-1.4,0);
        \draw[thick] (-2.2,0) circle (0.7);
        \node at (-2.2,0) {\small$\mathfrak{su}$};
        \draw[thick] (0,0.8)--(0,1.4);
        \draw[thick] (0,2.2) circle (0.7);
        \node at (0,2.2) {\small$\mathfrak{su}$};
        \draw[thick] (3,0)--(3.6,0);
        \draw[thick] (4.4,0) circle (0.7);
         \draw[thick] (-3,0)--(-3.6,0);
        \node at (4.4,0) {\small$\mathfrak{su}$};
         \draw[thick] (-4.4,0) circle (0.7);
        \node at (-4.4,0) {\small$\mathfrak{su}$};
         \draw[thick] (5.2,0)--(5.8,0);
            \draw[thick] (5.9,0.7)--(7.3,0.7)--(7.3,-0.7)--(5.9,-0.7)--cycle;
            \draw[thick] (-5.2,0)--(-5.8,0);
            \draw[thick] (-5.9,0.7)--(-7.3,0.7)--(-7.3,-0.7)--(-5.9,-0.7)--cycle;
            \draw[thick] (0.7,3.7)--(0.7,5.1)--(-0.7,5.1)--(-0.7,3.7)--cycle;
            \draw[thick] (0.7,-1.5)--(0.7,-2.9)--(-0.7,-2.9)--(-0.7,-1.5)--cycle;
            \draw[thick] (0,3.0)--(0,3.6);
            \draw[thick] (0,-0.8)--(0,-1.4);
        \node at (0,4.4) {\scalebox{1.2}{$f_4$}};
        \node at (6.6,0) {\scalebox{1.2}{$f_3$}};
        \node at (-6.6,0) {\scalebox{1.2}{$f_1$}};
        \node at (0,-2.2) {\scalebox{1.2}{$f_2$}};
         \node at (0,-3.8) {$\underbrace{\hspace{14 cm}}_{\scalebox{1}{\quad \qquad \scalebox{1.5}{ \text{Special unitary quiver phase}}}}$};
          
          
          \node at (16.75,0) {$\cdots$};
          \draw[thick,double] (20.2,0)--(21.5,0);
          \draw[thick,double] (16.2,0)--(15.2,0);
          %
          \draw[thick] (19.5,0) circle (0.7);
          \node at (19.5,0) {$\mathfrak{g}$};
          \draw[thick] (14.5,0) circle (0.7);
          \node at (14.5,0) {$\mathfrak{g}$};
          \draw[thick,double] (18.8,0)--(17.5,0);
          \draw[thick,double] (13.8,0)--(12.5,0);
        
        \draw[thick] (21.5,0.7)--(22.9,0.7)--(22.9,-0.7)--(21.5,-0.7)--cycle;
        \node at (22.2,0) {$\mathfrak{g}$};
        \draw[thick] (12.5,0.7)--(11.1,0.7)--(11.1,-0.7)--(12.5,-0.7)--cycle;
        \node at (11.8,0) {$\mathfrak{g}$};
       \node at (17,-2.3) {$\underbrace{\hspace{12 cm}}_{\scalebox{1}{\quad \qquad\scalebox{1.5}{\text{Generalized quiver phase}}}}$};
        \end{tikzpicture}}
          \caption{The resolution of the singular 5d CM threefolds either gives rise to a special unitary quiver phase (left), or to a linear generalized quiver phase (right). The two low-energy theories are UV dual. }
    \label{fig:resolution}
    \end{figure}
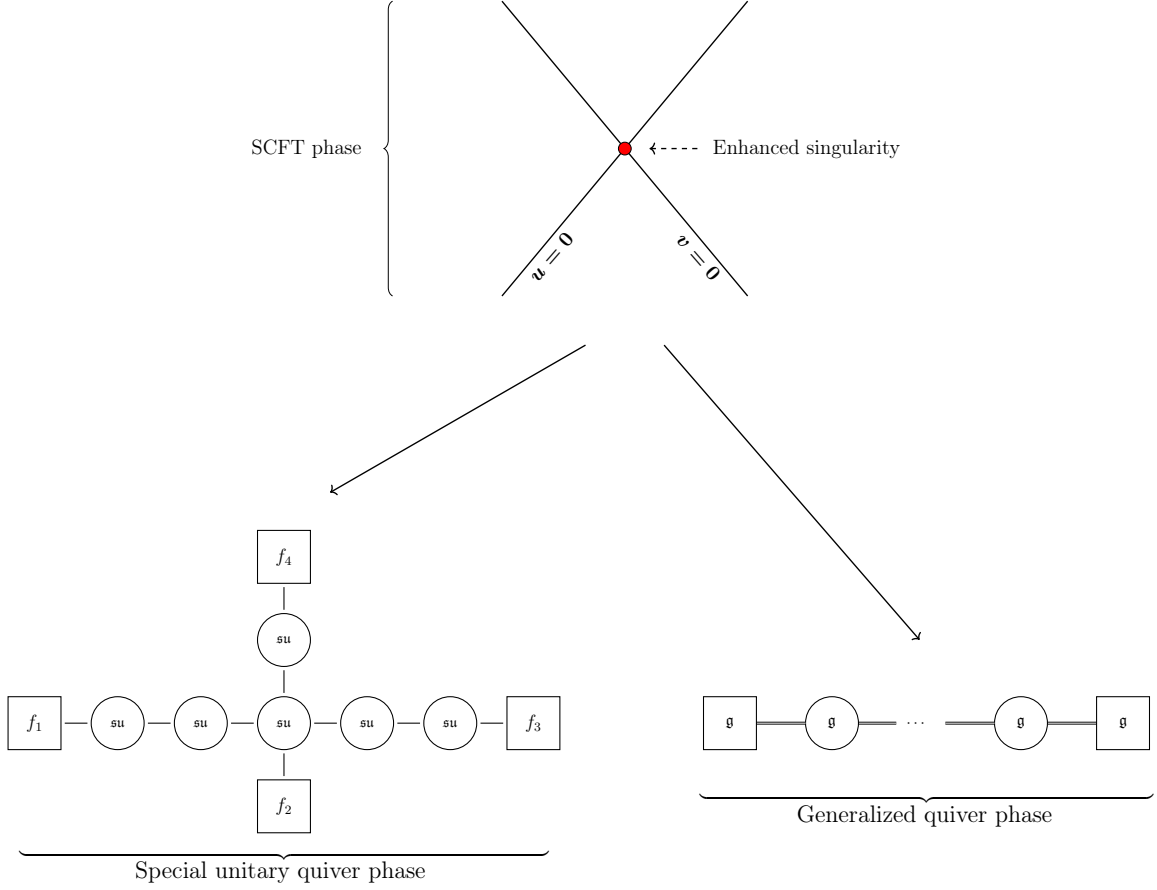

The singular CY3 \eqref{bifund 3folds} admits two preferred crepant resolutions \cite{DeMarco:2023irn}, which illustrate a novel 5d UV duality. On the one hand, they admit a low-energy gauge theory phase with special unitary gauge nodes and bifundamental hypermultiplets; on the other hand, they give rise to a generalized linear quiver with gauge nodes of type $\mathfrak{g}$, and edges labelled by a 5d CM theory. In order to distinguish the latter, we introduce quivers with double edges, so as to signify the fact that they do not correspond to hypermultiplets, but to full-fledged interacting CM theories. The two distinct low-energy (generalized) quiver gauge theories constitute a novel example of UV duality, as they both flow to the same 5d SCFT point\footnote{From the algebraic geometric point of view, this operation is equivalent to the blow-down of the resolved phase.}. A visual recap of the two resolution procedures is presented in Figure \ref{fig:resolution}.\\
\indent The special unitary quiver phase is extremely helpful to bolster the claim that the corresponding SCFT point enjoys $\mathfrak{g}\oplus\mathfrak{g}$ flavor symmetry. Indeed, the work of \cite{Yonekura} proves that any 5d low-energy special unitary quiver with balanced nodes, shaped like the $\mathfrak{g}$ Dynkin diagram, flows to a fixed point with \textit{at least} $\mathfrak{g}\oplus\mathfrak{g}$ flavor symmetry. This reasoning provides an independent field-theoretic check that the CY3 in \eqref{bifund 3folds} give rise to (at least) bifundamental CM theories. In general, the full flavor symmetry may contain additional $A$ factors.

Taking a look back at Table \ref{table:oscar}, in this section we have reviewed the existence of 5d CM SCFTs with at most \textit{two} flavor factors of type $D,E$, that we report here for convenience:        
 \renewcommand{\arraystretch}{1.1}
\begin{table}[H]
\centering
$
\begin{array}{|C||l|l|l|l|l|}
\hline
\text{Valency} & \boldsymbol{A_n} & \boldsymbol{D_{n}}  & \boldsymbol{E_6} & \boldsymbol{E_7} & \boldsymbol{E_8} \\
 \hline
 \hline
 2 & A^{\oplus 2}  & \begin{array}{l} D^{\oplus 2}\end{array} & \begin{array}{l}E_6^{\oplus 2} \\\end{array}& E_7^{\oplus 2} & \begin{array}{l}E_8^{\oplus 2} \\ \end{array}\\
  \hline
 3 & A^{\oplus 3} & \begin{array}{l} D^{\oplus 2}\oplus A\end{array}& E_6^{\oplus 2} \oplus A & \begin{array}{l} E_7^{\oplus 2}\oplus A\end{array}& E_8^{\oplus 2}\oplus A \\
  \hline
 4 &A^{\oplus 4}& \begin{array}{l}D^{\oplus 2} \oplus A^{\oplus 2}\end{array} & E_6^{\oplus 2} \oplus A^{\oplus 2}& E_7^{\oplus 2} \oplus A^{\oplus 2}& E_8^{\oplus 2} \oplus A^{\oplus 2}\\
  \hline
5 &A^{\oplus 5} & \begin{array}{l} D^{\oplus 2}\oplus A^{\oplus 3}\end{array} &E_6^{\oplus 2} \oplus A^{\oplus 3} &E_7^{\oplus 2} \oplus A^{\oplus 3} &E_7^{\oplus 2} \oplus A^{\oplus 3}  \\
  \hline
  6 &A^{\oplus 6} & \begin{array}{l}D^{\oplus 2}\oplus A^{\oplus 4}\end{array} &E_6^{\oplus 2} \oplus A^{\oplus 4}  & &  \\
  \hline
    \end{array}
    $
    \caption{Summary of $n$-valent 5d conformal matter theories that can be engineered with the technology of Section \ref{sec:5dbifundamental}. They can all be obtained by gauging of a finite set of families of 5d CM atoms. Notice that at most \textit{two} flavor factors of type $D,E$ singularities can be produced.}
    \label{table:2-fundamental CM}
\end{table}

\subsection{Novel Trinion and Tetraon 5d conformal matter of type \texorpdfstring{$D$}{D}}
\label{sec:5dtriniontetraon}
In the following, motivated by the review undertaken in the previous sections, we show how to construct novel 5d CM SCFTs which are:
\begin{itemize}
    \item non-toric;
    \item with 3-valent and 4-valent UV flavor symmetry of type $D$.
\end{itemize}
We shall do so employing the reduction of M-theory on non-compact CY3 with canonical singularities.
Abiding by the geometric engineering dictionary that we have already employed to construct bifundamental 5d CM in section \ref{sec:5dbifundamental}, in order to construct trinions (tetraons) we look for non-compact canonical CY3 with three (four) non-compact singular lines of type $D$. As we will see, this automatically gives rise to a UV flavor symmetry $F_{UV}$ of the following type:
\begin{equation}\label{UV flavor}
    \begin{split}
& \text{Trinions: } \quad F_{UV} \supseteq D_{2j+1}\oplus D_{k}\oplus D_{k}\\
& \text{Trinions: } \quad F_{UV} \supseteq E_7\oplus D_{k}\oplus D_{k}\\
& \text{Tetraons: } \quad F_{UV} \supseteq D_{2j}\oplus D_{k}\oplus D_{k}\oplus D_{k} \\
    \end{split},
\end{equation}
with $j\geq 2$ and $k\geq 4$.


Our starting building blocks are the Du Val singularities, that we have recalled in Table \ref{ADE sing}. Du Val singularities are hypersurfaces in $\mathbb{C}^3$: in order to build the desired threefolds we apply a suitable base-change, obtaining a hypersurface in $\mathbb{C}^4$.

Du Val singularities are classified as:
\begin{equation}
P_{\mathfrak f}(x_1,x_2,x_3) = 0 \qquad \mathfrak f \in ADE.
\end{equation}

A possible construction we might apply is the "base-change construction":
\begin{equation}
\label{eq:threefoldeqbifundamental}
\begin{cases}P_{\mathfrak f}(x_1,x_2,x_3) = 0, \\ x_i^n = Q_{\mathfrak{g}}(u,v), \end{cases}
\end{equation}
where $n \in\mathbb{N}$
and $Q_{\mathfrak{g}}(u,v)$ is one of the polynomials appearing \eqref{ADE sing}, with $x_1 = 0, x_2 = u, x_3 = v$.\\
Naturally, we must require that the threefold in \eqref{eq:threefoldeqbifundamental} is a canonical Calabi-Yau, i.e.\ it should satisfy the inequality \eqref{canonicity}. In turn, this imposes non-trivial stringent constraints on $n$ and on the shape of $Q(u,v)$.\\

The only allowed possibility is, provided that $Q_{\mathfrak{g}}(u,v) \neq uv$ (otherwise we recover a subcase of \eqref{bifund 3folds}): $n=1$, as well as $\mathfrak{f} = D_k$. In this case, $Q_{\mathfrak{g}}(u,v)$ can be factorized into either one, two or three factors. The cases that are consistent with the Calabi-Yau condition \eqref{canonicity} are as follows. In general, we make substitutions of the form:
\begin{equation}
    x_3 = Q_{\mathfrak{g}}(u,v),
\end{equation}
where we recall that $Q_{\mathfrak{g}}$ is the part of $P_{\mathfrak{g}}$ without the $x_1^2$ summand. The allowed quasi-homogeneous threefolds are collected in Table \ref{table:allthreefolds}. \footnote{For graphical ease, we have renamed $x_1 \rightarrow x, x_2 \rightarrow y, x_3 \rightarrow z$.} We shall often denote the threefolds as $X_k(\mathfrak g)$, since the only data specifying them is the choice of $D_k$ singularity and the factor $Q_{\mathfrak{g}}(u,v)$. Notice that trinion and tetraon 5d SCFTs of type $D$ with flavor symmetry as shown below Table \ref{tab:lacompletezzaperdio} appear, as advertised.\\
In the remainder of this work, we will especially be interested in the cases where $Q_{\mathfrak{g}}(u,v)$ has two or three factors, as they provide the most clear-cut examples of novel tri/tetra-valent conformal matter 5d SCFTs that go beyond the toric realm. We will indeed see that by increasing the number of irreducible factors of $Q_{\mathfrak g}(u,v)$, we also increase the number of singular curves in the blow-down phase of the CY3. In the next section we show how to perform a resolution of the singularities in Table \ref{table:allthreefolds}, that allows us to detect the rank of the Coulomb branch of the associated 5d SCFT.


\renewcommand{\arraystretch}{2}
\begin{table}[H]
\centering
\begin{equation*}
\begin{array}{|c|l|}
\hline
\hline
    \multicolumn{2}{|c|}{\textbf{$Q_{\mathfrak{g}}(u,v)$ has one factor}} \\
    \hline
    \hline

    \textbf{Bifundamental } \boldsymbol{(A_{2j},D_k)} \quad\quad & \begin{cases}
        x^2 +z y^2+z^{k-1} =0 \\
        z = u^2+v^{2j+1}
        \end{cases}
     \\
   
    \textbf{Bifundamental } \boldsymbol{(E_6,D_k)} \quad\quad & \begin{cases}
        x^2 +z y^2+z^{k-1} =0 \\
        z = u^3+v^4
        \end{cases}
     \\
   
    \textbf{Bifundamental } \boldsymbol{(E_8,D_k)} \quad\quad &\begin{cases}
        x^2 +z y^2+z^{k-1} =0 \\
        z = u^3+v^5
        \end{cases}
   \\
\hline
\hline
    \multicolumn{2}{|c|}{\textbf{$Q_{\mathfrak{g}}(u,v)$ has two factors}} \\
    \hline
    \hline

    \textbf{Trinion } \boldsymbol{(A_{2j+1},D_k,D_k)  } &\hspace{-0.8cm} \quad\quad \begin{cases}
        x^2 +z y^2+z^{k-1} =0 \\
        z = (u+v^{j+1})(u-v^{j+1})
        \end{cases}
    \\
   
    \textbf{Trinion } \boldsymbol{(D_{2j+1},D_k,D_k)  } &\hspace{-0.8cm} \quad\quad \begin{cases}
        x^2 +z y^2+z^{k-1} =0 \\
        z = u(u^{2j-1}+v^2)
        \end{cases}
    \\
   
    \textbf{Trinion } \boldsymbol{(E_7,D_k,D_k) }\quad\quad &\begin{cases}
        x^2 +z y^2+z^{k-1} =0 \\
        z = u(u^2+v^3)
        \end{cases}
    \\
\hline
\hline
    \multicolumn{2}{|c|}{\textbf{$Q_{\mathfrak{g}}(u,v)$ has three factors}} \\
    \hline
    \hline
    
    \textbf{Tetraon } \boldsymbol{(D_{2j},D_k,D_k,D_k) }\quad\quad & \begin{cases}
        x^2 +z y^2+z^{k-1} =0 \\
        z = u(u^{j-1}+v)(u^{j-1}-v)
        \end{cases}
    \\
    \hline
    \end{array}
    \end{equation*}
    \caption{Novel 2-, 3- and 4-valent 5d conformal matter theories. Notice that they are all of type $(\mathfrak{g},D_k^{\oplus n})$, for some appropriate $n$.}
    \label{table:allthreefolds}
\end{table}

\section{5d SCFT data for novel bifundamental/trinion/tetraon theories}
\label{sec:CBdata}
In this section, we wish to examine the CY3 in Table \ref{table:allthreefolds} corresponding to trinions and tetraons (as well as bifundamentals) in closer detail, in order to glean rigorous data labeling the corresponding 5d SCFTs. We will do so by resolving the CY3, determining in this way the rank of the corresponding 5d SCFT. As a bonus, the partially resolved phase, which can be presented in the form of a generalized quiver gauge theory, will allow us to present a non-trivial check of the rank of the UV flavor symmetry, hinting at a generalization of the flavor symmetry enhancement argument introduced by \cite{Yonekura}.\\

The non-compact CY3 in Table \ref{table:allthreefolds} sport a collection of two, three and four, respectively, non-compact singular lines, supporting Du Val singularities of type $A,D,E$:
\begin{equation}\label{singular threefolds}
\scalemath{0.8}{
\renewcommand{\arraystretch}{1.2}
\begin{array}{l|l|ll|l}
 \begin{array}{c}
    \textbf{Bifundamental }\\
    (A_{2j},D_k)\\
    \end{array}:\hspace{0.3cm} & \begin{cases}
        x^2 +z y^2+z^{k-1} =0 \\
        z = u^2+v^{2j+1}
        \end{cases} \hspace{0.3cm} &
    \textbf{Singularities: }\quad &
    \renewcommand{\arraystretch}{1}\begin{array}{l}
         x = y = u^2+v^{2j+1} = 0 \\
         x = u = v = 0\\
    \end{array}\quad &\begin{array}{l}
       D_k \text{ type} \\
       A_{2j} \text{ type} \\
    \end{array} \\
\hline
\hline
 \begin{array}{c}
    \textbf{Bifundamental }\\
    (E_6,D_k)\\
    \end{array}:\hspace{0.3cm} & \begin{cases}
        x^2 +z y^2+z^{k-1} =0 \\
        z = u^3+v^4
        \end{cases} \hspace{0.3cm} &
    \textbf{Singularities: }\quad &
    \renewcommand{\arraystretch}{1}\begin{array}{l}
         x = y = u^3+v^4 = 0 \\
         x = u = v = 0\\
    \end{array}\quad &\begin{array}{l}
       D_k \text{ type} \\
       E_6 \text{ type} \\
    \end{array} \\
\hline
\hline
 \begin{array}{c}
    \textbf{Bifundamental }\\
    (E_8,D_k)\\
    \end{array}:\hspace{0.3cm} & \begin{cases}
        x^2 +z y^2+z^{k-1} =0 \\
        z = u^3+v^5
        \end{cases} \hspace{0.3cm} &
    \textbf{Singularities: }\quad &
    \renewcommand{\arraystretch}{1}\begin{array}{l}
         x = y = u^3+v^5 = 0\\
         x = u = v = 0\\
    \end{array}\quad &\begin{array}{l}
       D_k \text{ type} \\
       E_8 \text{ type} \\
    \end{array} \\
\hline
\hline
 \begin{array}{c}
    \textbf{Trinion }\\
    (A_{2j+1},D_k,D_k)\\
    \end{array}:\hspace{0.3cm} & \begin{cases}
        x^2 +z y^2+z^{k-1} =0 \\
        z =(u+v^{j+1})(u-v^{j+1})
        \end{cases} \hspace{0.3cm} &
    \textbf{Singularities: }\quad &
    \renewcommand{\arraystretch}{1}\begin{array}{l}
         x = y = u+v^{j+1} = 0\\
         x = y = u-v^{j+1} = 0 \\
         x = u = v = 0\\
    \end{array}\quad &\begin{array}{l}
       D_k \text{ type} \\
       D_k \text{ type} \\
       A_{2j+1} \text{ type} \\
    \end{array} \\
\hline
\hline
\begin{array}{c}
    \textbf{Trinion }\\
    (D_{2j+1},D_k,D_k)\\
    \end{array}:\hspace{0.3cm} & \begin{cases}
        x^2 +z y^2+z^{k-1} =0 \\
        z = u(u^{2j-1}+v^2)
        \end{cases} \hspace{0.3cm} &
    \textbf{Singularities: }\quad &
    \renewcommand{\arraystretch}{1}\begin{array}{l}
         x = y = u = 0\\
         x = y = u^{2j-1}+v^2 = 0 \\
         x = u = v = 0\\
    \end{array}\quad &\begin{array}{l}
       D_k \text{ type} \\
       D_k \text{ type} \\
       D_{2j+1} \text{ type} \\
    \end{array} \\
\hline
\hline
\begin{array}{c}
    \textbf{Trinion }\\
    (E_7,D_k,D_k)\\
    \end{array}:\hspace{0.3cm} & \begin{cases}
        x^2 +z y^2+z^{k-1} =0 \\
        z = u(u^2+v^3)
        \end{cases} \hspace{0.3cm} &
    \textbf{Singularities: }\quad &
    \renewcommand{\arraystretch}{1}\begin{array}{l}
         x = y = u = 0\\
         x = y = u^2+v^3 = 0 \\
         x = u = v = 0\\
    \end{array}\quad & \begin{array}{l}
       D_k \text{ type} \\
       D_k \text{ type} \\
       E_7 \text{ type} \\
    \end{array} \\
    \hline
    \hline
\begin{array}{c}
     \textbf{Tetraon }\\
     (D_{2j},D_k,D_k,D_k)
     \end{array}: \hspace{0.3cm} & \begin{cases}
        x^2 +z y^2+z^{k-1} =0 \\
        z = u(u^{j-1}+v)(u^{j-1}-v)
        \end{cases}\hspace{0.3cm} &
     \textbf{Singularities: }\quad & \renewcommand{\arraystretch}{1}\begin{array}{l}
         x = y = u = 0\\
         x = y = u^{j-1}+v = 0 \\
         x = y = u^{j-1}-v = 0 \\
         x = u = v = 0\\
    \end{array} \quad & \begin{array}{l}
       D_k \text{ type} \\
       D_k \text{ type} \\
       D_k \text{ type} \\
       D_{2j} \text{ type} \\
    \end{array}
\end{array}}
\end{equation}
Given the pattern of singularities, it is immediate to observe that M-theory geometric engineering on \eqref{singular threefolds} produces 5d SCFTs with flavor symmetry as in \eqref{UV flavor}. Each flavor factor is supported on one of the non-compact singular lines.

\subsection{Crepant resolution and Coulomb branch data}\label{sec:resolution}

We now embark on the task of partially resolving the threefolds in \eqref{singular threefolds} via a resolution
\begin{equation}
    \pi_1 \circ \pi_2: \widetilde{\widetilde{X}} \to \widetilde{X} \to X,
\end{equation}
with $X$ the singular threefolds. This will allow us to detect the ranks of the corresponding 5d SCFTs. We will start by summing up the resolution procedure of \cite{DeMarco:2023irn} and apply it to the threefolds in Table \ref{table:allthreefolds}, obtaining partially resolved threefolds $\pi_1: \widetilde{X} \to X$. After this first step, the partially resolved threefolds will still display residual singularities that require a further resolution $\pi_2: \widetilde{\widetilde{X}} \to \widetilde{X}$ that inflates  \textit{compact} divisors. We will split the divisors inflated by $\pi_2$ into  "horizontal" and "vertical" divisors, according to whether they are contracted to points or complex curves by $\pi_2$. We will manage to take into account the contributions of both these kinds of divisors reducing, at the end of the section, the problem to the study of local models. 
Finally, we will combine these local contributions to extract the rank of M-theory on the threefolds in \cref{table:allthreefolds}.\\

\begin{figure}
    \centering
    \scalebox{0.8}{
    \begin{tikzpicture}
        \draw[thick] (0,0) circle (0.65);
        \node at (0,0) {\small$\mathfrak{g}$};
        \draw[thick,double] (0.7,0)--(1.3,0);
        \draw[thick] (2,0) circle (0.65);
        \node at (2,0) {\small$\mathfrak{g}$};
        \draw[thick,double] (-0.7,0)--(-1.3,0);
        \draw[thick,double] (2.7,0)--(3.3,0);
        \node at (4,0) {$\ldots$};
        \draw[thick,double] (4.7,0)--(5.3,0);
        \draw[thick] (6,0) circle (0.65);
        \draw[thick,double] (6.7,0)--(7.4,0);
        \draw[thick] (7.5,-0.5)--(8.5,-0.5)--(8.5,0.5)--(7.5,0.5)--cycle ;
        \node at (6,0) {\small${\mathfrak g}$};
        \node at (8,0) {\small${\mathfrak g}$};
        \node at (-2,0) {\small\text{smooth}};
        \node at (0,2) {\small\text{smooth}};
        \node at (0.5,1) {\small$\mathcal{T}_{\mathfrak{g}}$};
        \node at (-1,-0.5) {\small$\mathcal{T}_{\mathfrak{g}}$};
         \node at (1,-0.5) {\small$X_{\mathfrak{g}}^{(1)}$};
        \draw[thick] (0,2) circle (0.65);
        \node at (3,-0.5) {\small$X_{\mathfrak{g}}^{(1)}$};
        \node at (5,-0.5) {\small$X_{\mathfrak{g}}^{(1)}$};
        \node at (7,-0.5) {\small$X_{\mathfrak{g}}^{(1)}$};
        \draw[thick] (-2,0) circle (0.65);
        \draw[thick,double] (0,0.7)--(0,1.3);
        \end{tikzpicture}}
    \caption{Pictorial representation of $\widetilde{X}$. The compact curves are arranged like a $D_k$ Dynkin diagram.}
    \label{D4zfig}
    \end{figure}
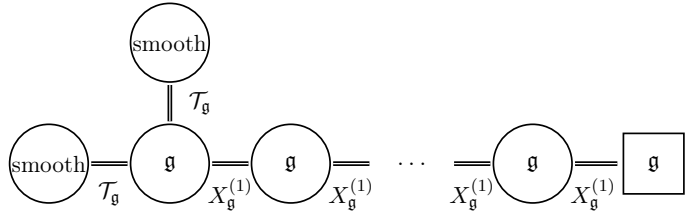

\indent Let us start by recalling the resolution procedure presented in\footnote{The atlas used for the $A_n$ resolution was, e.g., presented in \cite{CONSTELLATION}.} \cite{DeMarco:2023irn}: 
\begin{enumerate}
    \item Perform the resolution of the $P_{D_{k}}$ Du Val \textit{surface} singularity displayed in \eqref{ADE sing}, obtaining the following expressions: 
    \begin{eqnarray}
        \label{eq:resfrombasechange}
        \begin{cases}
        x_{1} = x_1(a_j,b_j), \nonumber \\
        x_{2} = x_{2}(a_j,b_j),\nonumber \\
        x_{3} = x_3(a_j,b_j),
        \end{cases}
    \end{eqnarray}
    where the coordinates $(a_j,b_j)  \in \mathbb C^2$ constitute an atlas of  the \textit{resolved} Du Val surface and the blow-down map is described by the functions $x_i(a_j,b_j)$.
    \item To produce the first partial resolution $\pi_1: \widetilde{X} \to X$, we set $x_3 = Q_{\mathfrak g}(u,v)$, obtaining the partial resolution of (respectively). We note that, at this point, the remaining equations $x_{k} = x_k(a_j,b_j)$, with $k \neq 3$, are spurious and can be dropped. 
\end{enumerate}
This procedure resolves the lines of $P_{D_k}$ of the threefold listed in \cref{table:allthreefolds} and corresponds to lifting to the \textit{resolved} Du Val surface $\widetilde{P}_{D_k}$ the base-change that we used to obtain the aforementioned threefolds from the \textit{singular} Du Val surface $P_{D_k}$. We call  $\widetilde{X}$ the partially resolved threefold obtained in this way and 
$\pi_1$ the partial resolution map 
\begin{equation}
\label{eq:partialresmap}
    \pi_1: \quad \widetilde{X} \longrightarrow X,
\end{equation}
where $X$ is one of the threefolds in \cref{table:allthreefolds}.
 The singular locus of $\widetilde{X}$ is the union of the subvariety $\pi_{1}^{-1}(0)$ (that is blown-down to the origin) with the non-compact singular line of type $\mathfrak g$ (associated to the polynomial $Q_{\mathfrak g}(u,v)$) inside the threefolds of \cref{table:allthreefolds}. $\pi_1^{-1}(0)$ consists of various $\mathbb{P}^1_i$'s, with $i = 1, \ldots, k$, as in Figure \ref{D4zfig}.
Each curve associated to a node of Figure \ref{D4zfig} labeled by $\mathfrak{g}$ supports a line of $P_{\mathfrak{g}}$ singularities.
 The edges are labeled as follows:
 \begin{itemize}
     \item The label $X_{\mathfrak{g}}^{(1)}$ keeps track of the fact that the edge is a non-trivial 5d \textit{bifundamental} conformal matter SCFT. Locally near the edge, the threefold takes the schematic form that we have reviewed in \eqref{elementary 3folds}:
     \begin{equation}\label{CM x}
         \left(x^2+Q_{\mathfrak{g}}(u,v) + \cdots\right)|_{x=ab}  =a^2b^2+Q_{\mathfrak{g}}(u,v) + \cdots = 0,
     \end{equation}
     with the dots standing for subleading terms that do not modify the CY3 geometry near the considered edge.\footnote{From now on, we will simply omit these subleading terms.} The properties of the corresponding SCFT were studied in detail in \cite{DeMarco:2023irn}.
       \item The label $\mathcal{T}_{\mathfrak{g}}$ keeps track of the fact that the edge is a non-trivial 5d \textit{1-valent} conformal matter SCFT\footnote{It would be interesting to appreciate the role of the $\mathcal{T}_{\mathfrak{g}}$ theories in more nuance. A suggestive hypothesis, that requires further investigation, is that they are the 5d parent of some 4d $\mathcal{N}=2$ $D_p(G)$ theories.}. Locally near the edge, up to subleading terms the threefold takes the schematic form:
     \begin{equation}\label{T_g eq}
         ab^2+Q_{\mathfrak{g}}(u,v) = 0.
     \end{equation}
 \end{itemize}

 
 The threefold $\widetilde{X}$ has residual singularities, concentrated along the lines corresponding to nodes denoted by $\mathfrak g$ in Figure \ref{D4zfig}, that enhance at the intersection points with \textit{all} the other nodes.\\
 \indent Let us move to the next step, and consider a resolution $\pi_2: \widetilde{\widetilde{X}}\to \widetilde{X}$, obtained with a sequence of blow-up whose centers "dominate\footnote{By this we mean the following: given $Y \subset X$, $\widetilde{Y} \subset \widetilde{X}$, and a map $\pi: \widetilde{X} \to X$, we say that $\widetilde{Y}$ dominates $Y$ via the blow-up map $\pi$ if $\pi(\widetilde{Y}) = Y$.}" the lines of Du Val singularities, and organize the possible divisors that appear in the $\pi_2$ resolution as: 
 \begin{itemize}
 \item "horizontal" divisors, if they dominate, via the map $\pi_2$, the complex curves associated to the nodes of Figure \ref{D4zfig};
 \item "vertical" divisors, if they are contracted to the intersection points between the curves in Figure \ref{D4zfig}.
 \end{itemize}
 We can now count the total number of compact divisors as follows: 
 \begin{itemize}
     \item $\text{rank}(\mathfrak g)$ horizontal divisors for each node labelled with $\mathfrak g$ in Figure \ref{D4zfig};
     \item no horizontal divisors for the nodes labeled by "smooth" in Figure \ref{D4zfig}, 
     \item $\text{rank}(X_{\mathfrak g}^{(1)})$ vertical divisors (following the notation of \cite{DeMarco:2025ugw}) for each intersection between two nodes labeled by $\mathfrak g$ in Figure \ref{D4zfig}. $\text{rank}(X_{\mathfrak g}^{(1)})$ is the rank of the 5d bifundamental CM labeled by the Lie algebra $\mathfrak{g}$, and given by the threefold $X_{\mathfrak{g}}^{(1)}$ in \eqref{elementary 3folds}, and it can be read off from \cite{DeMarco:2023irn};
     \item We are only left to count $r_{\mathfrak g}$, the number of vertical divisors associated with the theories $\mathcal{T}_{\mathfrak{g}}$. 
 \end{itemize}
We can already present, at this stage, the general formula for the rank of Coulomb branch of the 5d SCFT engineered by M-theory on the threefolds $X_k(\mathfrak g)$ in \cref{table:allthreefolds}:
\begin{equation}
    \label{eq:rankfinal}
    \text{rank}\Big(X_k(\mathfrak g)\Big) = 2 r_{\mathfrak g} +(k-2) \text{rank}(X_{\mathfrak g}^{(1)})+ (k-2) \text{rank}(\mathfrak g),
\end{equation}
where the first two summands are the contributions of the "vertical" divisors, while the last one is the contribution of the "horizontal" ones.\\
The values of $r_{\mathfrak{g}}$ can be explicitly computed following Appendix \ref{app:vertical}. The end result is:
\begin{eqnarray}
\label{eq:ranktopmaintext}
  r_{A_{k}} = 0,\quad  r_{D_{2j}} = j-1, \quad r_{D_{2j+1}} = j-1, \quad r_{E_6} = 1, \quad r_{E_7} = 3, \quad r_{E_8} = 4 
\end{eqnarray}
The expressions \eqref{eq:rankfinal} and \eqref{eq:ranktopmaintext} encode the dimensions of the Coulomb branch of the trinions and tetraons in Table \ref{table:allthreefolds}.\\ 

To conclude this section, we highlight that the resolution we just presented does not produce a weakly coupled 5d gauge phase. It is then natural to wonder whether we can find other resolutions associated with a Lagrangian description of the SCFT. Indeed, a necessary condition for this is the presence of a well-defined IIA limit in the associated chamber of the K\"ahler cone. For a Lagrangian phase with special unitary groups, this amounts to requiring that the threefold is $\mathbb C^*$-fibered, with the $\mathbb C^*$ acting with opposite weights on two of the ambient space coordinates. One can readily check that there is no such action for the threefolds of \eqref{singular threefolds} even in the singular phase, and then also in any chamber of the K\"ahler cone. Another option, producing a Lagrangian phase with symplectic and orthogonal gauge groups, requires (as a \textit{necessary} condition) that the threefold can be written as \cite{Collinucci_2009}
\begin{equation}
    \label{eq:colllinucciesole}
    0=x^2 + y^2 z(u,v) + ...,
\end{equation}
with "$...$" denoting a generic polynomial in $z(u,v)$.
 In this case, the IIA limit is realized (at the holomorphic level) as the projection onto the $(u,v)$ coordinates, and it contains stacks of D6-branes on top of O6-planes.\footnote{This means that the $\mathbb C^*$ fiber is described by varying $x,y$ at fixed $u,v$ in \eqref{eq:colllinucciesole}.} The threefolds in \eqref{singular threefolds} are indeed compatible with \eqref{eq:colllinucciesole}, however, this  is not \textit{sufficient} for them to admit such IIA limit. In fact, in the setup considered in \cite{Collinucci_2009}, there are no fixed values of $u,v$ such that the threefold \eqref{eq:colllinucciesole} develops a curve of singularities. In our case, instead, the CY3 of \eqref{singular threefolds} has a curve of singularities at $u = v = 0$, and so it is quantitatively different from the geometries considered in \cite{Collinucci_2009}. An intuitive explanation of why in this case we do not find such IIA limit is the following:  the d.o.f.\ living along the aforementioned curve cannot be encoded in the IIA limit of \cite{Collinucci_2009} as their support completely disappears when we forget about $x,y$ performing the projection onto $(u,v)$; this prevents us from reaching the associated Lagrangian phase.   

\subsection{Trinions and tetraons as molecules}
\label{sec:trinionsasmolecules}
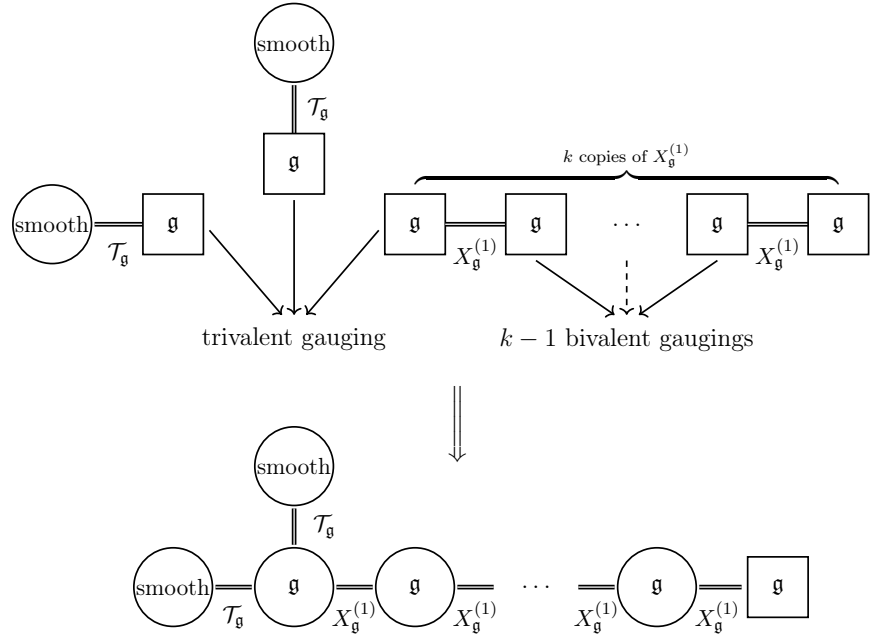
\begin{figure}[H]
    \centering
    \scalebox{0.8}{
    \begin{tikzpicture}
 \draw[thick] (-4,6) circle (0.65);
  \draw[thick,double] (-3.3,6)--(-2.5,6);
        \node at (-4,6) {\small smooth};
        \node at (-2,6) {\small$\mathfrak{g}$};
        \draw[thick] (-2.5,5.5)--(-1.5,5.5)--(-1.5,6.5)--(-2.5,6.5)--cycle ;
    \draw[thick] (0,9) circle (0.65);
  \draw[thick,double] (0,8.3)--(0,7.5);
        \node at (0,9) {\small smooth};
        \node at (0,7) {\small$\mathfrak{g}$};
        \draw[thick] (-0.5,6.5)--(0.5,6.5)--(0.5,7.5)--(-0.5,7.5)--cycle;
    \draw[thick] (1.5,5.5)--(2.5,5.5)--(2.5,6.5)--(1.5,6.5)--cycle;
  \draw[thick,double] (2.5,6)--(3.5,6);
        \node at (2,6) {\small $\mathfrak{g}$};
        \node at (4,6) {\small$\mathfrak{g}$};
        \draw[thick] (3.5,5.5)--(4.5,5.5)--(4.5,6.5)--(3.5,6.5)--cycle;
        \node at (5.5,6) {$\ldots$};
        \draw[thick] (6.5,5.5)--(7.5,5.5)--(7.5,6.5)--(6.5,6.5)--cycle;
  \draw[thick,double] (7.5,6)--(8.5,6);
        \node at (7,6) {\small $\mathfrak{g}$};
        \node at (9,6) {\small$\mathfrak{g}$};
        \draw[thick] (8.5,5.5)--(9.5,5.5)--(9.5,6.5)--(8.5,6.5)--cycle;
        \node at (5.5,7) {$\overbrace{\hspace{7cm}}^{k \text{ copies of } X_{\mathfrak{g}}^{(1)}}$};
        \node at (8,5.5) {\small$X_{\mathfrak{g}}^{(1)}$};
        \node at (3,5.5) {\small$X_{\mathfrak{g}}^{(1)}$};
        \node at (-2.9,5.5) {\small$\mathcal{T}_{\mathfrak{g}}$};
        \node at (0.4,7.9) {\small$\mathcal{T}_{\mathfrak{g}}$};
        \node at (0,4.1) {trivalent gauging};
        \draw[thick,<-] (0.2,4.5)--(1.4,5.9);
        \draw[thick,<-] (-0.2,4.5)--(-1.4,5.9);
        \draw[thick,->] (0,6.4)--(0,4.5);
        \node at (5.5,4.1) {$k-1$ bivalent gaugings};
        \draw[thick,->] (4,5.4)--(5.3,4.5);
        \draw[thick,dashed,->] (5.5,5.4)--(5.5,4.5);
        \draw[thick,->] (7,5.4)--(5.7,4.5);
        \node at (2.7,2.7)  {$\Bigg \Downarrow$};

        \draw[thick] (0,0) circle (0.65);
        \node at (0,0) {\small$\mathfrak{g}$};
        \draw[thick,double] (0.7,0)--(1.3,0);
        \draw[thick] (2,0) circle (0.65);
        \node at (2,0) {\small$\mathfrak{g}$};
        \draw[thick,double] (-0.7,0)--(-1.3,0);
        \draw[thick,double] (2.7,0)--(3.3,0);
        \node at (4,0) {$\ldots$};
        \draw[thick,double] (4.7,0)--(5.3,0);
        \draw[thick] (6,0) circle (0.65);
        \draw[thick,double] (6.7,0)--(7.4,0);
        \draw[thick] (7.5,-0.5)--(8.5,-0.5)--(8.5,0.5)--(7.5,0.5)--cycle ;
        \node at (6,0) {\small${\mathfrak g}$};
        \node at (8,0) {\small${\mathfrak g}$};
        \node at (-2,0) {\small\text{smooth}};
        \node at (0,2) {\small\text{smooth}};
        \node at (0.5,1) {\small$\mathcal{T}_{\mathfrak{g}}$};
        \node at (-1,-0.5) {\small$\mathcal{T}_{\mathfrak{g}}$};
         \node at (1,-0.5) {\small$X_{\mathfrak{g}}^{(1)}$};
        \draw[thick] (0,2) circle (0.65);
        \node at (3,-0.5) {\small$X_{\mathfrak{g}}^{(1)}$};
        \node at (5,-0.5) {\small$X_{\mathfrak{g}}^{(1)}$};
        \node at (7,-0.5) {\small$X_{\mathfrak{g}}^{(1)}$};
        \draw[thick] (-2,0) circle (0.65);
        \draw[thick,double] (0,0.7)--(0,1.3);
        \end{tikzpicture}}
    \caption{The novel bifundamental, trinion and tetraon theories are not irreducible: they arise as a trivalent gauging of $X_{\mathfrak{g}}^{(1)}$ bifundamental conformal matter and the $\mathcal{T}_{\mathfrak{g}}$ 1-valent theories.}
    \label{fig:gauging}
    \end{figure}

As we have hinted at in the Introduction, the novel bifundamental, trinion and tetraon theories, corresponding to the CY3 in Table \ref{table:allthreefolds}, \textit{are not irreducible}. Recall that these theories are labelled by their flavor group $(\mathfrak{g},D^{\oplus n})$ for some $n\leq 3$. As is evident from Figure \ref{D4zfig}, the CY3 at hand display a partial resolution which is a generalized quiver containing at least one gauge node of type $\mathfrak{g}$. Thus, they do not satisfy the definition of atom SCFT that we have established in the Introduction\footnote{Recall that an SCFT with flavor of type $(\mathfrak{g}_1,\ldots,\mathfrak{g}_k)$ is an atom if and only if it does not admit any low-energy (generalized) quiver phase with a gauge algebra among $\mathfrak{g}_i$, with $i=1,\ldots,k$.}. We can elucidate the non-atomic structure of the novel bifundamental, trinion and tetraon theories by ungauging appropriate nodes in the generalized quiver in Figure \ref{D4zfig}. Consider the case of the theory $(\mathfrak{g},D_k^{\oplus n})$, with $n\leq 3$, according to Table \ref{table:allthreefolds}. It can be noticed that it arises from the gauging of $k-2$ atoms of $X_{\mathfrak{g}}^{(1)}$ bifundamental conformal matter, with two 1-valent theories of type $\mathcal{T}_{\mathfrak{g}}$. The gauging of the $X_{\mathfrak{g}}^{(1)}$ atoms happens along the diagonal subgroup of two $\mathfrak{g}$ flavor factors, forming a bifundamental molecule. Then, the molecule is glued with the $\mathcal{T}_{\mathfrak{g}}$ theories by gauging three flavor nodes into a trivalent gauge node. The overall gauging procedure is visually summarized in Figure \ref{fig:gauging}.

It is evident that the gauging can also be exploited to produce the following construction: one can gauge the rightmost flavor node of the quiver at the bottom of Figure \ref{fig:gauging} with a bifundamental conformal matter of type $X_{\mathfrak{g}}^{(1)}$, producing a generalized quiver for the bifundamental/trinion/tetraon of type $(\mathfrak{g},D_{k+1}^n)$, since now the gauge nodes take the shape of the $D_{k+1}$ Dynkin diagram. This operation can be iterated an arbitrary number of times, producing generalized quivers of type $(\mathfrak{g},D_{k'}^n)$, for any $k'>k$. We will explore this avenue in deeper detail in Section \ref{sec:gluingtrinionstetraonsdnode}.

\subsection{Novel UV flavor enhancement from the generalized quiver phase}\label{sec:flavor enhancement}

The resolved phase depicted in Figure \ref{D4zfig} conveniently allows us to provide a check for the rank of flavor symmetry of the theories engineered by the singularities in Table  \ref{table:allthreefolds}. On the one hand, the pattern of non-isolated singular lines in the threefold gives a prediction for the flavor symmetry of the 5d theory at the fixed point. This is exactly what is summarized in Table \ref{table:allthreefolds}, where we obtain bifundamentals, trinions and tetraons of type $(\mathfrak{g},D_k^{\oplus n})$, for some suitable $n$.\\
\indent On the other hand, the field-theoretic interpretation of Figure \ref{D4zfig}, viewed as a generalized quiver gauge theory, tells us that:
\begin{itemize}
    \item each gauge node labeled by $\mathfrak{g}$ provides a topological $U(1)_T$ factor.
    \item Each edge labeled by $X_{\mathfrak{g}}^{(1)}$ corresponds to the 5d CM theory $X_{\mathfrak{g}}^{(1)}$. As it can be read off from Table 4 of \cite{DeMarco:2023irn}, such theories have flavor symmetry:
    \begin{equation}
        \mathfrak{g}\times \mathfrak{g}\times G_{rest}.
    \end{equation}
    Since the two $\mathfrak{g}$ factors are gauged in the quiver in Figure \ref{D4zfig} (apart from the rightmost node), each $X_{\mathfrak{g}}^{(1)}$ edge contributes with $G_{rest}$ to the flavor symmetry. The rightmost edge supplies $\mathfrak{g}\times G_{rest}$ to the flavor symmetry.
    \item Each edge labeled by $\mathcal{T}_{\mathfrak{g}}$ provides some flavor symmetry rank $f_{\mathfrak{g}}$. The rank $f_{\mathfrak{g}}$ can be explicitly computed analyzing the singular threefolds \eqref{T_g eq}, as shown in Appendix \ref{app:vertical}. The end result is:
    \begin{equation}\label{flavor rank T}
\scalemath{1}{
\renewcommand{\arraystretch}{1.2}
\begin{array}{c|c|c|c|c|c|c|c}
\boldsymbol{\mathfrak{g}}& A_{2j} & A_{2j+1}&  D_{2j} & D_{2j+1} & E_6 & E_7 & E_8\\
\hline
\hline
\boldsymbol{f_{\mathfrak{g}}} & 1 & 2 & 3 & 2 & 1 & 2 & 1\\
\end{array}
}
\end{equation}
\end{itemize}
Combining all the field-theoretic contributions, that we have just elucidated, to the UV flavor rank of the 5d SCFTs engineered by the CY3 in Table \ref{table:allthreefolds}, one finds complete agreement with what is expected from the non-isolated singularities in the Calabi-Yau threefold. Indeed, the expected total UV flavor rank of a theory of type $(\mathfrak{g},D_k^{\oplus n})$ is:
\begin{equation}
    \text{Total UV flavor rank} = \underbrace{k-2}_{U(1)_T \text{ contribution}} + \underbrace{(k-2)\text{rank}(G_{rest})+\text{rank}(\mathfrak{g}) }_{X_{\mathfrak{g}}^{(1)} \text{ contribution}}+\underbrace{2 f_{\mathfrak{g}}}_{\mathcal{T}_{\mathfrak{g}} \text{ contribution}},
\end{equation}
with $f_{\mathfrak g}$ and $G_{rest}$ being determined, in principle, by $\mathfrak g$ and $n$. In each case, we check that 
\begin{equation}
\label{eq:flavordataton}
   \text{rank}(G_{rest})-1 = f_{\mathfrak g} = n, 
\end{equation}
consistently with the M-theory top-down expectation.\\ 
\indent Let us summarize the relevant data for all the ADE algebras in Table \ref{flavor data}, where we highlight the total flavor rank of the bifundamental, trinion and tetraon theories of type $(\mathfrak{g},D_k^{\oplus n})$, with appropriate $n$, appearing in Table \ref{table:allthreefolds}.\\
 \begin{table}[H]
\renewcommand{\arraystretch}{1.2}
\begin{tabular}{|c|c|c|c|c|}
\hline
\diagbox{$\boldsymbol{\mathfrak{g}}$}{\footnotesize\makecell{\textbf{Flavor} \\\textbf{contribution}}}& $U(1)_T$ &  $X_{\mathfrak{g}}^{(1)}$ edges & $\mathcal{T}_{\mathfrak{g}}$ edges & \makecell{Total UV \\ flavor rank}\\
\hline
\hline
$\boldsymbol{A_{2j}}$ & $k-2$ & $0+\text{rank}(\mathfrak{g})$ & 2 & $k+\text{rank}(\mathfrak{g})$ \\
$\boldsymbol{A_{2j+1}}$ & $k-2$ & $(k-2)+\text{rank}(\mathfrak{g})$ & 4 & $2k+\text{rank}(\mathfrak{g})$\\
$\boldsymbol{D_{2j}}$ & $k-2$ & $2(k-2)+\text{rank}(\mathfrak{g})$ & 6 & $3k+\text{rank}(\mathfrak{g})$\\
$\boldsymbol{D_{2j+1}}$ & $k-2$ & $(k-2)+\text{rank}(\mathfrak{g})$ & 4 &$2k+\text{rank}(\mathfrak{g})$\\
$\boldsymbol{E_6}$ & $k-2$ & $0+\text{rank}(\mathfrak{g})$ & 2 & $k+\text{rank}(\mathfrak{g})$\\
$\boldsymbol{E_7}$ & $k-2$ & $(k-2)+\text{rank}(\mathfrak{g})$ & 4 & $2k+\text{rank}(\mathfrak{g})$\\
$\boldsymbol{E_8}$ & $k-2$ & $0+\text{rank}(\mathfrak{g})$& 2 & $k+\text{rank}(\mathfrak{g})$\\
\hline
\end{tabular}
\caption{Table summarizing the flavor rank contributions to the theories $(\mathfrak{g},D_k^{\oplus n})$ from the generalized quiver perspective, obtained from Figure \ref{D4zfig}. They precisely match the rank of the expected flavor symmetry detected from the corresponding CY3.}
\label{flavor data}
\end{table}
\indent We have shown the striking agreement on the rank of the UV flavor symmetry between the geometric CY3 point of view and the field-theoretic viewpoint on the generalized quiver in Figure \ref{D4zfig}. This observation clearly hints at a deeper interpretation in terms of instantonic symmetry enhancement, in the spirit of \cite{Yonekura}. In the cited work, indeed, it was shown that low-energy Lagrangian quivers with special unitary nodes assembled in the shape of a $\mathfrak{g}$ Dynkin diagram could exhibit non-trivial enhancement of their flavor symmetry to $\mathfrak{g}\oplus\mathfrak{g}$ in the UV, if some specific balancing conditions were met. In \cite{DeMarco:2023irn} we have constructed the corresponding singular CY3 geometries that engineer the UV fixed point. It would be interesting to pursue this further in the case of generalized quivers as in Figure \ref{D4zfig}, where, from the bottom-up perspective of the generalized quiver, we have up to now shown the agreement between the \textit{rank} of the flavor symmetry and the one of the UV fixed point engineered by M-theory. We can however assume a top-down approach: if one trusts the M-theory geometric engineering dictionary, the flavor symmetry of a $(\mathfrak g, D_k^{\oplus n})$ theory described by a generalized quiver like Figure \ref{D4zfig} \textit{is} at least 
\begin{equation}
    \label{eq:topdownstatement}
    \mathfrak g \oplus \underbrace{ D_{k} \oplus \dots \oplus D_{k}}_{n \text{ times}},
 \end{equation}
and hence we \textit{obtain} a non-Lagrangian generalization of the result of \cite{Yonekura}.\\

\indent One might wonder the extent to which this construction can be generalized: as we have mentioned, trinions and tetraons of type $E$ \textit{cannot} be engineered in such fashion. We flesh out the argument to exclude them in this context in Appendix \ref{app: no exceptional trinions}.

\section{The chemistry of novel bifundamentals/trinions/tetraons}\label{sec:gluingrules}
In \cref{sec:5dtriniontetraon} we have introduced a plethora of novel 5d bifundamental/trinion/tetraons, involving at least two flavor factors of type $D,E$: this is just half of the program to characterize these new 5d SCFTs. Indeed, we also need to identify the possible ways to glue them together in such a way to form \textit{5d generalised quivers}. We will see that these novel 5d SCFTs cannot be gauged together at will, but only admit a specific set of gluings.

\indent In this section, we analyze the gluing rules for the 5d CM theories listed in Table \ref{table:allthreefolds}, that correspond to a 5d generalized quiver theory once the CY3 employed to engineer them is suitably partially resolved.

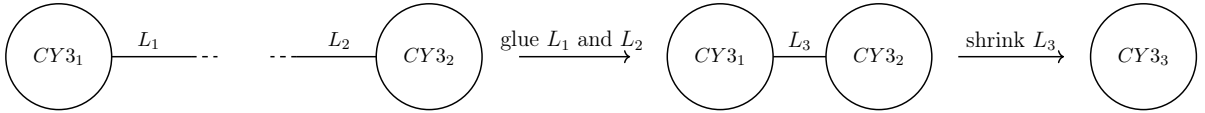
\begin{figure}[H]
    \centering
    \scalebox{0.7}{
    \begin{tikzpicture}
        \draw[thick] (0,0) circle (1);
        \node at (0,0) {\small$CY3_1$};
        \draw[thick] (1,0)--(2.5,0);
        \node at (1.7,0.3) {\small$L_1$};
        \draw[thick,dashed] (2.5,0)--(3,0);
        \draw[thick] (7,0) circle (1);
         \node at (7,0) {\small$CY3_2$};
        \draw[thick] (6,0)--(4.5,0);
        \node at (5.3,0.3) {\small$L_2$};
        \draw[thick,dashed] (4.5,0)--(4,0);
        \node at (9.7,0.3) {glue $L_1$ and $L_2$};
        \draw[thick,->] (8.7,0)--(10.8,0);
        \draw[thick] (12.5,0) circle (1);
        \draw[thick] (13.5,0)--(14.5,0);
        \draw[thick] (15.5,0) circle (1);
        \node at (12.5,0) {\small$CY3_1$};
        \node at (15.5,0) {\small$CY3_2$};
        \node at (14,0.3) {$L_3$};
        \draw[thick,->] (17,0)--(19,0);
        \node at (18,0.3) {shrink $L_3$};
        \draw[thick] (20.5,0) circle (1);
        \node at (20.5,0) {\small$CY3_3$};
        \end{tikzpicture}}
    \caption{Fusion of 5d SCFTs, from the CY3 perspective.}
    \label{fig:fusion}
    \end{figure}

Geometrically, fusion can be interpreted as follows. Consider two CY3's, each with at least a non-isolated singularity of type $\mathfrak{g}$ along a non-compact line, say $L_1$ and $L_2$, respectively. The fusion of the 5d SCFTs engineered by these CY3's is the 5d SCFT that arises from the gluing of $L_1$ with $L_2$ along a $\mathbb{P}^1$, once the volume of said $\mathbb{P}^1$ is sent to zero, recovering the SCFT phase. See Figure \ref{fig:fusion} for a schematic illustration of this surgery.

For the newly introduced trinion and tetraon 5d SCFTs, we investigate two classes of gluings:
\begin{itemize}
    \item gluing a trinion (tetraon) to a bifundamental 5d conformal matter SCFT, producing another trinion (tetraon). The gluing is performed in a \textit{partially resolved phase} of the CY3 singularities. Blowing down the geometry after the gluing recovers the final SCFT. 
    \item gluing a trinion/tetraon to a CY3 with at most two lines of type $D$ and no $E$ lines, producing a CY3 which is realized as a non-toric non-complete intersection. The gluing happens along a non-compact line \textit{in the singular phase}. The corresponding final 5d SCFT has flavor symmetry $(D^{\oplus 2},A^{\oplus n})$ (with $n\leq 3$). 
\end{itemize}

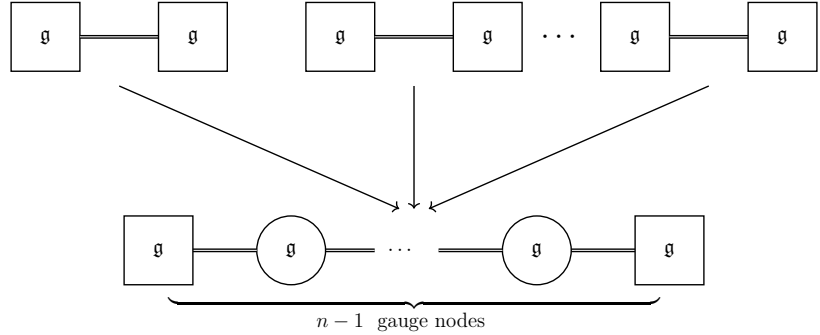
\begin{figure}[H]
    \centering
   \scalemath{0.65}{ \begin{tikzpicture}
 \draw[thick] (6.5,0.7)--(7.9,0.7)--(7.9,-0.7)--(6.5,-0.7)--cycle;
 \node at (7.2,0) {$\mathfrak{g}$};
 \draw[thick,double] (7.9,0)--(9.5,0);
 \draw[thick] (9.5,0.7)--(10.9,0.7)--(10.9,-0.7)--(9.5,-0.7)--cycle;
 \node at (10.2,0) {$\mathfrak{g}$};
  \node at (17.7,0) {\scalebox{1.5}{$\cdots$}};
 \draw[thick] (12.5,0.7)--(13.9,0.7)--(13.9,-0.7)--(12.5,-0.7)--cycle;
 \node at (13.2,0) {$\mathfrak{g}$};
 \draw[thick,double] (13.9,0)--(15.5,0);
 \draw[thick] (15.5,0.7)--(16.9,0.7)--(16.9,-0.7)--(15.5,-0.7)--cycle;
 \node at (16.2,0) {$\mathfrak{g}$};
 \draw[thick] (18.5,0.7)--(19.9,0.7)--(19.9,-0.7)--(18.5,-0.7)--cycle;
 \node at (19.2,0) {$\mathfrak{g}$};
 \draw[thick,double] (19.9,0)--(21.5,0);
 \draw[thick] (21.5,0.7)--(22.9,0.7)--(22.9,-0.7)--(21.5,-0.7)--cycle;
 \node at (22.2,0) {$\mathfrak{g}$};
 \draw[thick,->] (20.7,-1) -- (15,-3.5);
\draw[thick,->] (14.7,-1) -- (14.7,-3.5);
\draw[thick,->] (8.7,-1) -- (14.4,-3.5);
    \end{tikzpicture}}

\scalemath{0.65}{\begin{tikzpicture}
          %
          
          \node at (-2.25,0) {$\cdots$};
          \draw[thick,double] (1.2,0)--(2.5,0);
          \draw[thick,double] (-2.8,0)--(-3.8,0);
          %
          \draw[thick] (0.5,0) circle (0.7);
          \node at (0.5,0) {$\mathfrak{g}$};
          \draw[thick] (-4.5,0) circle (0.7);
          \node at (-4.5,0) {$\mathfrak{g}$};
          \draw[thick,double] (-0.2,0)--(-1.5,0);
          \draw[thick,double] (-5.2,0)--(-6.5,0);
        
        \draw[thick] (2.5,0.7)--(3.9,0.7)--(3.9,-0.7)--(2.5,-0.7)--cycle;
        \node at (3.2,0) {$\mathfrak{g}$};
        \draw[thick] (-6.5,0.7)--(-7.9,0.7)--(-7.9,-0.7)--(-6.5,-0.7)--cycle;
        \node at (-7.2,0) {$\mathfrak{g}$};
       \node at (-2,-1.3) {$\underbrace{\hspace{10 cm}}_{\scalebox{1}{\quad \qquad $n-1$ \text{ gauge nodes}}}$};
        \end{tikzpicture}}
          \caption{Linear gluings of 5d bifundamental CM produce generalized bifundamental 5d quivers, that admit a 5d SCFT phase. Double edges encode the presence of a non-trivial interacting SCFT.}
    \label{fig:gluing}
    \end{figure}

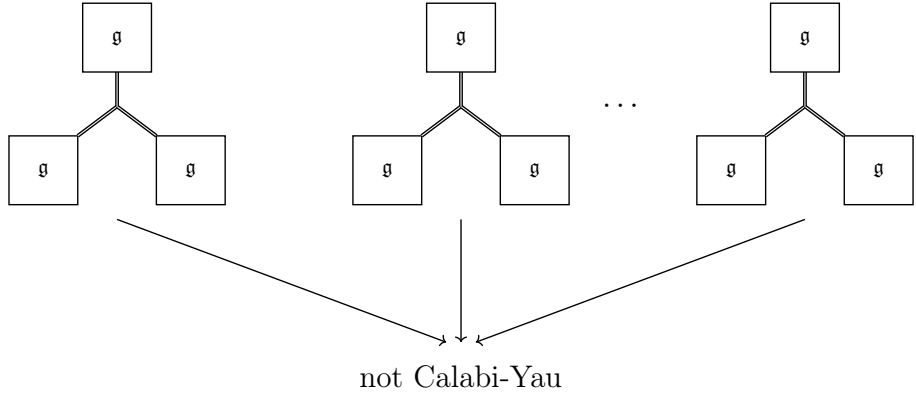
\begin{figure}[H]
    \centering
   \scalemath{0.65}{
    \begin{tikzpicture}
 \draw[thick] (6.5,0.7)--(7.9,0.7)--(7.9,-0.7)--(6.5,-0.7)--cycle;
 \node at (7.2,0) {$\mathfrak{g}$};
 \draw[thick,double] (7.9,0.7)--(8.7,1.3);
 \draw[thick,double] (9.5,0.7)--(8.7,1.3);

 \draw[thick] (9.5,0.7)--(10.9,0.7)--(10.9,-0.7)--(9.5,-0.7)--cycle;
\node at (10.2,0) {$\mathfrak{g}$};
 \draw[thick,double] (8.7,2)--(8.7,1.3);
 \draw[thick] (8,3.4)--(9.4,3.4)--(9.4,2.0)--(8,2.0)--cycle;
 \node at (8.7,2.7) {$\mathfrak{g}$};
 \draw[thick] (13.5,0.7)--(14.9,0.7)--(14.9,-0.7)--(13.5,-0.7)--cycle;
 \node at (14.2,0) {$\mathfrak{g}$};
 \draw[thick,double] (14.9,0.7)--(15.7,1.3);
 \draw[thick,double] (16.5,0.7)--(15.7,1.3);

 \draw[thick] (16.5,0.7)--(17.9,0.7)--(17.9,-0.7)--(16.5,-0.7)--cycle;
\node at (17.2,0) {$\mathfrak{g}$};
 \draw[thick,double] (15.7,2)--(15.7,1.3);
 \draw[thick] (15,3.4)--(16.4,3.4)--(16.4,2.0)--(15,2.0)--cycle;
 \node at (15.7,2.7) {$\mathfrak{g}$};
   \node at (19,1.3) {\scalebox{1.5}{$\cdots$}};
  \draw[thick] (20.5,0.7)--(21.9,0.7)--(21.9,-0.7)--(20.5,-0.7)--cycle;
 \node at (21.2,0) {$\mathfrak{g}$};
 \draw[thick,double] (21.9,0.7)--(22.7,1.3);
 \draw[thick,double] (23.5,0.7)--(22.7,1.3);

 \draw[thick] (23.5,0.7)--(24.9,0.7)--(24.9,-0.7)--(23.5,-0.7)--cycle;
\node at (24.2,0) {$\mathfrak{g}$};
 \draw[thick,double] (22.7,2)--(22.7,1.3);
 \draw[thick] (22,3.4)--(23.4,3.4)--(23.4,2.0)--(22,2.0)--cycle;
 \node at (22.7,2.7) {$\mathfrak{g}$};
 \draw[thick,->] (22.7,-1) -- (16,-3.5);
\draw[thick,->] (15.7,-1) -- (15.7,-3.5);
\draw[thick,->] (8.7,-1) -- (15.4,-3.5);
\node at (15.7,-4.2) {\scalebox{1.5}{not Calabi-Yau}};
    \end{tikzpicture}}
        
    \caption{Linear gluings of 5d non-toric trinion CM do not admit a generalized quiver phase that corresponds to a Calabi-Yau geometry. The same holds for non-toric tetraons. Double edges encode the presence of a non-trivial interacting SCFT.}
    \label{fig:gluingtrinions}
    \end{figure}

We hence see that the gluing rules for trinions and tetraons show that these new 5d SCFTs 
cannot be gauged together at will to form 5d SCFTs with a larger number of lines of type $D$ or $E$. In this respect, our novel trinions and tetraons bear a striking difference with respect to the bifundamental CM cases that were introduced in \cite{DeMarco:2023irn}, which can be bound together to form arbitrarily long molecules. In this regard, see Figure \ref{fig:gluing} and \ref{fig:gluingtrinions}.\\
\indent Physically, the rationale underlying this behaviour can be recast as follows:  gluings of $n$ copies of 5d CM with two $D$ ($E$)  flavor factors, which would produce CM theories with $n$ $D$ ($E$) flavor factors, are forbidden for $n>2$, as shown in \cite{DeMarco:2023irn}.
Therefore only linear gluings of these 5d CM theories produce canonical singular CY3, and they yield further 5d CM theories with at most two $D$ ($E$) factors (whose properties can be gleaned from the ones of their elementary building blocks). This constraint has one exception: one can consider \textit{trivalent} gluings of bifundamental CM with 1-valent theories of type $\mathcal{T}_{\mathfrak{g}}$, as shown in Figure \ref{fig:gauging}.\\
\indent On the other hand, linear gluings of trinions (tetraons) of type $D$ would produce $n$-valent 5d CM theories with $n>3$ ($n>4$) $D$ flavor factors. As we have reviewed, this is allowed for toric cases already explored in the literature. On the contrary, for the non-toric trinions and tetraons introduced in this work, linear gluings fail to produce generalized quivers that correspond to a consistent Calabi-Yau geometry. 
We redirect the reader to Figures \ref{fig:gluing} and \ref{fig:gluingtrinions} for a visual representation of these properties.
        
We begin in Section \ref{sec:gluingtrinionstetraonsdnode} explaining in detail why, as anticipated in \Cref{sec:trinionsasmolecules}, the novel bifundamentals/trinions/tetraons are molecules (rather than atoms or hybrids), building them constructively by trivalent and bivalent gauging of conformal matter theories with fewer flavor factors. Then, we study a specific class of gluings in the resolved phase of the corresponding CY3s, showing how  gauging a bifundamental/trinion/tetraon of type $(\mathfrak g,D_{k}^{\oplus n})$ (recall that $n\leq 3$) with a $(\mathfrak g,\mathfrak g)$ bifundamental produces a $(\mathfrak g,D_{k+1}^{\oplus n})$ bifundamental/trinion/tetraon: 
\begin{equation}
\label{eq:iterativegluing}
    D_k^{\oplus n} \oplus \text{\circled{$\mathfrak g$}}   \quad \text{\circled{$\mathfrak g$}} \oplus \mathfrak g \qquad  =  \qquad D_{k+1}^{\oplus n} \oplus \mathfrak g.
\end{equation}

After this, in Section \ref{sec:genquivers}, we will gauge one of the $D_k$ or $\mathfrak{g}$ factors of the flavor group of the fully singular geometries of Table \ref{table:allthreefolds}. 

Finally, in Section \ref{sec: blowdowns}, we specify which subset of the  generalized quiver theories of Section \ref{sec:genquivers} admits a flow to a 5d SCFT in the UV, that corresponds to a fully singular CY3.\\

\subsection{Gluing of the \texorpdfstring{$\mathfrak g$}{g} lines in the novel bifundamentals/trinions/tetraons}
\label{sec:gluingtrinionstetraonsdnode}
As anticipated in \Cref{sec:trinionsasmolecules}, the $(\mathfrak g,D_{k}^{\oplus n})$ theories presented in this work should be regarded as molecules. Indeed, the weak-coupling "ungluing" limit depicted in \Cref{fig:gluing} shows how such theories can be constructed with the following procedure. First, we start from the following generalized quiver 
\begin{equation}  
\label{eq:topolino1}
  \mathcal{T}_{\mathfrak g} = \begin{tikzpicture}[baseline=-0.5ex] 
     \draw[thick] (0,0) circle (0.7cm);
       \node at (-0.05,0) {\text{ \small{smooth}}};
     \draw[thick] (1.5,-0.5)--(2.5,-0.5)--(2.5,0.5)--(1.5,0.5)--cycle ;
       \node at (2,0) {$\mathfrak g$};
       \draw[thick,double] (0.7,0)--(1.5,0);
   \end{tikzpicture}
\end{equation}
defined as the patching of the affine varieties\footnote{The explicit shape of \eqref{eq:Tpequations} changes as we increase the rank of the $D_k$ singularity by subleading terms. However, near the origin the singularity \eqref{eq:Tpequations} these have no effect, so we can still represent it with \eqref{eq:topolino1}.}
\begin{eqnarray}
\label{eq:Tpequations}
    &&\frac{1}{2} i a_2 (a_2+i)^2
   b_2^2 + Q_{\mathfrak g}(u,v) = 0 \qquad  \subset \qquad  \mathbb C^4 \ni (a_2,b_2,u,v), \nonumber \\ 
    &&\frac{1}{2} i b_1 \left(a_1^2
   b_1+i\right)^2 + Q_{\mathfrak g}(u_1,v_1) = 0 \qquad \subset \qquad  \mathbb C^4 \ni (a_1,b_1,u_1,v_1),
\end{eqnarray}
with $i$ the imaginary unit, 
via the following transition functions 
\begin{equation}
\label{eq:trfunctTp}
   a_2 = b_1 a_1^2, \quad b_2 = \frac{1}{a_1}, \quad u = u_1, \quad v = v_1. 
\end{equation}
Let us comment on the singularity structure of \eqref{eq:Tpequations}: the circular node of \eqref{eq:topolino1} is the line $u = v = a_2 = 0$, spanned by the variable $b_2$, and, after passing to the $(a_1,b_1,u,v)$ patch it is isomorphic to a $\mathbb P^1$; the flavor node corresponds to the line $u = v = b_2 = 0$ (and its patching in the $(a_1,b_1,u_1,v_1)$ chart), which is non-compact and supports a Du Val singularity of type $\mathfrak{g}$. Finally, \eqref{eq:Tpequations} sports another line of Du Val singularities of type $\mathfrak{g}$, on $u = v  = a_2^2 b_2 + i = 0$; such curve is not included in the generalized quiver notation, as it does not intersect the curves associated with the round and the square node. Since the transition functions \eqref{eq:trfunctTp} for $u,v$ are trivial, we denote $u_1,v_1$ from now on with $u,v$. 

Along with \eqref{eq:topolino1}, we consider a $X_{\mathfrak g}^{(1)}$ bifundamental
\begin{equation}  
\label{eq:bifund1}
  X_{\mathfrak g}^{(1)} =  \begin{tikzpicture}[baseline=-0.5ex] 
    \draw[thick] (-0.5,-0.5)--(0.5,-0.5)--(0.5,0.5)--(-0.5,0.5)--cycle ;
     \draw[thick] (1.5,-0.5)--(2.5,-0.5)--(2.5,0.5)--(1.5,0.5)--cycle ;
       \node at (0,0) {$\mathfrak g$};
       \node at (2,0) {$\mathfrak g$};
       \draw[thick,double] (0.5,0)--(1.5,0);
   \end{tikzpicture}
\end{equation}
that we take to be defined as 
\begin{equation}
\label{eq:recallXg1}
    \frac{1}{2} a_5^2 \left(1-i
   b_5\right) b_5^2 + Q_{\mathfrak g}(u,v) = 0 \qquad \subset \qquad \mathbb C^4 \ni (a_5, b_5, u, v), 
\end{equation}
where, again, we wrote $u,v$ instead of $u_1,v_1$ as we will soon patch them trivially with the coordinates $u,v$ appearing in \eqref{eq:Tpequations}. We notice that we have included an unusual factor $\left(1-i
   b_5\right) $; this will help us to patch \eqref{eq:recallXg1}, and it does not change the singularity structure at $u = v = a_5 = b_5 = 0$ (as its non-constant part adds subleading monomials in a vicinity of the origin). Consequently, we can still represent \eqref{eq:recallXg1} with the generalized quiver \eqref{eq:bifund1}.  

We are now ready to perform the following gluing: 

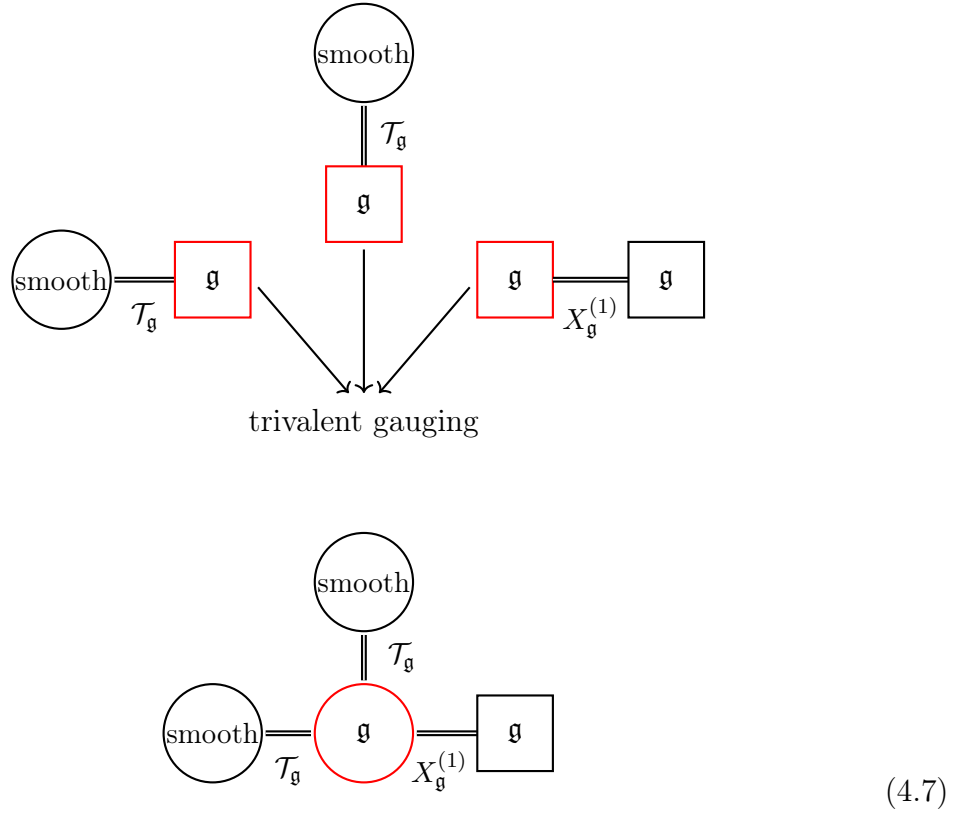
\begin{equation}
\label{eq:gluingfigureeq}
\begin{tikzpicture}
 \draw[thick] (-4,6) circle (0.65);
 \draw[thick,double] (-3.3,6)--(-2.5,6);
 \node at (-4,6) {\small smooth};
 \node at (-2,6) {\small$\mathfrak{g}$};
 \draw[thick,red] (-2.5,5.5)--(-1.5,5.5)--(-1.5,6.5)--(-2.5,6.5)--cycle ;

 \draw[thick] (0,9) circle (0.65);
 \draw[thick,double] (0,8.3)--(0,7.5);
 \node at (0,9) {\small smooth};
 \node at (0,7) {\small$\mathfrak{g}$};
 \draw[thick,red] (-0.5,6.5)--(0.5,6.5)--(0.5,7.5)--(-0.5,7.5)--cycle;

 \draw[thick,red] (1.5,5.5)--(2.5,5.5)--(2.5,6.5)--(1.5,6.5)--cycle;
 \draw[thick,double] (2.5,6)--(3.5,6);
 \node at (2,6) {\small $\mathfrak{g}$};
 \node at (4,6) {\small$\mathfrak{g}$};
 \draw[thick] (3.5,5.5)--(4.5,5.5)--(4.5,6.5)--(3.5,6.5)--cycle;

 \node at (3,5.5) {\small$X_{\mathfrak{g}}^{(1)}$};
 \node at (-2.9,5.5) {\small$\mathcal{T}_{\mathfrak{g}}$};
 \node at (0.4,7.9) {\small$\mathcal{T}_{\mathfrak{g}}$};

 \node at (0,4.1) {trivalent gauging};
 \draw[thick,<-] (0.2,4.5)--(1.4,5.9);
 \draw[thick,<-] (-0.2,4.5)--(-1.4,5.9);
 \draw[thick,->] (0,6.4)--(0,4.5);

 \draw[thick,red] (0,0) circle (0.65);
 \node at (0,0) {\small$\mathfrak{g}$};
 \draw[thick,double] (0.7,0)--(1.5,0);
 \draw[thick] (1.5,0.5)--(2.5,0.5)--(2.5,-0.5)--(1.5,-0.5)--cycle;
 \node at (2,0) {\small$\mathfrak{g}$};
 \draw[thick,double] (-0.7,0)--(-1.3,0);

 \node at (-2,0) {\small\text{smooth}};
 \node at (0,2) {\small\text{smooth}};
 \node at (0.5,1) {\small$\mathcal{T}_{\mathfrak{g}}$};
 \node at (-1,-0.5) {\small$\mathcal{T}_{\mathfrak{g}}$};
 \node at (1,-0.5) {\small$X_{\mathfrak{g}}^{(1)}$};

 \draw[thick] (0,2) circle (0.65);

 \draw[thick] (-2,0) circle (0.65);
 \draw[thick,double] (0,0.7)--(0,1.3);
\end{tikzpicture}
\end{equation}
Quantitatively, we introduce a new copy of \eqref{eq:Tpequations},
\begin{eqnarray}
\label{eq:Tpequationscopy}
    &&-\frac{1}{2} i a_3^2 b_3
   \left(b_3-i\right){}^2 + Q_{\mathfrak g}(u,v) = 0 \qquad  \subset \qquad  \mathbb C^4 \equiv U_4 \ni (a_3,b_3,u,v), \nonumber \\ 
    &&-\frac{1}{2} i a_4 \left(a_4
   b_4^2-i\right){}^2 + Q_{\mathfrak g}(u,v) = 0 \qquad \subset \qquad  \mathbb C^4 \equiv U_3 \ni (a_4,b_4,u,v),
\end{eqnarray}
with $U_3, U_4$ patched with the transition functions 
\begin{equation}
\label{eq:trfunctTpcopy}
   a_4 = b_3 a_3^2, \quad b_4 = \frac{1}{a_3}.
\end{equation}
The red curves in the upper part of \eqref{eq:gluingfigureeq}, supporting singularities of type $\mathfrak{g}$, are spanned by (respectively) the coordinates $a_2, b_3, b_5$. We can glue them together, to form the $\mathbb P^1$ corresponding to the red node of \eqref{eq:gluingfigureeq}, with the following transition functions: 
\begin{equation}
\label{eq:trivalentgluing}
    a_2 = 
   \frac{1}{b_3},\qquad b_2= a_3
   b_3^2 \qquad  a_2\ =  \frac{1}{b_5+i},\qquad b_2=  a_5 \left(b_5+i\right){}^2.
\end{equation}
 In this picture, the round black nodes of Figure \eqref{eq:gluingfigureeq} correspond to the loci $a_2 = 0$ and $b_3 = 0$ respectively, while the square black node corresponds to $b_5 = 0$. We now notice that the following functions: 
\begin{equation}\label{eq:blowdownchartsd4}
\scalemath{1}{
\begin{split}
&\varphi: U_2 \cong \mathbb{C}^2_{a_2,b_2} \times \mathbb C^2_{u,v} \xrightarrow{\ \varphi_2 \equiv \varphi|_{U_2}\ } \mathbb C^3_{x,y,z} \times \mathbb C^2_{u,v},\\
& (a_2,b_2,u,v)\ \longmapsto\ (x,y,z,u,v) =\Big(
\frac{1}{2} a_2
   \left(a_2^2+1\right)
   b_2^2,\sqrt{2} a_2
   b_2,\frac{1}{2} i
   a_2 \left(a_2+i\right){}^2
   b_2^2,u,v \Big),
   \end{split}}
\end{equation}
can be extended smoothly using the transition functions \eqref{eq:trivalentgluing} to all the open subsets $U_1, U_2, U_3, U_4, U_5$. This means that the full geometry in the lower part of \eqref{eq:gluingfigureeq} is mapped to a subvariety of $\mathbb C^3_{x,y,z} \times \mathbb C^2_{u,v}$, by evaluating for each of its points the functions $(x,y,z,u,v)$. We have called this map $\varphi$ in \eqref{eq:blowdownchartsd4}. 

To identify the equations of this subvariety, we notice that the functions $x,y,z$ satisfy the following relation: 
\begin{equation}
    \label{eq:d4fromtrivalentgluing}
    x^2+ z y^2 + z^2 = 0, 
\end{equation}
and that \eqref{eq:Tpequations} can be rewritten as 
\begin{equation}
\label{eq:basechanged4fromtrivgluing}
    z = Q_{\mathfrak g}(u,v). 
\end{equation}
We note that the zero loci of \eqref{eq:d4fromtrivalentgluing} and \eqref{eq:basechanged4fromtrivgluing} describe a geometry that lies in the affine space $\mathbb C^5 \ni (x,y,z,u,v)$, that is exactly the $(\mathfrak g,D_3^{\oplus n}) $ threefolds introduced in \Cref{sec:CBdata}. This shows that the novel 5d SCFT of type $(\mathfrak g,D_3^{\oplus n})$ is indeed a molecule, obtained via the gauging procedure outline in Figure \eqref{eq:gluingfigureeq}, with $a = 2, b = c =0$.

Instead of performing the blow-down \eqref{eq:blowdownchartsd4}, we can glue another bifundamental conformal matter of type $(\mathfrak{g},\mathfrak{g})$
\begin{equation}
\label{eq:recallXg1copy}
    \frac{1}{2} a_6^2 b_6^2
   \left(1-i a_6 b_6^2\right) + Q_{\mathfrak g}(u,v) = 0 \qquad \subset \qquad \mathbb C^4 \equiv U_6 \ni (a_6, b_6, u, v), 
\end{equation}
to the flavor node of the lower part of \eqref{eq:gluingfigureeq} using the following transition functions: 
\begin{equation}
    \label{eq:trfunctioniteration}
    a_4 = \frac{1}{b_5}, \qquad b_4 = b_5^2 a_5.  
\end{equation}
The gluing \eqref{eq:trfunctioniteration} corresponds, in the generalized quiver notation, to  
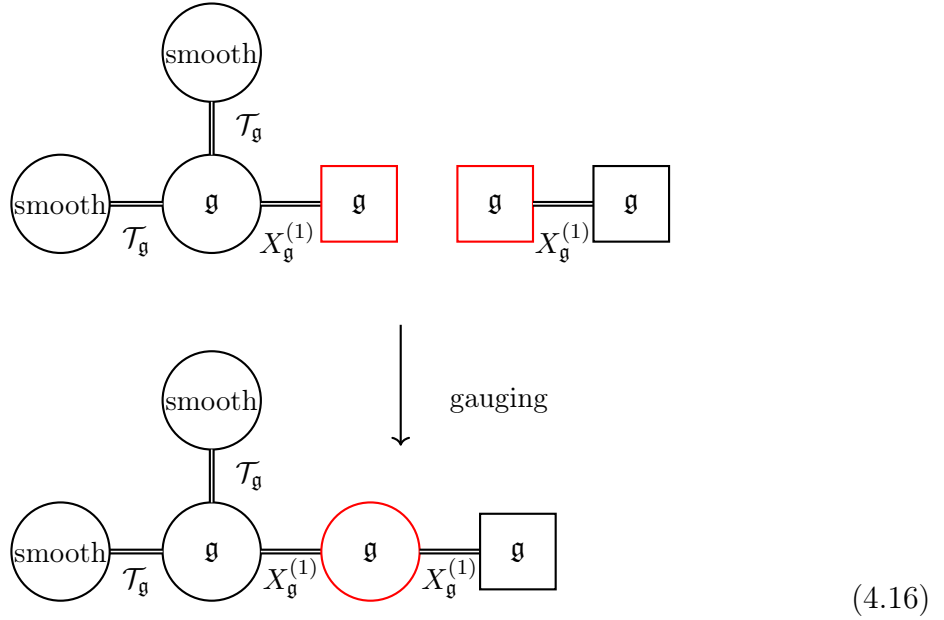
\begin{equation}
\label{eq:newgluingfigureeq}
\begin{tikzpicture}

 \begin{scope}[yshift=0cm]
 \draw[thick] (0,0) circle (0.65);
 \node at (0,0) {\small$\mathfrak{g}$};

 \draw[thick,double] (0.65,0)--(1.45,0);
 \draw[thick,red] (1.45,0.5)--(2.45,0.5)--(2.45,-0.5)--(1.45,-0.5)--cycle;
 \node at (1.95,0) {\small$\mathfrak{g}$};

 
 \draw[thick,red] (3.25,0.5)--(4.25,0.5)--(4.25,-0.5)--(3.25,-0.5)--cycle;
 \node at (3.75,0) {\small$\mathfrak{g}$};

 \draw[thick,double] (4.25,0)--(5.05,0);
  \node at (4.65,-0.5) {\small$X_{\mathfrak{g}}^{(1)}$};
 \draw[thick] (5.05,0.5)--(6.05,0.5)--(6.05,-0.5)--(5.05,-0.5)--cycle;
 \node at (5.55,0) {\small$\mathfrak{g}$};

 \draw[thick,double] (-0.65,0)--(-1.35,0);

 \node at (-2,0) {\small\text{smooth}};
 \node at (0,2) {\small\text{smooth}};
 \node at (0.5,1) {\small$\mathcal{T}_{\mathfrak{g}}$};
 \node at (-1,-0.5) {\small$\mathcal{T}_{\mathfrak{g}}$};
 \node at (1,-0.5) {\small$X_{\mathfrak{g}}^{(1)}$};

 \draw[thick] (0,2) circle (0.65);
 \draw[thick] (-2,0) circle (0.65);
 \draw[thick,double] (0,0.65)--(0,1.35);
\end{scope}

\begin{scope}[yshift=-4.6cm]
\draw[thick,->] (2.5,3)--(2.5,1.4);
 \node at (3.8,2) {\small{gauging}};
 \draw[thick] (0,0) circle (0.65);
 \node at (0,0) {\small$\mathfrak{g}$};

 \draw[thick,double] (0.65,0)--(1.45,0);
 \node at (1.05,-0.35) {\small$X_{\mathfrak g}^{(1)}$};

 \draw[thick,red] (2.10,0) circle (0.65);
 \node at (2.10,0) {\small$\mathfrak{g}$};

 \draw[thick,double] (2.75,0)--(3.55,0);
 \node at (3.15,-0.35) {\small$X_{\mathfrak g}^{(1)}$};

 \draw[thick] (3.55,0.5)--(4.55,0.5)--(4.55,-0.5)--(3.55,-0.5)--cycle;
 \node at (4.05,0) {\small$\mathfrak{g}$};

 \draw[thick,double] (-0.65,0)--(-1.35,0);

 \node at (-2,0) {\small\text{smooth}};
 \node at (0,2) {\small\text{smooth}};
 \node at (0.5,1) {\small$\mathcal{T}_{\mathfrak{g}}$};
 \node at (-1,-0.4) {\small$\mathcal{T}_{\mathfrak{g}}$};

 \draw[thick] (0,2) circle (0.65);
 \draw[thick] (-2,0) circle (0.65);
 \draw[thick,double] (0,0.65)--(0,1.35);
\end{scope}

\end{tikzpicture}
\end{equation}

Let us now blow down the geometry described by \eqref{eq:newgluingfigureeq}. We introduce the following three functions in the patch $U_2$:

\begin{equation}
\label{eq:blowdownchartsd5}
\scalemath{0.85}{
(a_2,b_2) \;\longmapsto\; (\widetilde{x},\widetilde{y},\widetilde{z}) =
\left(\frac{i a_2^2
   \left(a_2+i\right){}^2
   b_2^3}{\sqrt{2}},\frac{1}{2
   } a_2 \left(a_2^2+1\right)
   b_2^2,\frac{1}{2} i a_2
   \left(a_2+i\right){}^2
   b_2^2\right)}.
\end{equation}
One can check that they extend holomorphically with the transition functions \eqref{eq:trfunctioniteration}, and hence the geometry associated with the gluing \eqref{eq:newgluingfigureeq} is mapped to 
\begin{equation}
\label{eq:d5fromtrivalentgluingitetation}
    \widetilde{x}^2 + \widetilde{z} \widetilde{y}^2 + \widetilde{z}^3 = 0, \qquad \widetilde{z} = Q_{\mathfrak g}(u,v),
\end{equation}
inside $\mathbb C^5 \ni(\widetilde{x},\widetilde{y},\widetilde{z},u,v)$.

The previous derivation shows that the gluing \eqref{eq:newgluingfigureeq}  realizes the $(\mathfrak g,D_{4}^{\oplus n})$ theory as the gauging of $(\mathfrak g,D_{3}^{\oplus n})$ with $X_{\mathfrak g}^{(1)}$, proving \eqref{eq:iterativegluing} for $k = 3$.\footnote{Please note that, for the ease of notation, we denote $X_{\mathfrak g}^{(1)}$ as $\mathfrak g \oplus \mathfrak g$ in \eqref{eq:iterativegluing}.} 
Again, instead of performing the blow-down with \eqref{eq:blowdownchartsd5}, we can iterate the process, and glue another conformal matter bifundamental to $(\mathfrak g,D_{4}^{\oplus n} )$. The blow-down of the geometry proceeds analogously to what we showed, realizing \eqref{eq:iterativegluing} in full generality.

Summing up, the previous construction has simultaneously shown that: 
\begin{enumerate}
    \item The novel bifundamentals/trinions/tetraons presented in this work can be constructed via trivalent and bivalent gluings of the (bi)fundamentals theories described in \eqref{eq:topolino1} and \eqref{eq:recallXg1}. Equivalently, the novel bifundamentals/trinions/tetraons can be decomposed into (bi)fundamentals building blocks, ungauging the gauge $\mathfrak{g}$ nodes of \Cref{fig:mannaggia}. 
\item It is possible to gauge a theory of type $(\mathfrak g,D_k^{\oplus n})$ with $l$ copies of the bifundamental 5d conformal matter theory of type $(\mathfrak g,\mathfrak g)$, to produce a 5d SCFT of type $(\mathfrak g,D_{k+l}^{\oplus n})$. This is summarized in \eqref{eq:topdownstatement}. Both the theory $(\mathfrak g,D_k^{\oplus n})$ and $(\mathfrak g,D_{k+l}^{\oplus n})$ belong to the same row in Table \ref{table:allthreefolds}.
\end{enumerate}
We conclude with a remark: one can see that the the functions $(x,y,z)$ defined in \eqref{eq:blowdownchartsd4} do not extend to the full geometry \eqref{eq:newgluingfigureeq}. This means that, in particular, $(x,y,z)$ are not holomorphic functions of the  $(\widetilde{x},\widetilde{y},\widetilde{z})$ that appear in \eqref{eq:blowdownchartsd5}, as in general the associated poles are located also outside the locus contracted by $(\widetilde{x},\widetilde{y},\widetilde{z})$. It then follows that we could now have spotted them with the methods we will introduce in the next Section.

\subsection{Gluings of the novel bifundamentals/trinions/tetraons in the singular phase}
\label{sec:genquivers}
\indent In this Section we exhibit an alternative method to gauge a flavor factor of the novel bifundamental/trinion/tetraon theories. In particular, given a theory of type $(\mathfrak{g},D_k^{\oplus n})$ appearing in Table \ref{table:allthreefolds}, we will focus on gauging one of the non-compact singular lines, starting from the \textit{singular phase of the CY3}. This is in contrast to the approach of the previous Section, in which we performed the gluing in a partially resolved phase of the CY3, before flowing back to the SCFT fixed point.

\indent We can attack the problem with the following argument: gauging a diagonal $\mathfrak f$ factor between two different atoms of conformal matter amounts to gluing together two non-compact curves of $\mathfrak f$ singularities, producing a compact $\mathbb P^1$ where, nearby each point of this $\mathbb P^1$, the threefold locally looks like  $\mathbb C \times P_{\mathfrak f}$, with $P_{\mathfrak f}$ the $\mathfrak f$-type Du Val singularity. As the 5d SCFTs we are dealing with are defined by hypersurfaces in $\mathbb C^4$, in order to realize a large class of gluings we can consider an ambient space that is the most general rank three vector bundle over such $\mathbb P^1$. All these vector bundles split as 
\begin{equation}
\label{eq:rank3vectbundle}
    \mathcal O(-a) \oplus \mathcal O(-b) \oplus \mathcal O(-c),
\end{equation} with $a,b,c \in \mathbb Z$. This already captures a large class of possible gluings, but \textit{does not} capture gluings where a local neighbourhood of the  $\mathbb P^1$ that supports the gauged group can not be approximated by a rank-three vector bundle. Indeed, this is the case e.g. when its transition functions contain higher-order terms that make them non-binomial and that cannot be dropped using some scaling argument.  

All the curves of Du Val singularities appearing in the threefolds of \cref{sec:5dtriniontetraon} are (as is evident from \eqref{singular threefolds}) either
\renewcommand{\arraystretch}{1}
\begin{itemize}
    \item parametrized by the coordinate $y$;
    \item parametrized by the coordinate $v$  (up to linear coordinates redefinition between $u \leftrightarrow v$);
    \item aximally degenerating hyperelliptic curves, described as the zero loci of $u^j + v^2$.
\end{itemize}
We notice that, in the latter case, the curve itself is not smooth, with the singular point located where the 5d SCFT localizes. 

\indent Let us now state more precisely the gauging procedure anticipated around \eqref{eq:rank3vectbundle}. Such a procedure is rigorously realized via the gluing of two patches, say $U$ and $U'$, with local coordinates $(u,v,x,y)$ and $(u',v',x',y')$, respectively. In each local patch, the threefold is defined as a hypersurface:
\begin{equation}
  U: \quad  P(u,v,x,y) = 0, \quad\quad U': \quad  P'(u',v',x',y') = 0.
\end{equation}

\indent The gluing between patches happens along a compact curve isomorphic to $\mathbb{P}^1$. We use the following ambient space transition functions for the curve spanned by $u$:
\begin{equation}
\label{eq:ugluinggeneral}
    v = \frac{1}{v'}, \quad x = v'^a x', \quad y = v'^b y', \quad u = v'^c u'.
\end{equation}
Stated differently, the desired ambient space is obtained by gluing together two copies of $\mathbb C^4$, with coordinates (respectively) $(u,v,x,y)$ and $(u',v',x',y')$, with \cref{eq:ugluinggeneral}.\\

\indent We are going to use the following ambient space transition functions for the curve spanned by $y$:
\begin{equation}
\label{eq:ygluinggeneral}
    y = \frac{1}{y'}, \quad x = y'^a x', \quad v = y'^b v', \quad u = y'^c u'.
\end{equation}
We note that the ambient space, for generic choices of $a,b,c$, will not be CY. However, this will be balanced by picking the $P(u,v,x,y), P'(u',v',x',y')$ to be the local data of a section of the anticanonical bundle of the ambient space\footnote{This is a standard construction,  see e.g.\ the famous "quintic in $\mathbb P^4$".}. This is, by the adjunction formula, a necessary condition to trivialize the canonical bundle of the threefold, and in practice it boils down to defining
\begin{equation}
    P' = \frac{P\rvert_{\star}}{v'^{N}}, \quad N = a + b + c - 2  
\end{equation}
for the $v$-gluings \eqref{eq:ugluinggeneral}, where $\star$ means that we plugged in the transition functions \eqref{eq:ugluinggeneral} in $P$, and 

\begin{equation}\label{new patch}
    P' = \frac{P\rvert_{\ast}}{y'^{N}}, \quad N = a + b + c - 2,
\end{equation}
 where $\ast$ means that we inserted, inside $P$, the expressions \eqref{eq:ygluinggeneral} in spite of $(x,y,u,v)$. \\
 
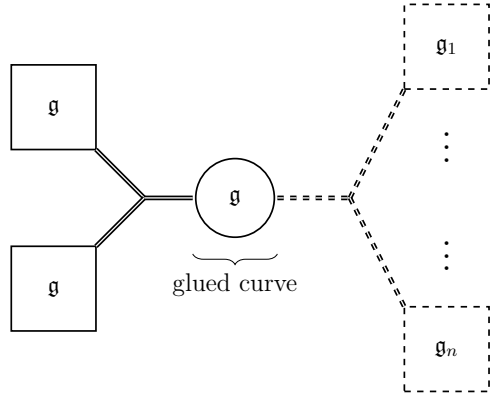
\begin{figure}[H]
    \centering
    \scalebox{0.8}{
    \begin{tikzpicture}
       \draw[thick] (0,0)--(1.4,0)--(1.4,1.4)--(0,1.4)--cycle;
       \draw[thick] (0,3)--(1.4,3)--(1.4,4.4)--(0,4.4)--cycle;
       \draw[thick,double] (1.4,3)--(2.2,2.2);
       \draw[thick,double] (1.4,1.4)--(2.2,2.2);
       \draw[thick,double] (2.2,2.2)--(3,2.2);
       \node at (0.7,0.7) {\small$\mathfrak{g}$};
       \node at (0.7,3.7) {\small$\mathfrak{g}$};
       \draw[thick] (3.7,2.2) circle (0.65);
        \node at (3.7,2.2) {\small$\mathfrak{g}$};
        \draw[thick,double,dashed] (4.4,2.2)--(5.6,2.2);
        \draw[thick,double,dashed] (5.6,2.2)--(6.5,4);
          \draw[thick,dashed] (6.5,4)--(7.9,4)--(7.9,5.4)--(6.5,5.4)--cycle;
          \draw[thick,double,dashed] (5.6,2.2)--(6.5,0.4);
          \draw[thick,dashed] (6.5,-1)--(7.9,-1)--(7.9,0.4)--(6.5,0.4)--cycle;
           \node at (7.2,4.7) {\small$\mathfrak{g}_1$};
       \node at (7.2,-0.3) {\small$\mathfrak{g}_n$};
          \node at (7.2,1.4) {\scalebox{1.5}{$\vdots$}};
          \node at (7.2,3.2) {\scalebox{1.5}{$\vdots$}};
           \draw [decorate,decoration={brace,amplitude=5pt,mirror,raise=4ex}]
  (3,2) -- (4.4,2) node[midway,yshift=-3em]{glued curve};
        \end{tikzpicture}}
    \caption{Gluing of a trinion theory $(\mathfrak{g},\mathfrak{g},\mathfrak{g})$ with a theory of type $(\mathfrak{g},\mathfrak{g},\mathfrak{g}_1,\ldots,\mathfrak{g}_n)$. The gluing produces the compact curve in the middle of the picture.}
    \label{fig:genquiver}
    \end{figure}

\indent  As a result of this gauging procedure, we have obtained a Calabi-Yau threefold $Y$ which is a section of the anticanonical bundle of the total ambient space:
 \begin{equation}\label{total space}
   \mathcal{A}:\quad \text{Tot}_{\mathbb{P}^1}\left( \mathcal{O}(-a)\oplus\mathcal{O}(-b)\oplus\mathcal{O}(-c)\right).
 \end{equation}
 The threefold $Y$ displays one compact curve,  the base $\mathbb{P}^1$ in \eqref{total space} and in the patch \eqref{new patch}, and $n$ new lines of singularities labelled by $\mathfrak{g}_i$ (with $i=1,\ldots,n$) can appear. From the physical point of view, we obtain a \textit{generalized quiver phase}, as depicted in Figure \ref{fig:genquiver} for the gluing of a trinion. We will return on the matter of whether these generalized quiver phases admit a 5d SCFT limit in Section \ref{sec: blowdowns}.\\
 \indent The remaining part of this Section is devoted to providing explicit examples of the above-mentioned gaugings, namely to identifying the allowed dashed flavor nodes in Figure \ref{fig:genquiver}.\\ 

\subsubsection{Gluings that admit a generalized quiver phase}\label{sec:gluingsgenquiv}

In this Section we identify all the possible gluings that admit a generalized quiver phase. As we will see in Section \ref{sec: blowdowns}, \textit{only a subset of these admit a completion to a UV 5d SCFT point}. We focus our attention on two of the theories from Table \ref{table:allthreefolds}: the $(D_5,D_5,D_5)$ trinion and the $(D_4,D_4,D_4,D_4)$ tetraon. The reasoning for the other singularities is completely analogous. From the standpoint of the 5d classification scheme, the most relevant remark is the following:\\

\textit{Given the novel trinions (tetraons) 5d SCFTs with 3 (4) flavor factors of $D$ type, the gluings in this Section never produce generalized quiver phases with more than 3 (4) flavor factors of type $D$.}\\

Namely, by gluing trinions (tetraons) of type $D$, a theory with an arbitrary number of $D$ flavor factors \textit{cannot} be obtained. This is in sharp contrast with the result reviewed in the $A$ case in Section \ref{sec:5dtoric}. This starkly distinguishes trinions of type $A$, which behave as genuine atom SCFTs, from $D$ trinions/tetraons which, as we have seen in Section \ref{sec:trinionsasmolecules}, are molecules themselves. \\

\noindent\textbf{$\boldsymbol{D_5}$ trinion}\\
In this section we list all the possible gluings of the $D_5$ trinion that give rise to a partially resolved Calabi-Yau threefold. The $D_5$ trinion reads:
\begin{equation}\label{D5 trinion}
\begin{cases}
    x^2+zy^2+z^4=0,\\
    z = u(u^3+v^2). \\
    \end{cases}
\end{equation}



\noindent In Table \ref{D5 trinion table} we list the degrees of the normal bundle $a,b,c$, given a gluing along $v$ or $y$. We will do so by discarding all the gluings that produce either denominators in the expression of $P'$, or for which $P'$ is not an irreducible polynomial, or that yield non-normal singularities. In the last column, we specify the singularities of the 5d conformal matter SCFT that is glued to the $D_5$ trinion, highlighting in a circle the singularities that are compactified by the gluing procedure.

\renewcommand{\arraystretch}{1.1}
\begin{table}[H]\centering
\begin{equation}
\scalemath{0.9}{
\begin{array}{|c|E|E|E|c|c|}
\hline 
 \multirow{2}{*}{\makecell{\textbf{Gluing}\\\textbf{coordinate}}} &
\multicolumn{3}{|c|}{\textbf{Degrees of line bundles}} & \multirow{2}{*}{\makecell{\textbf{Glued }\\ \textbf{singularities}}} \\
\cline{2-4}
 & \boldsymbol{a} & \boldsymbol{b} & \boldsymbol{c} & \\
\hline
\hline
v & -4 & -2 & 0 & \text{\circled{$D_5$}} \oplus D_5 \\
v & -3 & -3 & 0 & \text{\circled{$D_5$}} \oplus D_5 \\
v & -k-2 & -k-1 & 1 & \text{\circled{$D_5$}}\oplus A_1 \oplus D_{2k+1} \\
v & -k-2 & -k-2 & 1 & \text{\circled{$D_5$}} \oplus D_{2k+2} \\
v & -k & -k & 2 & \text{\circled{$D_5$}} \oplus A_{2k-1} \\
\hline
y & 0& 0& 0& \text{\circled{$D_5$}}\oplus D_5 \\
y & -1& 1& 0& \text{\circled{$D_5$}}\oplus D_5 \\
y & -2& -1& 1& \text{\circled{$D_5$}}\oplus D_7\oplus A_1 \\
y & -1& 0& 1& \text{\circled{$D_5$}}\oplus D_5\oplus A_1 \\
y & 0& 0& 1& \text{\circled{$D_5$}}\oplus D_4 \\
y & 0& 1& 1& \text{\circled{$D_5$}}\oplus A_3\oplus A_1 \\
y & 1& 1& 1& \text{\circled{$D_5$}}\oplus A_1\oplus A_1\\
y & 1& 2& 1& \text{\circled{$D_5$}}\oplus A_1 \\
y & -1& -1& 2& \text{\circled{$D_5$}}\oplus A_7 \\
y & 0& 0& 2& \text{\circled{$D_5$}}\oplus A_5 \\
y & 1& 1& 2& \text{\circled{$D_5$}}\oplus A_3 \\
y & 2& 2& 2& \text{\circled{$D_5$}}\oplus A_1 \\
y & 3& 3& 2& \text{\circled{$D_5$}} \\
\hline
\end{array}\nonumber
}
\end{equation}
\caption{Gluing data for 5d conformal matter trinion of type $D_5$. The labels $a,b,c$ identify the degrees of the normal bundle in \eqref{total space}.}\label{D5 trinion table}
\end{table}

\noindent\textbf{$\boldsymbol{D_4}$ tetraon}\\
In this section we exhaust all the gluings of the $D_4$ tetraon that satisfy the Calabi-Yau condition, using the same conventions as above.

The $D_4$ tetraon reads:
\begin{equation}
\begin{cases}
    x^2+zy^2+z^3=0, \\
    z = u(u-v)(u+v). \\
    \end{cases}
\end{equation}



\renewcommand{\arraystretch}{1.1}
\begin{table}[H]\centering
\begin{equation}
\scalemath{0.9}{
\begin{array}{|c|E|E|E|c|c|}
\hline 
 \multirow{2}{*}{\makecell{\textbf{Gluing}\\\textbf{coordinate}}} &
\multicolumn{3}{|c|}{\textbf{Degrees of line bundles}} & \multirow{2}{*}{\makecell{\textbf{Glued }\\ \textbf{singularities}}} \\
\cline{2-4}
 & \boldsymbol{a} & \boldsymbol{b} & \boldsymbol{c} & \\
\hline
\hline
v & -3 & -1 & 0 & \text{\circled{$D_4$}}\oplus D_4 \\
v & -2 & -2 & 0 & \text{\circled{$D_4$}}\oplus D_4 \\
v & -k-2 & -k-1 & 1 & \text{\circled{$D_4$}}\oplus D_{2k+2} \oplus A_1 \\
v & -k-1 & -k-1 & 1 & \text{\circled{$D_4$}}\oplus D_{2k+1} \\
v & -k & -k & 2 &\text{\circled{$D_4$}} \oplus A_{2k-1} \\
\hline
y & -2 & 1 & -1 &\text{\circled{$D_4$}}\oplus D_6 \oplus A_1 \\
y & -1 & 2 & -1 & \text{\circled{$D_4$}}\oplus A_5\\
y & -1 & 1 & -1 &\text{\circled{$D_4$}}\oplus D_5 \\
y & 0 & 0 & 0 & \text{\circled{$D_4$}} \oplus D_4 \\
y & -1 & 1 & 0 & \text{\circled{$D_4$}} \oplus D_4 \oplus A_1\\
y & 0 & 1 & 0 & \text{\circled{$D_4$}} \oplus A_3 \\
y & 0 & 2 & 0 &\text{\circled{$D_4$}}\oplus A_3 \\
y & -1 & 0 & 1 & \text{\circled{$D_4$}} \oplus D_4 \\
y & 0 & 1 & 1 & \text{\circled{$D_4$}} \oplus A_1\oplus A_1\oplus A_1 \\
y & 1 & 1 & 1 & \text{\circled{$D_4$}}\\
y & 1 & 2 & 1 & \text{\circled{$D_4$}} \oplus A_1\\
y & 2 & 2 & 2 & \text{\circled{$D_4$}}\\
\hline
\end{array}\nonumber
}
\end{equation}
\caption{Gluing data for 5d conformal matter tetraon of type $D_4$. The labels $a,b,c$ identify the degrees of the normal bundle in \eqref{total space}.}\label{D4 tetraon table}
\end{table}

\noindent\textbf{Infinite series and general result}\\
In this short section, we are going to list the infinite series of gluing for the conformal matter theories of type $(D_{2r+1}, D_k, D_k, D_k)$ and $(D_{2r}, D_k, D_k, D_k)$. \\
Let us circle the compactified line, that produces the gauging with one of the $D_k$ legs of the aforementioned singularities. Gluing with transition functions as in \eqref{eq:ugluinggeneral},
we get the following three infinite series:
\begin{itemize}
\item $a = b, c= 1$; this gives a gluing with the bifundamental  $(D_{-a-b + (3-k)}, \text{\circled{$D_{j}$}})$;

\item $a = b-1, c= 1$; giving $(A_1, D_{-a-b + (3-k)}, 
\text{\circled{$D_{k}$}})$;

\item $a = b, c= 2$; $(A_{-a-b -1}, \text{\circled{$D_{k}$}})$;
\end{itemize}

We notice the following facts:
\begin{enumerate}
    \item 
For even $k$, we have   $(D_{odd},\text{\circled{{$D_k$}}})$, $(D_{even},A_1,\text{\circled{$D_k$}})$, while for odd $k$, the first two series are $(D_{even},\text{\circled{$D_k$}})$ and $(D_{even},A_1,\text{\circled{$D_k$}})$.

\item All the results are independent from $r$, that simply changes the type of $D$-singularity on the "accidental" line that, in \cref{sec:1classification}, was at $x = u = v = 0$.
\item All the other gluings produce non-normal or non Calabi-Yau threefolds. 
\end{enumerate}

\subsubsection{Gluings that admit a UV complete generalized quiver phase}\label{sec: blowdowns}

In Section \ref{sec:gluingrules} we have laid down the geometrical rules to perform gauging operations between trinion and tetraon theories of type $D$, crepantly compactifying a line supporting Du Val singularities. In that framework, the end result is a 5d generalized quiver theory encoded by a CY3 $X$ in a partially resolved phase, where a collection of non-compact lines supporting Du Val singularities intersect on top of a $\mathbb{P}^1$ with non-vanishing volume. Algebraically, the CY3 engineering such gauge theory is embedded as a section of the anticanonical bundle in the ambient space.\\
\indent In this Section, we wish to identify which (if any) of the 5d generalized quiver phases obtained in such fashion admit a 5d SCFT phase. Geometrically, this corresponds to characterizing which partially resolved CY3 $Y$ can be blown down in a crepant fashion, thus eliminating all scales from the related theory. In the fully singular phase, the compact curve pictorially shown in Figure \ref{fig:genquiver}, along which the gluing happens, must shrink to zero size. Namely, this implies proving the existence of a crepant resolution map $f$ between a canonical maximally singular threefold $X$ and our threefold $Y$:
\begin{equation}\label{map f}
    f: \hspace{0.2cm} Y \longrightarrow X.
\end{equation}
 In this sense, $Y$ is a partial resolution of $X$, with $f$ \textit{at least} a small resolution that inflates the base $\mathbb{P}^1$ of \eqref{total space}.
 Thus, we say that $X$ is the \textit{singular phase} of $Y$, and that M-theory geometric engineering on $X$ produces a genuine 5d SCFT.\\ 
We would like to exhibit a necessary condition for the map $f$ in \eqref{map f} to exist. I.e.\ we wish to show how to explicitly construct the blow-down maps that produce the maximally singular phase.\\

\indent Let us start by recalling that $Y$ is covered by two charts related by \eqref{eq:ugluinggeneral} (or by \eqref{eq:ygluinggeneral}, depending on which non-compact line has been compactified). $Y$ lives in the total space \eqref{total space}, that admits the (GLSM) toric action:\footnote{To be precise, when one among $a,b,c$ is negative, the toric ambient space would be bigger, and contains the rank-three vector bundle over $\mathbb P^1$ as a Zariski open. However, this is equivalent for the purpose of our analysis.}
\begin{equation}\label{toric action 2}
    \begin{array}{ccccc}
      \lambda_0   &  \lambda_1  &\lambda_2 & 
    \lambda_3  & \lambda_4   \\
    \hline
      1 & -a  & -b & -c & 1 \\
    \end{array}.
\end{equation}
For convenience (but without loss of generality), we work in the patch where $\lambda_4$ has been fixed to 1. With this choice of coordinates, $\lambda_0$ corresponds to the coordinate on the $\mathbb{P}^1$, namely $v'$ or $y'$, while $\lambda_1$ corresponds to the $x'$ direction,\footnote{We recall that $\lambda_2, \lambda_3$ corresponds to other directions in the normal bundle of the compactified curve.} according to the transition functions \eqref{eq:ugluinggeneral} and \eqref{eq:ygluinggeneral}. It is easy to check that the possible terms involving $x'$ in the partially resolved threefold equation $Y$ are always of the form:
\begin{equation}\label{x terms}
    x'^2, \quad x'^2 y'^{k_1}, \quad x'^2 v'^{k_2},
\end{equation}
for some non-negative integers $k_1$ and $k_2$.
In order to construct the crepant blow-down map $f$, the monomials involving $x$ in \eqref{x terms} must be rewritten in terms of coordinates that are \textit{invariant} under the toric action \eqref{toric action 2}. It is evident that in order for this to be possible, $a$ must be greater than or equal to zero. Therefore, we can readily rule out the existence of the singular phase for all the gluings with $a<0$ from Tables \ref{D5 trinion table} and \ref{D4 tetraon table}. Furthermore, the threefolds with $a = 0$ are excluded if a monomial of type $x'^2 y'^{k_1}$ or $x'^2 v'^{k_2}$ (with $k_1,k_2\neq 0$) appears, as they can never be invariant under the toric action \eqref{toric action 2}.\\
\indent Following this reasoning, only a finite set of the partially resolved threefolds given by the gluing data in Table \ref{D5 trinion table} and \ref{D4 tetraon table} admit a singular phase. The latter can be straightforwardly obtained in a three step process:
\begin{itemize}
    \item choose a basis of invariant coordinates for the toric action \eqref{toric action 2};
    \item find a minimal set of independent relations between the invariant coordinates;
    \item the singular threefold is given by the intersection of the relations with the resolved threefold equation, rewritten in terms of the invariant coordinates.
\end{itemize}
Carefully carrying out the outlined reasoning, the 5d generalized quiver phases described by the ambient space data in Tables \ref{D5 trinion table} and \ref{D4 tetraon table} that admit a flow to a SCFT phase are respectively summarized in Tables \ref{SCFT D5 trinion table} and \ref{SCFT D4 tetraon table}.\\
\indent As an illustrative example, in Appendix \ref{app:sing geometries} we explicitly write down the singular geometries giving rise to the 5d SCFTs obtained from gluing $D_5$ trinions. Notice that many instances of singular CY3 are affine varieties which are \textit{non-toric non-complete intersections}. This is a realm that has so far been scarcely explored in the 5d SCFT literature.\\

\renewcommand{\arraystretch}{1.1}
\begin{table}[H]\centering
\begin{equation}
\scalemath{0.9}{
\begin{array}{|c|E|E|E|c|c|c|}
\hline 
 \multirow{2}{*}{\makecell{\textbf{Gluing}\\\textbf{coordinate}}} &
\multicolumn{3}{|c|}{\textbf{Degrees of line bundles}} & \multirow{2}{*}{\makecell{\textbf{Glued }\\ \textbf{singularities}}}  & \multirow{2}{*}{\makecell{\textbf{Glued singular CY3}\\}}\\
\cline{2-5}
 & \boldsymbol{a} & \boldsymbol{b} & \boldsymbol{c} & \\
\hline
\hline
v & 0 & 0 & 2 &  \text{\circled{$D_5$}} & \scalemath{0.8}{x^2+u^4 \left(-1+u^3 v^8\right)^4+u \left(-1+u^3 v^8\right) y^2=0}\\
\hline
y & 0& 1& 1& \text{\circled{$D_5$}}\oplus A_3\oplus A_1  & \scalemath{0.8}{x^2+u^4 y^{12} \left(v^2-u^3 y\right)^4+u y \left(-v^2+u^3
   y\right)=0}\\
y & 1& 1& 1&  \text{\circled{$D_5$}}\oplus A_1\oplus A_1 & \scalemath{0.8}{x^2 y-u v^2+u^4 y+u^4 y^{11} \left(v^2-u^3 y\right)^4=0}\\
y & 1& 2& 1&  \text{\circled{$D_5$}}\oplus A_1 &\scalemath{0.8}{ x^2 + u^4 - u v^2 y + y^{14} (u^4 - u v^2 y)^4 = 0}\\
y & 0& 0& 2&  \text{\circled{$D_5$}}\oplus A_5 & \scalemath{0.8}{x^2-u v^2+u^4 y^6+y^8 \left(u v^2-u^4 y^6\right)^4= 0} \\
y & 1& 1& 2&  \text{\circled{$D_5$}}\oplus A_3 & \scalemath{0.8}{x^2-u v^2+u^4 y^4+u^4 y^{14} \left(v^2-u^3 y^4\right)^4=0}\\
y & 2& 2& 2&  \text{\circled{$D_5$}}\oplus A_1 & \scalemath{0.8}{x^2-u v^2+u^4 y^2+\left(u v^2 y^5-u^4 y^7\right)^4 = 0}\\
y & 3& 3& 2&  \text{\circled{$D_5$}} & \scalemath{0.8}{x^2-u v^2+u^4+\left(u^4-u v^2\right)^4 y^{26} = 0}\\
\hline
\end{array}\nonumber}
\end{equation}
\caption{Gluing data for 5d conformal matter trinion of type $D_5$ that admits a SCFT phase. The circled algebras correspond to the singular lines which are compactified during the gluing.}\label{SCFT D5 trinion table}
\end{table}

\renewcommand{\arraystretch}{1.1}
\begin{table}[H]\centering
\begin{equation}
\scalemath{0.85}{
\begin{array}{|c|E|E|E|c|c|c|}
\hline 
 \multirow{2}{*}{\makecell{\textbf{Gluing}\\\textbf{coordinate}}} &
\multicolumn{3}{|c|}{\textbf{Degrees of line bundles}} & \multirow{2}{*}{\makecell{\textbf{Glued }\\ \textbf{singularities}}}  &\multirow{2}{*}{\makecell{\textbf{Glued singular CY3}\\}} \\
\cline{2-4}
 & \boldsymbol{a} & \boldsymbol{b} & \boldsymbol{c}  \\
\hline
\hline
v & 0 & 0 & 2 &  \text{\circled{$D_4$}}&  \scalemath{0.8}{x^2+u^3 \left(-1+u^2 v^6\right)^3+u \left(-1+u^2 v^6\right) y^2=0}\\
\hline
y & 0 & 2 & 0 & \text{\circled{$D_4$}} \oplus A_3 & \scalemath{0.8}{x^2-u^2 v+v^3 y^4+y^6 \left(-u^2 v+v^3 y^4\right)^3}\\
y & 0 & 1 & 1 & \text{\circled{$D_4$}} \oplus A_1\oplus A_1\oplus A_1 & \scalemath{0.8}{x^2+v (-u+v) (u+v) y+\left(-u^2 v+v^3\right)^3 y^9=0} \\
y & 1 & 1 & 1 & \text{\circled{$D_4$}} & \scalemath{0.8}{x^2 y+(-u+v) (u+v)+\left(-u^2 v+v^3\right)^3 y^8=0} \\
y & 1 & 2 & 1 & \text{\circled{$D_4$}} \oplus A_1& \scalemath{0.8}{x^2+v (-u+v y) (u+v y) \left(1+v^2 y^{10} \left(u^2-v^2
   y^2\right)^2\right)=0}\\
y & 2 & 2 & 2 & \text{\circled{$D_4$}} &  \scalemath{0.8}{x^2-u^2 v+v^3+\left(-u^2 v+v^3\right)^3 y^{14}=0} \\
\hline
\end{array}\nonumber}
\end{equation}
\caption{Gluing data for 5d conformal matter tetraon of type $D_4$ that admits a SCFT phase. The circled algebras correspond to the singular lines which are compactified during the gluing.}\label{SCFT D4 tetraon table}
\end{table}

A remark is in order: we note that a good rule of thumb to sum up the gluing rules is that the "molecule" that is obtained after the gluing should not exhibit a  flavor group different from those listed in \cref{sec:1classification}, up to $A$ factors.\\ 
\indent Stated differently, we could predict the fact that the gauging of two tetraons of type $D$ is \textit{not} allowed, as the final molecule would correspond to a 5d SCFT whose flavor group contains 6 factors of $D$ type. Indeed, this would arise from (after the contraction of the compact curve associated with the gauging) a singularity with 6 curves of $D$ type singularities intersecting at a point, that would very likely be, after the analysis we carried out in this work, not at finite distance in the CY moduli space. This also forbids the gauging of two trinions of $D$ type, as the resulting singularity would have 4 lines of $D$-type singularities intersecting at a single point, two of them being of $D_k$ type and two of them being of type $D_{2r+1}$. This setup is again in tension with the list of \cref{sec:1classification}, where we showed that these kind of singularities do not exist in the hypersurface case. 

\indent In the same spirit, we can follow the same reasoning to exclude the possibility of iterating the previous gauging procedure, giving rise to molecules composed of three or more atoms. Indeed, these would correspond to singularities with at least the same number of $D$ lines than the ones of \cref{sec:1classification}.\\

We can sum-up the previous argument as follows:  the explicit analysis of this section shows that  we can not produce, via gluings, new singularities with three or more lines of type $D$, extending in this way the list presented in \cref{sec:1classification}. 
 This limited set of gluings is a striking difference between 5d CM theories with at most two $D,E$ flavor factors, and the 5d CM trinions and tetraons of type $D$ that we are examining in this work. The former can be conformally gauged to produce infinitely long molecules that still correspond to singularities with at most two lines of type $D,E$, plus some additional $A$ lines. The latter, analogously, admit only gluings that preserve, or reduce, the number of $D$ factors in the flavor group. 

\section{A sporadic trinion case}\label{sec:sporadic}
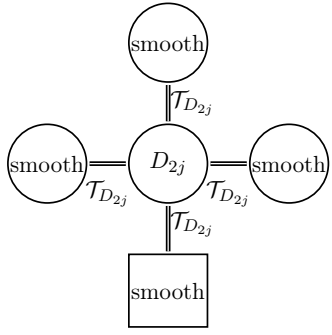
\begin{figure}[H]
    \centering
    \scalebox{0.8}{
    \begin{tikzpicture}
        \draw[thick] (0,0) circle (0.65);
        \node at (0,0) {\small$D_{2j}$};
        \draw[thick,double] (0.7,0)--(1.3,0);
        \draw[thick] (2,0) circle (0.65);
        \node at (2,0) {\small\text{smooth}};
        \draw[thick,double] (-0.7,0)--(-1.3,0);
        \draw[thick] (-0.65,-2.7)--(0.65,-2.7)--(0.65,-1.5)--(-0.65,-1.5)--cycle ;
        \node at (-2,0) {\small\text{smooth}};
        \node at (0,2) {\small\text{smooth}};
        \node at (0.4,1) {\small$\mathcal{T}_{D_{2j}}$};
        \node at (-1,-0.5) {\small$\mathcal{T}_{D_{2j}}$};
         \node at (1,-0.5) {\small$\mathcal{T}_{D_{2j}}$};
         \node at (0.4,-1) {\small$\mathcal{T}_{D_{2j}}$};
        \draw[thick] (0,2) circle (0.65);
        \draw[thick] (-2,0) circle (0.65);
        \draw[thick,double] (0,0.7)--(0,1.3);
        \draw[thick,double] (0,-0.7)--(0,-1.45);
        \node at (0,-2.1) {\small\text{smooth}};
        \end{tikzpicture}}
    \caption{Pictorial representation of the partial resolution of \eqref{sporadic D4}. The compact curves are arranged like a $D_4$ Dynkin diagram.}
    \label{D4sporadicfig}
    \end{figure}

In this Section we briefly comment on a trinion of type $(D_4^{\oplus 3})$ that arises from a hypersurface singularity that eludes the classification of Table \ref{table:allthreefolds}. Explicitly, it is engineered by the threefolds:
\begin{equation}\label{sporadic D4}
    \begin{cases}
        x^2+zy^2+z^3 = 0\\
        y=u(u^{j-1}+v)(u^{j-1}-v)\\
    \end{cases},
\end{equation}
which sports three non-compact singular lines of type $D_4$.
It is \textit{sporadic} in the sense that it does not come equipped with relatives arranged in infinite classes: indeed any similarly constructed threefold, e.g.\ obtained substituting the Du Val equation of type $D_4$ in \eqref{sporadic D4} with a $D_5$, would either saturate or violate the canonicity bound \eqref{canonicity}.\footnote{The only two cases that exactly saturate the bound are: \begin{equation}
    (D_5^{\oplus 3}): \hspace{0.2cm} \begin{cases}
        x^2+zy^2+z^4=0\\
        y=u(u+v)(u-v)\\
    \end{cases},\quad (E_6^{\oplus 3}): \hspace{0.2cm} \begin{cases}
        x^2+y^3+z^4=0\\
        z=u(u+v)(u-v)\\
    \end{cases}.
\end{equation} It is straightforward to check that no crepant resolution of these singularities exist, compatibly with the fact that they are \textit{not} canonical.}\\
\indent In order to inspect the properties of the 5d SCFT corresponding to \eqref{sporadic D4}, we resolve the singularity in a fashion completely analogous to the one laid out in Section \eqref{sec:resolution}. The end result of the partial resolution is the generalized quiver in Figure \ref{D4sporadicfig}. 

Notice that the partially resolved phase depicted in Figure \ref{D4sporadicfig} is obtained as the gauging of four theories of type $\mathcal{T}_{D_{2j}}$, which have been defined in \eqref{T_g eq}, and whose features have been classified in \eqref{eq:ranktopmaintext} and Table \ref{flavor data}. Hence, given the generalized quiver phase, we can extract the gauge and flavor rank of the 5d SCFT engineered by \eqref{sporadic D4}:
\begin{equation}
    \text{rank(gauge)} = 6j-4, \quad\quad \text{rank(flavor)} = 13.
\end{equation}
The rank of the flavor symmetry is compatible with the generalized flavor enhancement mechanism introduced in Section \ref{sec:flavor enhancement}, heavily hinting at the fact that the UV flavor symmetry is \textit{at least} $D_4^{\oplus 3}\oplus \mathfrak{u}(1)$. 

\section{Discussion and outlook}
\label{sec:conclusions}
In this work we have expanded the classification of 5d $\mathcal{N}=1$ SCFTs by introducing novel 5d fixed points with three or four UV flavor factors of type $D$, geometrically engineered by M-theory on non-toric CY3 hypersurfaces. We have also exhibited bifundamentals and trinions involving at most one $E$ and at least one $D$ flavor factor. We have clarified the physics of such fixed points, computing their CB dimensions, and proposing a novel flavor symmetry enhancement criterium to determine their flavor symmetry, employing a $D_k$-shaped generalized quiver description of the theory.  Furthermore, we have shown that appropriate gluings of these elementary blocks admit a 5d SCFT description, often in terms of M-theory on non-complete intersection CY3. These gluings always display an equal amount or less flavor factors of type $D,E$ than the original theories, unlike the well-known toric $A$ cases. 
Nonetheless, we have shown that the newly introduced bifundamentals/trinions/tetraons come in classes: each element of the class can be obtained as a gauging of some other 5d SCFT in the same class with a bifundamental 5d conformal matter SCFT.\\

Much room is left for further exploration of the landscape of 5d conformal matter theories, in order to elucidate and expand the five-dimensional atomic classification scheme.\\
  \indent Our argument against the existence of singularities with more than three or four lines of $D$ singularities hinges on the gluings of the finite list of families that can be written down as hypersurface singularities, listed in \cref{sec:5dtriniontetraon}.  A similar reasoning, of course, can be made for candidate singularities exhibiting multiple lines of type $E_6,E_7,E_8$, which cannot be constructed in our limited setups with CY3 exhibiting singularities at finite distance.\\
  \indent It is clear, though, that the geometric engineering techniques employed so far, however powerful, are starting to show their shortcomings: most of the literature has focused on toric CY3s, $\mathbb C^3$-orbifolds and complete-intersection singularities (as well as bottom-up constructions for low-rank theories \cite{Jefferson:2018irk}). 
Given the evidence collected in this work, it is still plausible that theories with flavor group at least $\mathfrak g^{\oplus n}$, with $\mathfrak g = D$ or $E$ and $n>2$ might not belong to these classes. In particular, non complete-intersection singularities that are both non-toric and not orbifolds of $\mathbb C^3$ are likely the vast majority in the landscape of CY3\footnote{In this context, it would be extremely interesting to lay down a precise criterion to establish whether a non-toric non-complete intersection singularity is at finite distance in the moduli space of the CY3. Relatedly, it is not known whether some (all?) 5d SCFTs engineered by non-toric non-complete intersection CY3 admit an equivalent description in terms of a complete intersection CY3.}. A motivation for this claim is their very natural appearance in Section \eqref{sec: blowdowns} and Appendix \ref{app:sing geometries}. We can however check, within the range accessible with the techniques presented in Section \ref{sec:gluingrules}, that such putative non-complete intersections of type $\mathfrak{g}^{\oplus n}$ (different from the ones appearing in this work) \textit{cannot} interact with the $n$-valent 5d CM that we presented so far. Stated differently, if these theories exist, they would be completely invisible to the analysis of the present work, and would thus span a "\textit{dark sector}" of $n$-valent 5d conformal matter, possibly admitting its own class of gluings. This demands the development of new and more intrinsic approaches for the analysis of 5d CM theories, and offers a challenging (and potentially highly rewarding) path for future work. It is however worth noticing that the aforementioned issues do not originate from the physics of M-theory, but rather from the limited set of geometric techniques that are available to define a non-compact CY3. Indeed, the generalized quivers studied in this work offer an alternative novel language to present and organize the CY3 landscape, beyond the limited corners that we can currently explore. In fact, without the generalized quivers, it would be very difficult to select, among all the possible non-complete intersections, the singularities presented in \Cref{app:sing geometries}, and to understand their interesting physical and geometrical properties. In this sense, exhibiting a bottom-up proof of the non-abelian enhancement of the flavor symmetries of the generalized quiver theories in \Cref{sec:CBdata} would provide a quick way to understand how many (and which types) of lines of Du Val singularity intersects at the maximum singular locus of the contraction of the generalized quiver. E.g.\ constructing generalized quivers with $E$ shape and $E$ gauge nodes could provide a partially resolved phase for a candidate trinion (or tetraon) UV fixed point of exceptional type. Crepantly shrinking such resolved phase, and thus identifying the fixed point, remains a tough unexplored challenge.  \\
  \indent Another possible starting point to map the landscape of 5d SCFT could be grounded on the faith in the conjecture of \cite{Jefferson:2017ahm}, claiming that all 5d SCFTs descend from 6d $\mathcal{N}=(1,0)$ SCFTs (or, in a refined version, from 6d LSTs \cite{Bhardwaj:2019hhd}), possibly with holonomies along the circle employed for the dimensional reduction.
   Geometrically, this is reflected in the following question: "does M-theory geometric engineering on a \textit{generic} canonical CY3 produces physical phenomena that cannot be obtained with F-theory geometric engineering on \textit{elliptic} CY3?" In this work, by leveraging the fact that we can consider also non-elliptically fibered\footnote{The answer is trivial in the case of elliptically fibered threefolds, as it consists only in taking the M-theory limit of F-theory.} CY3, we focused on one such peculiar feature: the presence of a factor of the form (at least) $D^{\oplus 3}$ and $E_7\oplus D^{\oplus 2}$ in the flavor group of the SCFT. The precise 6d origin of such theories remains to be understood. 
  The validity of the conjecture would readily rule out $n$-valent 5d CM with $n\geq 3$ and exceptional flavor factors $E_i^{\oplus n}$ (with $i=6,7,8$), as they would not fit in the largest bifundamental 6d $\mathcal{N}=(1,0)$ CM theory with $E_8^{\oplus 2}$ flavor symmetry. From this perspective, the lack of evidence for the existence of these theories offered by our work further reinforces the data in support of the conjecture.\\
\indent 
Additionally, the identification of the 6d origin of the theories $\mathcal{T}_{\mathfrak{g}}$ defined in Section \ref{sec:CBdata} would shed some further light on the 5d landscape of SCFTs with $\mathfrak{g}\oplus\mathfrak{g}'$ flavor group.\\
\indent Finally, we can regard the result of \Cref{sec:CBdata} as a generalization of the bottom-up instantonic symmetry enhancement of \cite{Tachikawa,Yonekura}. Our result is based on the top-down M-theory geometric engineering dictionary, and applies also to the realm of 5d SCFTs that do not admit a Lagrangian phase. It would be interesting to double-check it with bottom-up methods, which until now have only allowed us to match the expected flavor rank, stating precisely the link between our analysis and the physics of 5d $D$-type instantons. On top of that, the result of \cite{Tachikawa,Yonekura} is powerful enough to investigate the case where some of the weakly-coupled gauge groups appearing in an IR phase of the SCFT have non-zero CS levels. It would be interesting to further explore this direction for the generalized quivers that we introduced in this work.

\section*{Acknowledgments}
We would like to thank Andr\'es Collinucci for  discussions. 
The research of MDM is funded through an ARC advanced project, and further supported by IISN-Belgium (convention 4.4503.15). The work of MDZ is supported by the European Research Council (ERC) under the European Union’s Horizon Europe research and innovation program (grant agreement
No.\ 101171852) and by the VR project grant No.\ 2023-05590. MDZ  and AS acknowledge support from
the VR Centre for Geometry and Physics (VR grant No.\ 2022-06593). MDM, MDZ and AS have also received funding from the European Research Council (ERC) under the European Union’s Horizon 2020 research and innovation program (grant agreement No.\ 851931) during the initial stage of the work. MG is member of the GNSAGA - INdAM. MDZ also acknowledges support from the Simons Foundation Grants \#888984 (Simons Collaboration on Global Categorical Symmetries).

\appendix

\section{Ruling out a potential class of geometries}\label{sec:rulout?}
In this Appendix we present two no-go arguments: the first in order to show that a natural construction inspired by class $\mathcal{S}$ theories cannot straightforwardly produce trinions and tetraons in M-theory; the second to exclude the possibility to construct trinions and tetraons of exceptional type using the generalized quiver phases in Figure \ref{fig:mannaggia}.

\subsection{5d SCFTs and class \texorpdfstring{$\mathcal{S}$}{S}}
 A natural choice to consider in the context of the M-theory origin of 5d SCFTs are those geometries that in type IIB give rise to class $\mathcal S$ theories of type $\mathfrak{g}$, which are obtained from CY 3-fold singularities of type \cite{DelZotto:2015rca}
 \begin{equation}\label{eq:cumrunmagical}
     f_{ADE}(w_1(z_1,z_2,z_3,z_4),w_2(z_1,z_2,z_3,z_4),w_3(z_1,z_2,z_3,z_4)) = 0, 
 \end{equation}
 where $f_{ADE}=0$ is a du Val singularity of ADE type and the locus $w_1 = w_2 = w_3 =0$ is the UV curve of the corresponding class $\mathcal S$ theory \cite{DelZotto:2015rca}. In IIB the ADE singularities corresponds to 6d (2,0) SCFTs, and these geometries correspond to the class $\mathcal S$ construction \cite{Gaiotto:2009gz,Gaiotto:2009we}. In this appendix we give a short argument to argue that considering these same geometries in M-theory won't work. 
 
 \medskip
 
 Based on the analysis in \cite{Jefferson_2018} we expect shrinkable geometries in M-theory require surfaces intersecting along rational curves, in general the locus $w_1=w_2=w_3=0$ is not rational, hence naively we expect from it an obstruction to shrinkability.
 
 \medskip
 
 This can be explained by the following heuristic argument: if the geometry in \eqref{eq:cumrunmagical} were to give a 5d SCFT, its circle reduction would correspond to a 4d SCFT (upon suitable decoupling of the KK towers). Now, M-theory on an ADE singularity is not an SCFT, rather it is a 7d gauge theory of type ADE. Upon circle reduction one has a duality with Type IIA superstrings. From such duality it is natural to argue that the circle reduction of the 7d gauge theory is nothing but a 6d (1,1) little string theory.

 \medskip
 
In Type IIA the geometry in Equation \eqref{eq:cumrunmagical} corresponds to a reduction of this 6d little string on the Riemann surface $w_1 = w_2= w_3 = 0$. In the limit in which the corresponding Riemann surface shrinks to zero size, along each plumbing cylinder that shrinks as we take the volume of the Riemann surface to zero, infinite towers of winding modes for the LST become light, effectively enforcing a T-dual description where the Riemann surface becomes large. This mechanism obstructs the SCFT in 4d, and hence we do not expect to have a corresponding 5d SCFT to begin with. For this reason, we expect that the resulting geometries, while interesting, should not be shrinkable in M-theory, and do not give rise to 5d SCFTs.

\subsection{Generalized quivers and the absence of exceptional trinions and tetraons}\label{app: no exceptional trinions}
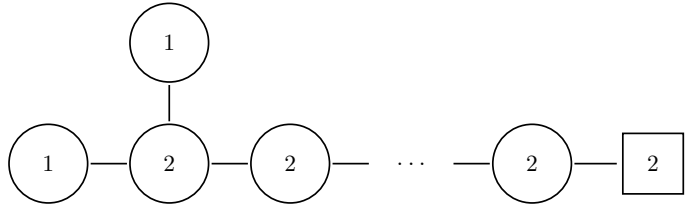
\begin{figure}[H]
    \centering
    \scalebox{0.8}{
    \begin{tikzpicture}
        \draw[thick] (0,0) circle (0.65);
        \node at (0,0) {\small$2$};
        \draw[thick] (0.7,0)--(1.3,0);
        \draw[thick] (2,0) circle (0.65);
        \node at (2,0) {\small$2$};
        \draw[thick] (-0.7,0)--(-1.3,0);
        \draw[thick] (2.7,0)--(3.3,0);
        \node at (4,0) {$\ldots$};
        \draw[thick] (4.7,0)--(5.3,0);
        \draw[thick] (6,0) circle (0.65);
        \draw[thick] (6.7,0)--(7.4,0);
        \draw[thick] (7.5,-0.5)--(8.5,-0.5)--(8.5,0.5)--(7.5,0.5)--cycle ;
        \node at (6,0) {\small$2$};
        \node at (8,0) {\small$2$};
        \draw[thick] (-2,0) circle (0.65);
        \draw[thick] (0,2) circle (0.65);
        \node at (-2,0) {\small $1$};
        \node at (0,2) {\small $1$};
        \draw[thick] (0,0.7)--(0,1.3);
        \end{tikzpicture}}
 \caption{Balanced quiver of type $D$, labelled by the weight in \eqref{D weight}.}
    \label{fig:quiverweights}
    \end{figure}

From the geometric point of view, our techniques for constructing trinions and tetraons rely on the generalized quiver realization of the type depicted in Figure \ref{fig:mannaggia}, that can then be crepantly shrinked to a 5d SCFT phase engineered by a canonical CY3. The crucial reason why the quivers admit a UV completion is that the edges are labelled by $\mathcal{T}_{\mathfrak{g}}$ and $X_{\mathfrak{g}}^{(1)}$. In a local patch, these can be written as canonical CY3 of the form:
\begin{equation}
\begin{split}
    & \mathcal{T}_{\mathfrak{g}}: \hspace{0.5cm}\quad ab^2 = Q_{\mathfrak{g}}(u,v)\\
    & X_{\mathfrak{g}}^{(1)}: \quad a^2b^2 = Q_{\mathfrak{g}}(u,v)\\
    \end{split},
\end{equation}
with $Q_{\mathfrak{g}}(u,v)$ a polynomial labelling the $\mathfrak{g}$ Du Val singularity, obtained from Table \ref{tab:DAA CM} and setting $x_1 = 0$. In turn, the multiplicities of $a$ and $b$ define a labelling of the nodes in the quiver, according to Figure \ref{fig:quiverweights}. 
Notice that the quiver can be labelled by the following weight of the Lie algebra of type $D$:
\begin{equation}\label{D weight}
\renewcommand{\arraystretch}{0.8}
   \scalemath{0.8}{ \begin{bmatrix}
            0&&&&\\
            &0&\dots&0&2\\
           0 &&&&
        \end{bmatrix}}
\end{equation}
For a generic algebra of type $D$, this is the only weight that corresponds to a balanced quiver with gauge multiplicities 1 or 2\footnote{With one exception for the $D_4$ case, that we will explore in Section \ref{sec:sporadic}.}. Any other weight produces at least one gauge node with multiplicity 3. This would result in a local patch of the generalized quiver being described by the threefold
\begin{equation}
    a^2b^3 = Q_{\mathfrak{g}}(u,v),
\end{equation}
which is clearly \textit{not canonical}, since it has a singular line of non-Du Val type \cite{reid1980canonical}.\\
\indent Now comes the argument for excluding trinions and tetraons of exceptional type: a candidate generalized quiver phase with shape $E_6,E_7,E_8$ must be labelled by a weight of the corresponding Lie algebra. \textit{All} such weights label at least one of the gauge nodes with multiplicity greater than 2, thus stumbling into a non-canonical threefold in at least one local patch of the generalized quiver. Hence, within the reach of our techniques, exceptional trinions and tetraons would correspond to generalized quivers that do not admit a flow to a UV fixed point, and are thus excluded.

\section{Singular geometries for gluings of trinions and tetraons}
\label{app:sing geometries}
In this Appendix we schematically collect the singular geometries for the gluings of trinions and tetraons that admit a SCFT phase, collected in Tables \ref{SCFT D5 trinion table} and \ref{SCFT D4 tetraon table}. In general, one proceeds in the fashion outline in Section \ref{sec: blowdowns}.\\
\indent For concreteness, let us recall the pivotal equations in the case of a trinion singularity\footnote{The same reasoning pulls through for the tetraon $D_{2n}$ case.}. The starting point is the singular equation of a trinion
\begin{equation}
    \begin{cases}
        x^2+zy^2+z^{2n-2} = 0,\\
        z = u(u^{2n-3}+v^2), \\
    \end{cases}
\end{equation}
displaying three non-compact lines of $D_{2n-1}$ singularities along
\begin{equation}\label{sing lines}
    x = y = u = 0, \quad x = u = v = 0, \quad x = y = v^2+u^{2n-3} =0.
\end{equation}
In Section \ref{sec:gluingrules} we have shown the most general way to compactify the lines parameterized by $v$ and $y$ in \eqref{sing lines}, respectively, while preserving the Calabi-Yau condition. Physically, this operation corresponds to gauging the flavor group associated to one of the non-compact lines, introducing possibly extra flavor groups, i.e.\ non-compact lines of singularity passing through the point at infinity that has been added for the compactification. In full generality, the gauged threefold $Y$ is a section in $\Gamma(\mathcal A, -K_{\mathcal{A}})$ in the ambient space:
\begin{equation}\label{total space 2}
   \mathcal{A}:\quad \text{Tot}_{\mathbb{P}^1} \left(\mathcal{O}(-a)\oplus\mathcal{O}(-b)\oplus\mathcal{O}(-c)\right).
 \end{equation}
The two charts covering $\mathcal{A}$ are then related as:
\begin{equation}\label{charts P1}
\begin{split}
& v \text{ line}: \quad   v = \frac{1}{v'}, \quad x = v'^a x', \quad y = v'^b y', \quad u = v'^c u',\\
 & y \text{ line:} \quad   y = \frac{1}{y'}, \quad x = y'^a x', \quad v = y'^b v', \quad u = y'^c u',
 \end{split}
\end{equation}
depending on the line that has been compactified.

\begin{figure}[H]
    \centering
    \scalebox{0.8}{
    \begin{tikzpicture}
        \draw[thick] (-2,0)--(2,0);
        \draw[thick] (0,-1.5)--(0,1.5);
        \draw[thick] (-1.3,-1.3)--(1.3,1.3);
        \draw[<->,thick]  (1.8,0.3)--(2.5,0.3);
        \node at (2.8,0.7) {$\mathcal{O}(0)$ direction};
        \node at (-1.5,0.3) {$D_5$ line};
        \node at (-0.8,1.2) {$D_5$ line};
        \node at (-2,-1) {\circled{$D_5$} line};
        %
        \draw[->,thick] (6,0)--(7.5,0);
        \node at (6.75,0.3) {blow-down};
        \draw[thick] (9.5,0)--(13.5,0);
        \draw[thick] (11.5,-1.5)--(11.5,1.5);
        \draw[thick,red,fill=black] (11.5,0) circle (0.07);
        \node at (10,0.3) {$D_6$ line};
        \node at (10.7,-1) {$D_5$ line};
        \end{tikzpicture}}
    \caption{Pictorial representation of the partially resolved and singular phase of the $D_5$ trinion with compactified $u$ coordinate and normal bundle $\mathcal{O}(0)\oplus\mathcal{O}(0)\oplus\mathcal{O}(-2)$. The black lines are non-compact, the line labelled by a "circled" algebra is compact. Notice that the $D_5$ line in the $\mathcal{O}(0)$ direction is enhanced to $D_6$ in the singular phase.}
    \label{fig: D5 blowdown}
    \end{figure}
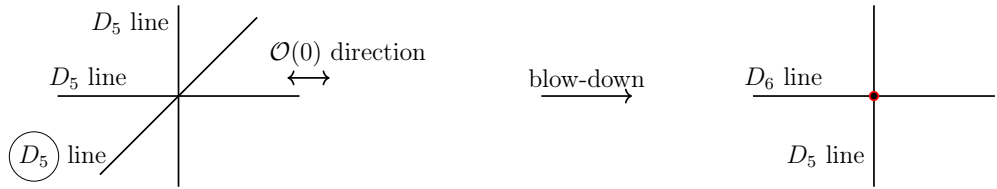

Let us show how the singular phase corresponding to the gluings in Table \ref{SCFT D5 trinion table} can be explicitly retrieved. As a warm-up example, we consider a case in which the ambient space \eqref{total space 2} is Calabi-Yau. We wish to compactify the $v$ coordinate, with normal bundle labelled by:
\begin{equation}\label{A1 normal 1}
    \{a,b,c\} = \{-k,-k,2\}\big|_{k=0} = \{0,0,2\}.
\end{equation}
We remark here that the labels $a,b,c$ identify the degrees of the normal bundle that specify the ambient space \eqref{total space 2}.
The ambient space containing the CY threefold is then:
\begin{equation}\label{A1 total}
       \mathcal{A}:\quad  \text{Tot}_{\mathbb{P}^1}\left(\mathcal{O}(0)\oplus\mathcal{O}(0)\oplus\mathcal{O}(-2)\right).
\end{equation}
In this case, it is easy to construct the blow-down map, noticing that \eqref{A1 normal 1} is precisely the normal bundle of the resolved $A_1 \times \mathbb{C}^2$. Hence, the blow-down map is crepant by construction and reads:
\begin{equation}\label{D5 blowdown}
 x=x, \quad y = y,  \quad A_1 = u v^2, \quad A_2 = u, \quad A_3 = vu.
\end{equation}
The resulting singular Calabi-Yau threefold reads:
\begin{equation}\label{D5 sing}
    \begin{cases}
        A_1 A_2 = A_3^2,\\
        x^2+y^2(A_1+A_2^4)+(A_1+A_2^4)^4 =0.
    \end{cases}
\end{equation}
Let us cross-check the validity of our procedure, investigating the singular locus of \eqref{D5 sing}:
\begin{equation}
\begin{array}{ccc}
  x= A_1 = A_2 = A_3 =0    &   \longleftrightarrow & D_6 \text{ singularity} \\
  x= y = A_1+A_2^4 =A_3^2+A_2^5=0    &   \longleftrightarrow &  D_5 \text{ singularity}\\
\end{array}
\end{equation}
As expected from Table \ref{SCFT D5 trinion table}, we end up with only two singular non-compact lines, because one of the $D_5$ lines in the partially resolved phase has been compactified according to \eqref{A1 normal 1}, and no $A$ line is present because we have chosen $k=0$. What is puzzling is the presence of a non-compact line of $D_6$ singularities, that did not manifestly appear in the resolved phase. On closer inspection, though, its presence is completely natural: the blow-down map \eqref{D5 blowdown} does not only shrink the base of the ambient space \eqref{A1 total} to zero-volume, but also all the fibers of a $\mathbb{P}^1$-fibered non-compact 4-cycle, locally given by $\mathbb{P}^1\times \mathbb{C}$: the base of this 4-cycle corresponds to one of the $\mathcal{O}(0)$ directions in \eqref{A1 total}. The base of \eqref{A1 total} is the central fiber of the 4-cycle, and it supports a $D_5$ singularity. This feature of the blow-down map effectively enhances one of the $D_5$ singular lines that were present in the partially resolved phase to a $D_6$ singular line in the maximally singular phase. No scale remains in the resulting threefold, thus producing the sought-after 5d SCFT in the transverse directions. We pictorially represent the blow-down operation in Figure \ref{fig: D5 blowdown}.

We can now safely move to a more involved case, in which the ambient space is not Calabi-Yau. For the sake of concreteness, consider the case where the line parameterized by $y$ is glued according to:
\begin{equation}\label{A1 normal 2}
    \{a,b,c\} = \{1,1,1\}.
\end{equation}
Namely, the ambient space containing the CY threefold is then:
\begin{equation}\label{conifold total}
       \mathcal{A}:\quad  \text{Tot}_{\mathbb{P}^1}\left(\mathcal{O}(-1)\oplus\mathcal{O}(-1)\oplus\mathcal{O}(-1)\right).
\end{equation}
In order to access the singular phase, notice that the total space \eqref{conifold total} comes naturally equipped with a toric action. Calling the toric coordinates $\lambda_0$ with $i = 0,\ldots,4$, the explicit action reads:
\begin{equation}\label{toric action}
    \begin{array}{ccccc}
      \lambda_0   &  \lambda_1  &\lambda_2 & 
    \lambda_3  & \lambda_4   \\
    \hline
      1 & -1  & -1 & -1 & 1 \\
    \end{array}.
\end{equation}
It is convenient to focus on the patch with $\lambda_4 = 1$ (the one with $\lambda_4 = 0$ lends itself to a completely analogous analysis). Then, it is easy to check that the ring of invariant operators under \eqref{toric action} is generated by:
\begin{equation}\label{inv coord}
\begin{array}{cc}
     A_1 = \lambda_0 \lambda_1,\quad & B_1 = \lambda_1,  \\
   A_2 = \lambda_0 \lambda_2,\quad   & B_2 = \lambda_2, \\ 
   A_3 = \lambda_0 \lambda_3,\quad  &  B_3 = \lambda_3. \\
\end{array}    
\end{equation}
The invariant coordinates \eqref{inv coord} satisfy relations that come from the toric action, as well as from the glued equations \eqref{D5 trinion}. The relations can be presented as:
\begin{equation}\label{sing 3fold}
\begin{cases}
   A_1B_2-A_2B_1=0, \\  
    A_1B_3-A_3B_1=0, \\  
     A_2B_3-A_3B_2=0, \\
A_1B_1+B_2(B_3^2+A_2B_2^2)+B_2A_2^3(A_3^2+A_2^3)^4 =0.\\
\end{cases}
\end{equation}
\eqref{sing 3fold} is a \textit{non-complete intersection} singular threefold, and it can be straightforwardly checked that it displays two lines of $D_5$ singularities, as well as two lines of $A_1$ singularities, as expected from Table \ref{SCFT D5 trinion table}. Notice that the invariant locus \eqref{sing 3fold} is not irreducible. Indeed, other than the threefold of our interest, also the coordinate plane $H=\{B_1=B_2=B_3=0\}$ is invariant under the torus action \eqref{toric action}. Therefore, in order to focus exclusively on the singular threefold we saturate \eqref{sing 3fold}, see \cite[Theorem 10 Ch. 4 \S4]{coxlittleshea},  with respect to the defining ideal of $H$ and we obtain the following singular phase:
\begin{equation}
    \begin{cases}
        A_1B_2-A_2B_1=0, \\  
    A_1B_3-A_3B_1=0, \\  
     A_2B_3-A_3B_2=0, \\
A_1^2+A_2 \left(A_2^3 \left(A_2^3+A_3^2\right)^4+A_2
   B_2^2+B_3^2\right)=0, \\
   A_1 B_1+A_2 B_2 \left(A_2^{14}+B_2^2\right)+A_2^4 A_3
   \left(4 A_2^9+6 A_2^6 A_3^2+4 A_2^3 A_3^4+A_3^6\right)
   B_3+B_2 B_3^2= 0, \\
    \end{cases}
\end{equation}
which correctly exhibits only two lines of $D_5$ singularities along with two lines supporting $A_1$ singularities. A similar procedure can be followed in all cases in Table \ref{SCFT D5 trinion table} in order to retrieve the singular expression without additional planes of singularities.

In Table \ref{tab:resthreefolds} we list all the threefolds corresponding to the gaugings in Table \ref{SCFT D5 trinion table}, that admit a singular phase which is a 5d SCFT.\\

\renewcommand{\arraystretch}{1.2}

\begin{table}[H]\centering
\begin{equation}
\scalemath{0.8}{
\begin{array}{c|c|l|c}
\makecell{\textbf{Glued}
\\\textbf{coordinate}} & \textbf{Ambient space} & \textbf{Singular equation} & \textbf{Lines of singularity} \\
\hline 
\hline
v & \{0,0,2\} & \scalemath{0.8}{\begin{cases}A_1 A_2=A_3^2,\\
A_4^2+A_5^2(A_1+A_2^4)+(A_1+A_2^4)^4 = 0. \\
\end{cases} }& D_5\oplus D_6\\
\hline
y & \{0,0,2\} &\scalemath{0.8}{ \begin{cases}
  A_1A_2 =A_3^2,  \\
    A_4^2-A_5^2 A_2+A_1^3 A_2+A_1^4 \left(A_5^2-A_1^3\right)^4.\\
\end{cases} }& D_5\oplus D_6 \oplus A_5\\
\hline
y & \{0,1,1\} & \scalemath{0.8}{\begin{cases}
    A_1A_2 -A_3A_4 = 0, \\
    A_5^2-A_3 \left(A_4^2-A_3 A_2^2\right)+A_3^4
   \left(A_1^2-A_3^3\right)^4.
\end{cases}} & D_5\oplus D_5 \oplus A_3 \oplus A_1\\
\hline
y & \{1,1,1\} & \scalemath{0.8}{\begin{cases}
   A_1B_2-A_2B_1=0, \\  
    A_1B_3-A_3B_1=0, \\  
     A_2B_3-A_3B_2=0, \\
A_1B_1+B_2(B_3^2+A_2B_2^2)+B_2A_2^3(A_3^2+A_2^3)^4 =0. \\
\end{cases} }&  D_5\oplus D_5 \oplus A_1 \oplus A_1\\
\hline
y & \{1,2,1\} &\scalemath{0.8}{\begin{cases}A_2 B_2-C_1^2=0,\quad\quad A_2 B_1-A_1 C_1=0,\\
A_3 B_1-A_1 B_3=0,\quad\quad A_1 B_2-B_1 C_1=0,\\
A_2 B_3-A_3 C_1=0,\quad\quad A_3 B_2-B_3 C_1=0,\\ 
A_3^2 B_3^2 \left(A_3^3-A_2^2\right)^4-A_3 B_2^2+B_1^2+B_3^4 =0.\\
\end{cases}} & D_5\oplus D_5 \oplus A_1\\
\hline
y & \{1,1,2\} &\scalemath{0.8}{ \begin{cases}A_3 B_3-C_1^2=0,\quad\quad A_2 B_1-A_1 B_2=0,\\
A_3 B_1-A_1 C_1=0,\quad\quad A_1 B_3-B_1 C_1=0,\\
A_3 B_2-A_2 C_1=0,\quad\quad A_2 B_3-B_2 C_1=0,\\
A_3^3 B_3 \left(A_2^2-A_3^3\right)^4+B_1^2-B_2^2 B_3+C_1^4=0. \\
\end{cases} }& D_5\oplus D_5\oplus A_3\\
\hline
y & \{2,2,2\} & \scalemath{0.8}{\begin{cases}
C_1^2-A_1 B_1= 0, \quad\quad
C_2^2-A_2 B_2= 0,\\
C_3^2-A_3 B_3= 0, \quad\quad
A_2 B_1-A_1 B_2= 0,\\
A_2 C_1-A_1 C_2= 0, \quad\quad
A_3 B_1-A_1 B_3= 0,\\
A_3 C_1-A_1 C_3= 0, \quad\quad
C_1 C_2-A_1 B_2= 0,\\
C_1 C_3-A_1 B_3= 0, \quad\quad
A_3 B_2-A_2 B_3= 0,\\
A_3 C_2-A_2 C_3= 0, \quad\quad
C_2 C_3-A_2 B_3= 0,\\
B_2 C_1-B_1 C_2= 0, \quad\quad
B_3 C_1-B_1 C_3= 0,\\
B_3 C_2-B_2 C_3= 0,\\
\left(A_2^2 C_3-A_3^3 C_3\right)^5+B_1^2-B_2^2 B_3+B_3^2 C_3^2 = 0. \\
\end{cases}} & D_5\oplus D_5\oplus A_1 \\
\hline 
y & \{3,3,2\} & \scalemath{0.8}{ \begin{cases}
    A_2 C_1-A_1 C_2 = 0,\quad\quad A_3
   C_1-A_1 C_3 = 0,\\
   A_3 C_2-A_2
   C_3 = 0,\quad\quad
   B_2 D_1-B_1
   D_2 = 0,\\ 
   B_1 C_3-B_3
   D_1 = 0,\quad\quad
   B_2 C_3-B_3
   D_2 = 0,\\
   A_2 D_1-A_1
   D_2 = 0,\quad\quad
   C_1^2-A_1 D_1 = 0,\\
   C_1
   C_2-A_1 D_2 = 0, \quad\quad
   A_1 B_3-A_3
   D_1 = 0,\\
   A_1 B_3-C_1
   C_3 = 0, \quad\quad
   A_1 B_1-C_1
   D_1 = 0,\\
   C_2^2-A_2 D_2 = 0, \quad\quad
   A_2
   B_3-A_3 D_2 = 0,\\
   A_2 B_3-C_2
   C_3 = 0, \quad\quad
   A_2 B_1-A_1
   B_2 = 0,\\
   A_1 B_2-C_1
   D_2 = 0, \quad\quad
   A_1 B_2-C_2
   D_1 = 0,\\
   A_2 B_2-C_2
   D_2 = 0, \quad\quad
   A_3 B_3-C_3^2=0,\\
   A_3
   B_1-B_3 C_1 = 0, \quad\quad
   A_3 B_1-C_3
   D_1 = 0,\\
   A_3 B_2-B_3
   C_2 = 0, \quad\quad
   A_3 B_2-C_3
   D_2 = 0,\\
   B_1 C_1-D_1^2 = 0, \quad\quad
   B_2
   C_1-B_1 C_2 = 0,\\
   B_1 C_2-D_1
   D_2 = 0, \quad\quad
   B_2
   C_2-D_2^2 = 0,\\
   B_1^2-B_2^2
   B_3+B_3^4+A_3^3 B_3 \left(-A_2
   C_2+A_3^2 C_3\right)^4 = 0.\\
   \end{cases}}& D_5\oplus D_5 \\
\hline
\end{array}\nonumber}
\end{equation}

\vspace{-0.5cm}

\caption{Singular phases of gauged 5d conformal matter trinions of type $D_5$.}\label{tab:resthreefolds}
\end{table}

\section{Computation of \texorpdfstring{$\boldsymbol{r_{\mathfrak{g}}}$}{rg}, \texorpdfstring{$\boldsymbol{f_{\mathfrak{g}}}$}{fg}}
\label{app:vertical}
\indent In this appendix, we show in full details the computation of the number of vertical divisors $r_{\mathfrak{g}}$ and of homologically independent vertical curves $f_{\mathfrak g}$, that are needed to characterize the conformal matter theories presented in this work.\\

\indent Consider the curve $\mathcal C \cong \mathbb P^1$ associated with the only \textit{trivalent} node in Figure \ref{D4zfig}. This curve intersects the two curves associated with the "smooth" labels in Figure \ref{D4zfig} in two points $p_1$, $p_2$. In a neighborhood $U$ of, e.g., $p_1$, the threefold is modeled by the following local equation: 
\begin{equation}
\label{eq:localmodel}
Q_{\mathfrak g}(u,v) = a b^2 , \qquad p_1 = \left\{u = v = a = b = 0\right\},
\end{equation}
where $(a,b) \in \mathbb C^2$ are coordinates in the corresponding chart of the atlas spanned by $(a_i,b_i) \in \mathbb C^2$, that covers $\widetilde{X}$. Equation \eqref{eq:localmodel} displays a line of singularities of type $P_{\mathfrak g}$ along $u = v = b = 0$, parametrized by the coordinate $a \in \mathbb C \subset \mathcal C$; this corresponds to a patch inside the line corresponding to the trivalent node in \Cref{D4zfig}. If we instead fix $b \neq 0$, we see a $\mathbb C^2$; hence, we can conclude that we have a $\mathbb C^2$-bundle over a complex line spanned by $b$; this corresponds to a patch inside the compact curve labeled by "smooth" in \Cref{D4zfig}.  We observe that it is not possible, for $\mathfrak g \neq A_1$ and for fixed $b$, to put  \eqref{eq:localmodel} in a single-center Taub-NUT form, signaling that there is no standard IIA limit for this line. 

To physically characterize the 5d SCFTs presented in this work, we need two pieces of data about \eqref{eq:localmodel}: 
\begin{enumerate}
    \item The rank of the Coulomb Branch $r_{\mathfrak g}$ of the SCFT engineered by of M-theory on \eqref{eq:localmodel}.
    \item The rank of the flavor group of M-theory on \eqref{eq:localmodel}, that we compute as
    \begin{equation}
    \label{eq:flavorgrouplocalmodel}
        \text{rank}(\mathcal F) = \text{rank}\left(H_2(T_{\mathfrak g},\mathbb Z)\right) - \text{rank}\left(H_4(T_{\mathfrak g},\mathbb Z\right))  = \text{rank}(\mathfrak g) + f_{\mathfrak g}, 
    \end{equation}
    where in the last step we isolated the non-trivial contribution to the flavor group, that cannot be inferred simply looking at the non-compact Du Val $P_{\mathfrak g}$ line of \eqref{eq:localmodel}. 
\end{enumerate}

We will obtain $r_{\mathfrak g}, f_{\mathfrak g}$ performing a resolution of \eqref{eq:localmodel}, starting by resolving the line of $P_{\mathfrak g}$ singularities by subsequently blowing up curves that dominate, via the $\pi_2$ map,  $\mathcal C$.
\\\\Let's consider, for example, $\mathfrak g = D_4$ in \eqref{eq:localmodel}. It is convenient to introduce the following polynomial
\begin{equation}
    \label{eq:d4localmodel}
    T_{D_4}(a,b,u,v) \equiv uv(u+v) - a b^2, 
\end{equation}
and to describe \eqref{eq:localmodel} for $\mathfrak g = D_4$ as its zero-locus:
\begin{equation}\label{zero locus}
   Q_{D_4}(u,v)-ab^2 =  T_{D_4}(a,b,u,v) = 0.
\end{equation}

\begin{itemize}
    \item \textbf{First blow-up $\boldsymbol{\pi_1}$}
\end{itemize}
We can now blow-up the singular line $ \mathcal C \equiv \left\{u = v = b = 0 \right\}$ of \eqref{zero locus}. To do so, we leverage the technology of toric ambient space blow-up, introducing a new coordinate $\delta_1$ and modding the new ambient space by a $\mathbb C^*$-action. This is nothing but the usual blow-up of the ambient $\mathbb C^4 \ni (a,b,u,v)$ along $\mathcal C$. The blow-up map reads 

\begin{equation}
    \label{eq:d4firstblow-up}
 \pi_1:\quad    (a,b,u,v) = (a, b_1 \delta_1, u_1 \delta_1,v_1\delta_1), 
\end{equation}
with $(a,b_1,u_1,v_1,\delta_1)$ belonging to the following toric (ambient) space: 
\begin{equation}
  \label{eq:d4firstblambient}
\begin{array}{c|cccccc}
 & a & b_1 & u_1 & v_1 & \delta_1  & \text{FI} \\
\hline
 \mathbb C^*_1&  0 & 1 & 1 & 1 & -1 & \xi_1  \\
\end{array}
\end{equation}
where, from now on, we assume $\xi_i > 0$. 
We are interested in the proper transform of \eqref{eq:d4localmodel},\footnote{The proper transform is defined as follows: let $V \subset \mathcal A$ be a variety of a certain ambient space $\mathcal A$. Take a blow-up of $\mathcal A$ with center $B$ (namely, we blow-up $B$ inside $\mathcal A$), then the proper transform of $V$ is defined as follows: 1) take $\hat{V} = V \setminus (B \cap V)$, 2) take the preimage $V_{open} \subset \text{Bl}_{B}(\mathcal A)$ of $\hat{V}$ w.r.t. the blow-up map, 3) take the Zariski closure of $V_{open}$. This procedure outputs the proper transform, that is the (closure) of the preimage under the blow-up, with the exceptional divisor excised. This operation is obtained dividing by  the appropriate power of $\delta_1$ in \eqref{eq:d4threefold1} (an operation that restricts us to $\delta_1 \neq 0$) and then extending the resulting variety to $\delta_1 = 0$.} described as the zero locus of
\begin{equation}
    \label{eq:d4threefold1}
    T^{(1)}(a,b_1,u_1,v_1,\delta_1) \equiv \frac{T(a,b,u,v)\rvert_{\eqref{eq:d4firstblow-up}}}{\delta_1^2}  = a b_1^2 + u_1 v_1 (u_1 + v_1) \delta_1. 
\end{equation}

Let us describe the geometry of \eqref{eq:d4threefold1}: the ambient space coordinates $(b_1, u_1, v_1)$ are the homogeneous coordinates of a  $\mathbb P^2$, located, in the ambient space, at $\delta_1 = 0$. Indeed, thinking of \eqref{eq:d4firstblambient} as a GLSM charge matrix, the D-term associated to $\mathbb C^*_1$ reads: 
\begin{equation}
    \label{eq:dtermfirstblow-up}
    |b_1|^2 + |u_1|^2 + |v_1|^2 - |\delta_1|^2  = \xi_1,   
\end{equation}
and setting $\delta_1 = 0$ we see that,\footnote{The ideal locus we exclude by consistency of the \eqref{eq:dtermfirstblow-up} is what is called, in the toric GIT literature, the Stanley-Reisner ideal.} after imposing \eqref{eq:dtermfirstblow-up}, \begin{equation}\label{eq:auxP2description}
(b_1,u_1,v_1) \in \frac{\mathbb C^3 \setminus \left\{ b_1 = u_1 = v_1 = 0 \right\}}{\mathbb C^*_1} = \mathbb P^2,
\end{equation}
We now have to intersect the toric ambient space with the equation of the threefold \eqref{eq:d4threefold1}:  the $\mathbb P^2$ in \eqref{eq:auxP2description} is sliced along the following line for $a \neq 0$, 
\begin{equation}
    \label{eq:noncompacttrivalentP1bundle}
    b_1 = \delta_1 = 0
\end{equation}

This is the $\mathbb P^1$ that partially resolves the line $\mathcal C$ as 
\begin{equation}
\label{eq:firstpartialresD4}
 A_1 \oplus A_1 \oplus A_1 \rightarrow D_4,
\end{equation}
and, varying the value $a$, we obtain a non-compact "horizontal" divisor, that is a $\mathbb P^1$ bundle over $\mathcal C$. 
On the other hand, for $a = 0$ \eqref{eq:auxP2description} is fully contained in the threefold. We hence conclude that the first blow-up \eqref{eq:d4firstblow-up} inflates a compact divisor, that contracts to the origin of \eqref{zero locus} (and hence is what we called a "vertical" divisor in the main text). As we will shortly see, there are no other such compact vertical divisors in the resolution of \eqref{eq:d4localmodel}. Hence, we get our first result: M-theory on \eqref{eq:d4localmodel} is a rank one 5d SCFT (since, as we will see momentarily, no other compact divisor will be inflated in the subsequent blow-ups):
\begin{equation}
    r_{D_4} = 1,
\end{equation}with  the K\"ahler volume $\xi_1$ of \eqref{eq:auxP2description} being the real scalar parametrizing the Coulomb Branch.

Notice that \eqref{eq:auxP2description} intersects the horizontal divisor \eqref{eq:noncompacttrivalentP1bundle} along the locus $a = b_1 = \delta_1 = 0$, that is a specific representative  $\mathcal L$ of the hyperplane bundle of \eqref{eq:auxP2description}: 
\begin{equation}
\label{eq:specialhyperplanesec}
    \left\{a = b_1 = \delta_1 = 0\right\} \cong \mathbb P^1 \equiv \mathcal L \subset \mathbb P^2.
\end{equation}
Let us conclude the analysis of \eqref{eq:d4threefold1} by describing its residual singularities: we still see three non-compact lines of $A_1$ singularities: 
\begin{equation}
 \label{eq:residualA1liensD4}
   \delta_1 =  b_1 = u_1 = 0, \quad \delta_1 = b_1 = v_1 = 0, \quad \delta_1 = b_1 = u_1 + v_1 = 0,  
\end{equation}
that are associated with the r.h.s. of \eqref{eq:firstpartialresD4}. These three lines intersect the $\mathcal L \subset \mathbb P^2$ in the following three points: 
\begin{equation}
\label{eq:threepointsd4}
    a = b_1 = \delta_1 = u_1 = 0, \quad  a = b_1 = \delta_1 = v_1 = 0, \quad  a = b_1 = \delta_1 = u_1 + v_1 = 0. 
\end{equation}

\begin{itemize}
    \item \textbf{Second blow-up $\boldsymbol{\pi_2}$}
\end{itemize}
We now proceed by blowing up the residual lines in \eqref{eq:residualA1liensD4}. The most convenient way to do so is to blow-up the non-Cartier divisor 
\begin{equation}
    \label{eq:noncartierdiv}
    b_1 = \delta_1 = 0
\end{equation}
that contains all the three non-compact lines \eqref{eq:residualA1liensD4}: 
\begin{equation}
    \label{eq:d4secondblow-up}
 \pi_2:\quad   (a,b_1,u_1,v_1,\delta_1) =  (a, b_2 \delta_2, u_1, v_1, D_1 \delta_2), 
\end{equation}
where the ambient space coordinates $(a,b_2,u_1,v_1,D_2,\delta_2)$ belong to 
\begin{equation}
\label{eq:d4secondblambient}
\begin{array}{c|cccccc|c}
 & a & b_2 & u_1 & v_1 & D_1  &\delta_2 & \text{FI} \\
\hline
 \mathbb C^*_1&  0 & 1 & 1 & 1 & -1 & 0 & \xi_1  \\
  \mathbb C^*_2&  0 & 1 & 0 & 0 & 1 & -1 & \xi_2  \\
\end{array}
\end{equation}
The proper transform of \eqref{eq:d4threefold1} is
\begin{equation}
\label{eq:d4threefold2}
   T^{(2)}(a,b_2,u_1,v_1,D_1,\delta_2) = \frac{T^{(1)}(a,b_1,u_1,v_1,\delta_1)\rvert_{\eqref{eq:d4secondblow-up}}}{\delta_2} = a b_2^2 \delta_2 +  D_1 u_1 v_1 (u_1 + v_1).
\end{equation}
The proper transform of \eqref{eq:auxP2description} is  
\begin{equation}
\label{eq:propertrsecondP2}
   a = D_1 = 0, 
\end{equation}
and the proper transform of $\mathcal L$ is 
\begin{equation}
\label{eq:propertrL2}
   \mathcal L^{(2)} = \left\{a = D_1 = \delta_2 = 0 \right\}.
\end{equation}
The exceptional curves of the blow-up \eqref{eq:d4secondblow-up} are, for each $a \neq 0$, the preimages of the singular points \eqref{eq:residualA1liensD4}:
\begin{equation}
    \label{eq:exccurvesecond}
    \delta_2 = u_1 =0,  \quad \delta_2 = v_1 = 0, \quad \delta_2 = u_1  + v_1=  0.
\end{equation}
Imposing \eqref{eq:exccurvesecond}, \eqref{eq:d4threefold2} is automatically satisfied for each value of $a \neq 0$, and hence \eqref{eq:exccurvesecond} span three non-compact $\mathbb P^1$ bundles that complete the partial resolution of the $P_{D_4}$ line \eqref{eq:firstpartialresD4}: 
\begin{equation}
   \text{smooth} \to  A_1 \oplus A_1 \oplus A_1  \to D_4 .
\end{equation}
These three $\mathbb P^1$ bundles intersect \eqref{eq:propertrsecondP2}, \eqref{eq:propertrL2} in the following three points (rather than in curves): 
\begin{equation}
\label{eq:pointsandflopsoutsecondbl}
    \delta_2 = u_1 = a = D_1 = 0, \quad  \delta_2 = v_1 = a = D_1 = 0, \quad  \delta_2 = u_1 + v_1 = a = D_1 = 0; 
\end{equation}
This was indeed expected: we have blown-up a non-Cartier divisor that intersects the compact divisor \eqref{eq:auxP2description} exactly along $\mathcal L$; since $\mathcal L$ is a smooth codimension one locus of \eqref{eq:auxP2description},\footnote{The threefold can be singular along points of  $\mathcal L$, but \eqref{eq:auxP2description} is smooth as a surface.} the blow-up map \eqref{eq:d4secondblow-up} is an isomorphism once we restrict to the divisor in \eqref{eq:auxP2description}, and the proper transform \eqref{eq:propertrsecondP2} is again isomorphic to $\mathbb P^2$.

We conclude the analysis of \eqref{eq:d4threefold2} by noticing that its singular locus consists of the following three conifold points: 
\begin{equation}
\label{eq:leftoverconifolds}
    \delta_2 = u_1 = a = D_1 = 0, \quad  \delta_2 = v_1 = a = D_1 = 0, \quad  \delta_2 = u_1 + v_1 = a = D_1 = 0,
\end{equation}
which precisely coincide with \eqref{eq:pointsandflopsoutsecondbl}.

\begin{itemize}
    \item \textbf{Third blow-up $\boldsymbol{\pi_3}$}
\end{itemize}
To resolve the conifold points \eqref{eq:leftoverconifolds}, we blow-up the divisor \eqref{eq:propertrsecondP2}, that from the point of view of the threefold \eqref{eq:d4threefold2} is a non-Cartier divisor containing all the \eqref{eq:leftoverconifolds}.
The blow-up map is 
\begin{equation}
    \label{eq:d4thirdblow-up}
 \pi_3: \quad  (a,b_2,u_1,v_1,D_1,\delta_2) =  (A \delta_3 ,b_2,u_1,v_1,D_2 \delta_3,\delta_2), 
\end{equation}
with $(A,b_2,u_1,v_1,D_2,\delta_2, \delta_3)$ in the following ambient space: 
\begin{equation}
\label{eq:d4thirdblambient}
\begin{array}{c|ccccccc|c}
 & A & b_2 & u_1 & v_1 & D_2  &\delta_2 & \delta_3 &\text{FI} \\
\hline
 \mathbb C^*_1&  0 & 1 & 1 & 1 & -1 & 0 & 0& \xi_1  \\
  \mathbb C^*_2&  0 & 1 & 0 & 0 & 1 & -1 & 0& \xi_2  \\
   \mathbb C^*_3&  1 & 0 & 0 & 0 & 1 & 0 & -1& \xi_3  \\
\end{array} 
\end{equation}

The proper transform of \eqref{eq:d4threefold2} is
\begin{equation}
\label{eq:d4threefold3}
   T^{(3)}(A,b_2,u_1,v_1,D_2,\delta_2,\delta_3) = \frac{T^{(2)}(a,b_2,u_1,v_1,D_1,\delta_2)\rvert_{\eqref{eq:d4thirdblow-up}}}{\delta_3} = A b_2^2 \delta_2 + D_2 u_1 v_1 (u_1 + v_1), 
\end{equation}
the proper transform of \eqref{eq:propertrsecondP2} is 
\begin{equation}
\label{eq:propertrthirdP2}
    \delta_3 = T^{(3)}=0, 
\end{equation}
and the one of \eqref{eq:propertrL2} is \begin{equation}
\label{eq:propertrL3}
   \mathcal L^{(3)} \equiv \left\{ \delta_3 = T^{(3)}=  b_2 = 0\right\}.  
\end{equation}

The exceptional curves of the blow-up \eqref{eq:d4thirdblow-up} are the preimages of the three conifold points \eqref{eq:leftoverconifolds}, namely 
\begin{equation}
\label{eq:conifoldcurves}
    L_{u} = \left\{u_1 = \delta_2 = \delta_3 = 0\right\}, \quad   L_{v} = \left\{v_1 = \delta_2 = \delta_3 = 0\right\}, \quad   L_{u+v} = \left\{u_1 +v_1 = \delta_2 = \delta_3 = 0\right\}, 
\end{equation}
that belong to \eqref{eq:propertrthirdP2}, and intersect $\mathcal L^{(3)}$ in three points. This is not surprising, as the conifold points \eqref{eq:leftoverconifolds} were on \eqref{eq:propertrL2}.

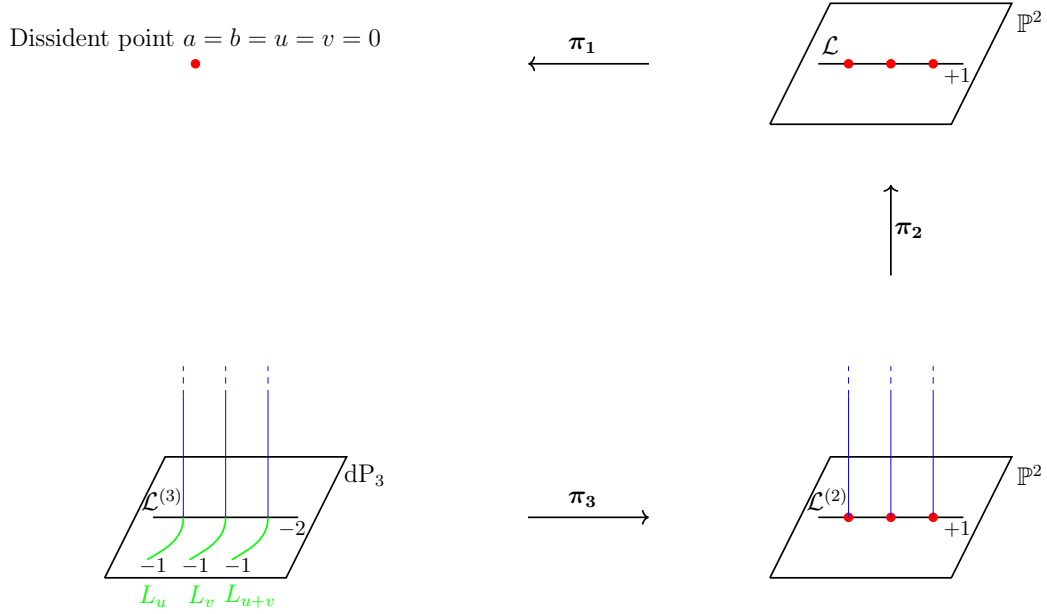
\begin{figure}[H]
    \centering
    \scalebox{0.8}{
    \begin{tikzpicture}
       \node at (0,0.4) {Dissident point $a = b = u = v = 0$};
       \draw[red,fill=red] (0,0) circle (0.07);
        \draw[<-,thick] (5.5,0)--(7.5,0);
        \node at (6.4,0.3) {$\boldsymbol{\pi_1}$};
       \draw[thick] (9.5,-1)--(12.5,-1)--(13.5,1)--(10.5,1)--(9.5,-1);
       \draw[thick] (10.3,0)--(12.7,0);
        \node at (10.5,0.3) {$\mathcal{L}$};
       \draw[red,fill=red] (10.8,0) circle (0.07);
       \draw[red,fill=red] (11.5,0) circle (0.07);
       \draw[red,fill=red] (12.2,0) circle (0.07);
       \node at (12.6,-0.2) {\footnotesize$+1$};
       \node at (13.8,0.7) {$\mathbb{P}^2$};
        \draw[<-,thick] (11.5,-2)--(11.5,-3.5);
        \node at (11.8,-2.75) {$\boldsymbol{\pi_2}$};
         \draw[thick] (9.5,-8.5)--(12.5,-8.5)--(13.5,-6.5)--(10.5,-6.5)--(9.5,-8.5);
         \node at (12.6,-7.7) {\footnotesize$+1$};
         \node at (13.8,-6.8) {$\mathbb{P}^2$};
         \draw[thick] (10.3,-7.5)--(12.7,-7.5);
          \node at (10.45,-7.2) {$\mathcal{L}^{(2)}$};
         \draw[red,fill=red] (10.8,-7.5) circle (0.07);
       \draw[red,fill=red] (11.5,-7.5) circle (0.07);
       \draw[red,fill=red] (12.2,-7.5) circle (0.07);
       \draw[blue] (10.8,-7.5)--(10.8,-5.5);
       \draw[blue] (11.5,-7.5)--(11.5,-5.5);
       \draw[blue] (12.2,-7.5)--(12.2,-5.5);
       \draw[dashed,blue] (10.8,-5.5)--(10.8,-5);
       \draw[dashed,blue] (11.5,-5.5)--(11.5,-5);
       \draw[dashed,blue] (12.2,-5.5)--(12.2,-5);
        \draw[<-,thick] (7.5,-7.5)--(5.5,-7.5);
        \node at (6.4,-7.2) {$\boldsymbol{\pi_3}$};
         \draw[thick] (-1.5,-8.5)--(1.5,-8.5)--(2.5,-6.5)--(-0.5,-6.5)--(-1.5,-8.5);
         \node at (1.6,-7.7) {\footnotesize$-2$};
         \node at (2.8,-6.8) {dP$_3$};
         \draw[thick] (-0.7,-7.5)--(1.7,-7.5);
          \node at (-0.55,-7.2) {$\mathcal{L}^{(3)}$};
       \draw[blue] (-0.2,-7.5)--(-0.2,-5.5);
       \draw[blue] (0.5,-7.5)--(0.5,-5.5);
       \draw[blue] (1.2,-7.5)--(1.2,-5.5);
       \draw[dashed,blue] (-0.2,-5.5)--(-0.2,-5);
       \draw[dashed,blue] (0.5,-5.5)--(0.5,-5);
       \draw[dashed,blue] (1.2,-5.5)--(1.2,-5);
        \node at (-0.7,-8.8) {\textcolor{green}{$L_u$}};
         \node at (0.1,-8.8) {\textcolor{green}{$L_v$}};
          \node at (0.9,-8.8) {\textcolor{green}{$L_{u+v}$}};
       \draw[thick,green] (-0.2,-7.5) to [out=-90,in=30] (-0.8,-8.2);
       \draw[thick,green] (0.5,-7.5)  to [out=-90,in=30] (-0.1,-8.2);
       \draw[thick,green] (1.2,-7.5) to [out=-90,in=30] (0.6,-8.2);
       \node at (-0.7,-8.3) {\footnotesize$-1$};
       \node at (0,-8.3) {\footnotesize$-1$};
       \node at (0.7,-8.3) {\footnotesize$-1$};
        \end{tikzpicture}}
    \caption{Pictorial representation of the crepant blow-up of \eqref{zero locus}, focusing on the pre-image of the origin. The red points are singular points.}
    \label{fig: D4 exceptional}
    \end{figure}

We pictorially represent the resolution procedure in Figure \ref{fig: D4 exceptional}.

\begin{itemize}
    \item \textbf{Summary of the blow-up sequence}
\end{itemize}
We can recap the results of the blow-up procedure as follows:
\begin{itemize}
    \item The first blow-up  \eqref{eq:d4firstblow-up} inflates a compact vertical divisor \eqref{eq:auxP2description}. Since the other blow-ups do not inflate other vertical divisors, M-theory on \eqref{eq:d4localmodel} is a 5d SCFT of rank-one. 
    \item The aforementioned exceptional divisor is a $\mathbb P^2$ after both the first \eqref{eq:d4firstblow-up} and the second \eqref{eq:d4secondblow-up} blow-ups. After the third blow-up \eqref{eq:d4thirdblow-up}, since the vertical divisor itself is the non-Cartier divisor of the small resolution of three conifold points \eqref{eq:leftoverconifolds}, it gets blown-up in three points and becomes a non-generic\footnote{Indeed, the three points that we have blown-up lie, on the same line \eqref{eq:specialhyperplanesec}, contained in \eqref{eq:auxP2description}.} $\text{dP}_3$. This means that the three new curves \eqref{eq:conifoldcurves} lie inside the vertical divisor, and are "flopped-in" by \eqref{eq:d4thirdblow-up}. 
    \item To compute $f_{\mathfrak g}$, and hence extract the rank of the flavor group \eqref{eq:flavorgrouplocalmodel}, we must carefully study the curves appearing in the resolution, to avoid double-counting in homology. 
    
    First, consider the fibers of the non-compact $\mathbb P^1$ bundles fibered over $a \neq 0$. These are the proper transforms under the various blow-ups of \eqref{eq:noncompacttrivalentP1bundle} and \eqref{eq:exccurvesecond}; they generate four homology classes in $H_2$, that account for the rank of the expected $D_4$ flavor factor in \eqref{eq:d4localmodel}. We call the subspace of $H_2$ generated by these four classes $H_{2,hor} \cong \mathbb Z^4$. These four non-compact divisors intersect the vertical divisor along $\mathcal L^{(3)}$, that hence has to be thought of as an element of $H_{2,hor}$. We will check this in the next step, computing the normal bundle of $\mathcal L^{(3)}$ and checking that it admits a flat direction, so that $\mathcal{L}^{(3)}$ can freely move away from the origin of \eqref{eq:d4localmodel}.

    \item We now focus on the curves $H_{2,vert}$, that are rigidly contained in the vertical divisor \eqref{eq:propertrthirdP2}. Stated differently, the elements of $H_{2,vert}$ are those curves that cannot be translated along the resolved non-compact $D_4$ line. Namely, their normal bundle does not admit flat directions. \eqref{eq:propertrthirdP2} contains five homologically independent curves. We have \eqref{eq:conifoldcurves}, with normal bundle $\mathcal O(-1) \oplus \mathcal O(-1)$, $\mathcal L^{(3)}$, with normal bundle $\mathcal O(-2) \oplus \mathcal O$ and the proper transform of the hyperplane section $\mathcal H$ of \eqref{eq:auxP2description} (with normal bundle $\mathcal O(1) \oplus O(-3)$).\footnote{We notice that, contrary to the cases studied in \cite{Collinucci:2022rii,DeMarco:2021try,Collinucci:2021ofd,DeMarco:2022dgh,Collinucci:2021wty}, we can not contract such curve without contracting the full vertical divisor, and hence we can not consider them to be the exceptional locus of a small resolution, generating higher-length flops.} The normal bundle of \eqref{eq:conifoldcurves} is $\mathcal O(-1) \oplus \mathcal O(-1)$ because the curves are contracted to conifold points. To compute the normal bundle of $\mathcal L^{(3)}$, we notice that it coincides with the hyperplane section of \eqref{eq:auxP2description} before performing the last two blow-ups, and hence its normal bundle within $\eqref{eq:auxP2description}$ is $\mathcal O(1)$. By applying the standard blow-up formula (see e.g. \cite{hartshorne2013algebraic}), since $\mathcal L^{(3)}$ is blown-up by \eqref{eq:d4thirdblow-up} in the three points \eqref{eq:leftoverconifolds}, its self-intersection within the compact vertical divisor decreases by three. Hence, $\mathcal L^{(3)}$ has normal bundle $\mathcal O(-2)$ within \eqref{eq:propertrthirdP2}, and, imposing the Calabi-Yau condition, $\mathcal O(-2) \oplus \mathcal O$ within the fully resolved threefold.
    
     We conclude that the "vertical" homology, that contracts on the origin of \eqref{eq:d4localmodel}, is 
    \begin{equation}
        \label{eq:verticalhom}
        H_{2,vert}= \langle \mathcal H, \mathcal L_u, \mathcal L_v, \mathcal L_{u+v} \rangle_{\mathbb Z} \cong \mathbb Z^4, 
    \end{equation}
    where we did not count, \textit{crucially}, $\mathcal L^{(3)}$ inside $H_{2,vert}$: it can be translated outside $a = 0$, as it is in the same homology class of the trivalent $\mathbb P^1$ resolving the $P_{D_4}$ line of \eqref{eq:d4localmodel}, and hence it is \textit{already} accounted for in $H_{2,hor}$.
\end{itemize}
Consequently, we have 
    \begin{equation}
        \text{rank}(\mathcal F) = \text{rank} \left(H_{2,vert}\right) + \text{rank} \left(H_{2,hor}\right) - \text{rank}(H_4) = 4 + 4 - 1 = 7 =   \text{rank}(D_4) + f_{D_4}, 
    \end{equation}
    and hence 
\begin{equation}
    f_{D_4} = 3.
\end{equation}

We can conclude the analysis of \eqref{eq:d4localmodel} noticing  that  $\mathcal L^{(3)}$ is a curve with normal bundle $\mathcal O(-2) \oplus \mathcal O(0)$ that lies at the intersection of two divisors of the threefold. This means that the curve cannot be flopped into a line with normal bundle $\mathcal O(-1) \oplus \mathcal O(-1)$, and hence M-theory on a threefold containing \eqref{eq:d4localmodel} as a local patch does not admit a weakly coupled gauge theory description. We can draw this conclusion by analyzing the local patch because the normal bundle of a $\mathbb P^1$ is a local object, and to determine it (and its possible flops) we can use just two charts that encompass the full $\mathbb P^1$.

We can use a similar procedure also to analyze other $\mathfrak g$ in \eqref{eq:localmodel}. We observe that each time that we blow-up a line of $\mathfrak g$, with $\mathfrak g \neq A_k$, we obtain a vertical divisor isomorphic to $\mathbb P^2$.\footnote{As in the $D_4$ case that we just analyzed, subsequent blow-ups can change this, producing del Pezzo's. This however does not change their contribution to $r_{\mathfrak g}$.} Furthermore, there will always be, after the complete resolution of the $\mathfrak g$ line of \eqref{eq:localmodel}, one of the vertical divisors containing the $\mathbb P^1$ associated to the trivalent node of the $\mathfrak g$ singularity. This means that, after the complete resolution of the $\mathfrak g$ line of \eqref{eq:localmodel}, such vertical divisor  will be a non-generic Del Pezzo, namely a $\mathbb P^2$ blown-up at three points that lie on the same line $\mathcal L_{\text{trivalent}}$,\footnote{In expression \eqref{eq:d4localmodel}, $\mathcal L_{trivalent} = \mathcal L^{(3)}$.} and hence M-theory on such threefolds does not admit a weakly coupled IIA phase.

Running such analysis for all the types of $\mathfrak g$ singularities, we obtain the following results for the rank of M-theory on \eqref{eq:localmodel}: 
\begin{eqnarray}
    r_{A_{j}} = 0,\quad r_{D_{2j}} = j-1, \quad r_{D_{2j+1}} = j-1, \quad r_{E_6} = 1, \quad r_{E_7} = 3, \quad r_{E_8} = 4.
\end{eqnarray}
For the flavor group data, we obtain instead that \textit{$f_{\mathfrak g}$ always coincides with the number of factors of $Q_{\mathfrak g}(u,v)$.} We have collected the results in Table \ref{flavor rank T}. The result can be understood as follows:  the analogue of \eqref{eq:d4threefold2} will display a number of conifold points equal to the number of factors of $Q_{\mathfrak g}(u,v)$. In all the cases, the (generic) hyperplane sections $\mathcal H_i$, with $i = 1,...,r_{\mathfrak g}$ will contribute to $\text{dim}(H_2(T_{\mathfrak{g}},\mathbb{Z}))$ in such a way to exactly cancel with $\text{dim}(H_4(T_{\mathfrak{g}},\mathbb{Z}))$ in \eqref{eq:flavorgrouplocalmodel}. Finally, we can always exhibit a flop chamber where the only intersection between the non-compact $\mathbb P^1$-bundles and the vertical divisor will be $\mathcal L_{trivalent}$, that in all the cases has to correctly be subtracted from the vertical homology.

\newpage

\bibliographystyle{at}
\bibliography{bibliography.bib}
\end{document}